\documentclass[nofootinbib,preprint,preprintnumbers,a4paper,11pt]{revtex4}
\usepackage{epsfig}
\usepackage{footnote}
\usepackage{ulem}
\usepackage{color}
\usepackage{array}
\usepackage{amssymb}
\usepackage{amsmath}
\usepackage{graphicx}
\usepackage{subfigure}
\usepackage{longtable}
\usepackage{verbatim}
\usepackage{amsfonts}
\usepackage{hyperref}
\usepackage{multirow}
\usepackage{cancel}

\begin{document}

\title{Isospin sum rules in charmed baryon weak decays}

\author{Jin-Feng Luo$^{1}$}

\author{Di Wang$^{1}$}\email{wangdi@hunnu.edu.cn}

\address{%
$^1$Department of Physics, Hunan Normal University, and Key Laboratory of Low-Dimensional Quantum Structures and Quantum Control of Ministry of Education, Changsha 410081, China}

\begin{abstract}
Isospin symmetry is the most precise flavor symmetry.
The effective Hamiltonian of charm quark weak decay is zero under the isospin lowering operators $I_-^n$, which permits us to generate isospin sum rules through several master formulas.
In this work, we derive the master formulas of isospin sum rules for the two- and three-body non-leptonic decays of singly and doubly charmed baryons.
Hundreds of isospin sum rules are derived to test of isospin symmetry and provide hints for the new decay modes.
The isospin sum rules for multi-body decays are not broken by the intermediate resonances and hence can be used to study the isospin partners of exotic hadrons.
\end{abstract}

\maketitle


\section{Introduction}

Charmed baryon decays provide laboratories to study strong and weak interactions in heavy-to-light baryonic transitions.
Many measurements for the singly and doubly charmed baryon decays were performed by LHCb \cite{LHCb:2023eeb,LHCb:2019epo,LHCb:2019ybf,LHCb:2019qed,LHCb:2021eaf,LHCb:2022rpd,Aaij:2018wzf,LHCb:2018pcs,Aaij:2017ueg,Aaij:2017nsd,Aaij:2017xva},
Belle (II) \cite{Belle:2021crz,Belle:2021dgc,Belle:2021btl,Berger:2018pli,Zupanc:2013iki,Yang:2015ytm,Pal:2017ypp},
and BESIII \cite{BESIII:2023vfi,BESIII:2023ooh,BESIII:2023rky,BESIII:2023iwu,BESIII:2022udq,BESIII:2022izy,BESIII:2022vrr,BESIII:2022xne,BESIII:2022onh,BESIII:2022aok,BESIII:2022bkj,BESIII:2021fqx,BESIII:2020kap,BESIII:2020cpu,BESIII:2019odb,BESIII:2018qyg,
Ablikim:2018jfs,Ablikim:2018woi,Ablikim:2018bir,Ablikim:2015prg,Ablikim:2015flg,Ablikim:2016tze,Ablikim:2016mcr,
Ablikim:2016vqd,Ablikim:2017ors,Ablikim:2017iqd} in the recent years.
Flavor symmetry, which has been widely studied in the literature \cite{Gronau:2018vei,Hsiao:2023mvw,Geng:2023pkr,Xing:2023bzh,Qin:2021zqx,Qin:2020zlg,He:2018joe,Wang:2022wrb,Hsiao:2020iwc,Zhao:2018mov,Groote:2021pxt,Han:2021azw,
Chau:1995gk,Kohara:1991ug,Geng:2019awr,He:2021qnc,Li:2021rfj,Geng:2020zgr,Geng:2018rse,Geng:2019bfz,Geng:2019xbo,
Hsiao:2019yur,Jia:2019zxi,Wang:2019dls,Savage:1989qr,Sheikholeslami:1991ab,
Sharma:1996sc,Verma:1995dk,Wang:2017azm,Shi:2017dto,Wang:2018utj,Lu:2016ogy,
Geng:2017esc,Geng:2017mxn,Wang:2017gxe,Geng:2018plk,Geng:2018bow}, is a powerful tool to analyze heavy baryon weak decays.
Isospin symmetry is the most precise flavor symmetry.
Isospin sum rules could be used to extract useful information about hadronic dynamics.
For instance, isospin sum rule of the $D\to \pi\pi$ system plays an important role in studying the penguin amplitude of $D$ meson decays \cite{Gavrilova:2023fzy,Wang:2022nbm}.
And two isospin sum rules of the $\Lambda_c\to \Delta K$ system can help us to reduce the hadronic $SU(3)$ breaking effects in analyzing CP asymmetry in the chain decay of $\Lambda_c^+\to \Delta^+ K(t)(\to \pi^+\pi^-)$ \cite{Wang:2022cfs}.

The isospin sum rules are usually derived by writing the isospin amplitudes and combining several modes to form a polygon in the complex plane.
Inspired by Refs.~\cite{Grossman:2012ry,Wang:2022kwe}, we propose a simple approach to generate isospin sum rules for the $c$- and $b$-hadron decays without relying on isospin amplitudes \cite{Wang:2023pnb}.
The effective Hamiltonian of heavy hadron weak decay is zero under a series of isospin lowering operators $I_-^n$, which allows us to generate isospin sum rules from several master formulas.
In this work, we apply the approach to the charmed baryon decays.
The master formulas of isospin sum rules for the two- and three-body non-leptonic decays of singly and doubly charmed baryon are derived.
Many isospin sum rules are derived, which could provide hints for exploration of new decay modes and serve as examination of the isospin symmetry.
It is found the isospin sum rules for multi-body decays are not broken by intermediate resonances, and those ones with two particles within one isospin multiplet cannot be tested by the total branching fractions without partial wave analysis.

The rest of this paper is structured as follows.
The theoretical framework of deriving isospin sum rules is presented in Sec. \ref{tf}.
The master formulas of isospin sum rules for the non-leptonic decays of singly and doubly charmed baryons are derived in Sec. \ref{sumrule}.
Phenomenological analysis for the isospin sum rules is shown in  Sec. \ref{pa}.
And Sec. \ref{summary} is a brief summary.
The isospin sum rules of singly and doubly charmed baryon decays derived by the master formulas are listed in Appendices \ref{singlyx} and \ref{doublyx} respectively.

\section{Theoretical framework}\label{tf}

Taking the two-body $D$ meson decays as examples, the basic idea of generating isospin sum rules by $I_-^n$ is shown as follows.
In the $SU(3)$ picture, the effective Hamiltonian of charm decay can be written as \cite{Wang:2020gmn}
\begin{equation}\label{h}
 \mathcal H_{\rm eff}=\sum_{i,j,k=1}^3 H^{ij}_{k}O^{ij}_{k},
\end{equation}
where $O^{ij}_{k}$ denotes the four-quark operator and $H$ is a $3\times 3\times 3$ coefficient matrix.
The initial and final states of charm decay can be written as
\begin{align}
  |M_\alpha\rangle = (M_\alpha)^{i}_{j}|M^{i}_{j} \rangle,
\end{align}
where $|M^{i}_{j} \rangle$ is the quark composition of the meson state, $|M^{i}_{j} \rangle = |q_i\bar q_j\rangle$, and $(M_\alpha)$ is the coefficient matrix.
For instance, the $|M_{\pi^0}\rangle$ state is written as
\begin{align}
|M_{\pi^0}\rangle = (M_{\pi^0})^i_j|M^{i}_{j} \rangle,
\qquad \text {and} \qquad (M_{\pi^0}) = \left( \begin{array}{ccc}
  \frac{1}{\sqrt{2}} & 0 & 0 \\
  0 & -\frac{1}{\sqrt{2}} & 0 \\
0 & 0 & 0 \\
 \end{array}
\right).
\end{align}
In this way, the information of initial/final state is absorbed into the coefficient matrix $(M_\alpha)$ and $|M^{i}_{j} \rangle$ is universal under the flavor symmetry.
Decay amplitude of the $D_\gamma\to M_\alpha M_\beta$ mode is constructed as
\begin{align}\label{amp}
\mathcal{A}(D_\gamma\to M_\alpha M_\beta)& = \langle M_\alpha M_\beta |\mathcal{H}_{\rm eff}| D_\gamma\rangle\nonumber\\&~=\sum_{\omega}\,(M_\alpha)^n_m\langle M^n_m|(M_\beta)^s_r\langle M^s_r||H^{jk}_lO^{jk}_l||(D_\gamma)_i|D_i\rangle\nonumber\\& ~~=\sum_\omega\,\langle M_{m}^{n} M^s_r |O^{jk}_{l} |D_{i}\rangle \times (M_\alpha)_{m}^{n}(M_\beta)_{r}^{s} H^{jk}_{l}(D_\gamma)_i\nonumber\\& ~~~= \sum_\omega X_{\omega}(C_\omega)_{\alpha\beta\gamma},
\end{align}
where $X_\omega = \langle M_{m}^{n} M^s_r |O^{jk}_{l} |D_{i}\rangle$ is the reduced matrix element and $(C_\omega)_{\alpha\beta\gamma}=(M_\alpha)_{m}^{n}(M_\beta)_{r}^{s} H^{jk}_{l}(D_\gamma)_i$ is the Clebsch-Gordan (CG) coefficient \cite{Eckart30,Wigner59}.

The isospin lowering operator $I_-$ is
\begin{eqnarray}
 I_-=  \left( \begin{array}{ccc}
   0   & 0  & 0 \\
     1 &   0  & 0 \\
    0 & 0 & 0 \\
  \end{array}\right).
\end{eqnarray}
If the effective Hamiltonian \eqref{h} is fully invariant under $I_-^n$, i.e., $I_-^n\,H=0$, it follows that
\begin{equation}\label{rule}
  \langle M_\alpha M_\beta |I_-^n\,\mathcal{H}_{\rm eff}| D_\gamma\rangle = \sum_\omega\,\langle M_{m}^{n} M^s_r |O^{jk}_{l} |D_{i}\rangle \times (M_\alpha)_{m}^{n}(M_\beta)_{r}^{s} (I_-^n\,H)^{jk}_{l}(D_\gamma)_i = 0.
\end{equation}
One can apply $I_-^n$ to the initial/final states and expand the results using the initial/final states as bases. 
The LHS of Eq.~\eqref{rule} is then expressed as a sum of several decay amplitudes and the RHS of Eq.~\eqref{rule} remains zero.
The sum of decay amplitudes generated by $I_-$ for the two-body decays of $D$ mesons is written as \cite{Wang:2023pnb}
\begin{align}\label{rulex}
{ SumI_-}\,[\gamma, \alpha,\beta]= \sum_\mu\left[\{[I_-]_{M_8}\}_\alpha^\mu \mathcal{A}_{ \gamma \to \mu \beta} +  \{[I_-]_{M_8}\}_\beta^\mu \mathcal{A}_{\gamma\to \alpha\mu } + \{[I_-]_{D}\}_\gamma^\mu \mathcal{A}_{\mu\to \alpha \beta }\right],
\end{align}
in which $[I_-]_{M_8}$ and $[I_-]_{D}$ are the coefficient matrices found in Ref.~\cite{Wang:2023pnb}, and $\mathcal{A}_{ \gamma \to \mu \beta}$, $\mathcal{A}_{\gamma\to \alpha\mu }$, $\mathcal{A}_{\mu\to \alpha \beta }$ are the decay amplitudes of $D_\gamma\to M_\mu M_\beta$, $D_\gamma\to M_\alpha M_\mu$, $D_\mu\to M_\alpha M_\beta$ respectively.
It was found in Ref.~\cite{Wang:2023pnb} that the effective Hamiltonian of the Cabibbo-favored (CF) transition is zero under $I_-^n$ with $n\geq 1$.
And the effective Hamiltonian of the singly Cabibbo-suppressed (SCS) and doubly Cabibbo-suppressed (DCS) transitions is zero under $I_-^n$ with $n\geq 2$.
Thus, we can use Eq.~\eqref{rulex} to generate isospin sum rules for the CF decays directly.
For instance, choosing $\{\gamma, \alpha, \beta\} = \{D^0, \pi^+,\overline K^0\}$ in Eq.~\eqref{rulex} yields the isospin sum rule of the $D\to K\pi$ system,
\begin{align}
{ SumI_-}\,[D^0, \pi^+,\overline K^0]=-\sqrt{2}\,\mathcal{A}(D^0\to \pi^0\overline K^0)-\mathcal{A}(D^0\to \pi^+K^-)+\mathcal{A}(D^+\to \pi^+\overline K^0)=0.
\end{align}
Isospin sum rules for the SCS and DCS transitions are generated by operating Eq.~\eqref{rulex} at least twice.
For example, choosing $\{\gamma, \alpha, \beta\} = \{D^0, \pi^+,\pi^+\}$ yields the isospin sum rule of the $D\to \pi\pi$ system,
\begin{align}
{ SumI_-^2}\,[D^0, \pi^+,\pi^+]&=-\sqrt{2}\,{ SumI_-}\,[D^0, \pi^+,\pi^0]-\sqrt{2}\,{ SumI_-}\,[D^0, \pi^0,\pi^+]+{ SumI_-}\,[D^+, \pi^+,\pi^+]\nonumber\\&=4\,\big[ \mathcal{A}(D^0\to \pi^0 \pi^0)-\mathcal{A}(D^0\to \pi^+\pi^-)-\sqrt{2}\,\mathcal{A}(D^+\to \pi^+ \pi^0)\big]=0.
\end{align}
It is stressed that the equations $I_-\,H_{\rm CF}=0$, $I_-^2\,H_{\rm SCS}=0$ and $I_-^2\,H_{\rm DCS}=0$ are derived without approximating the CKM matrix \cite{Wang:2023pnb}.
So the isospin sum rules generated through Eq.~\eqref{rulex} do not rely on any approximation of the CKM matrix.
In the rest of this paper, we extend this approach to the two- and three-body non-leptonic decays of singly and doubly charmed baryons.
The advantage of our method is that it avoids the Wigner-Eckart invariants.
It provides a simple and systematic method for generating isospin sum rules.

\section{Isospin sum rules of charmed baryon decays}\label{sumrule}

\subsection{Singly charmed baryon decays}\label{singly}

To obtain isospin sum rules for the $\mathcal{B}_{c\overline 3}\to M_8 \mathcal{B}_8$ decays, we apply the isospin lowering operator $I_-$ to the initial and final states.
The charmed baryon anti-triplet is
\begin{eqnarray}
 [\mathcal{B}_{c\overline 3}]=  \left( \begin{array}{ccc}
   0   & \Lambda_c^+  & \Xi_c^+ \\
    -\Lambda_c^+ &   0   & \Xi_c^0 \\
    -\Xi_c^+ & -\Xi_c^0 & 0 \\
  \end{array}\right),
\end{eqnarray}
which can be expressed via the Levi-Civita tensor as
\begin{eqnarray}
[\mathcal{B}_{c\overline 3}]_{ij}=\epsilon_{ijk}[\mathcal{B}_{c\overline 3}]^{k}\qquad {\rm with}\qquad [\mathcal{B}_{c\overline 3}]^{k}=\left( \begin{array}{ccc}
     \Xi_c^0 \\
    -\Xi_c^+  \\
    \Lambda_c^+ \\
  \end{array}\right).
\end{eqnarray}
Under the isospin lowering operator $I_-$, the charmed baryon anti-triplet is transformed as
\begin{align}\label{tc3}
&I_-|[\mathcal{B}_{c\overline 3}]_\alpha\rangle = I_-\cdot |[\mathcal{B}_{ c\overline3}]_\alpha\rangle= \sum_\beta([I_-]_{\mathcal{B}_{ c\overline 3}})^{\beta}_{\alpha}|[\mathcal{B}_{c\overline 3}]_\beta\rangle.
\end{align}
The matrix element $([I_-]_{\mathcal{B}_{c\overline 3}})^{\beta}_{\alpha}$ is coefficient of
$I_-|[\mathcal{B}_{c\overline 3}]_\alpha\rangle$ with $|[\mathcal{B}_{ c\overline 3}]_\beta\rangle$ as bases.
If we define the charmed baryon anti-triplet bases as
 $|[\mathcal{B}_{c\overline 3}]_\beta\rangle = ( |\Xi^0_c\rangle,\,\, |\Xi^+_c\rangle ,\,\, |\Lambda^+_c\rangle )$,
 the coefficient matrix $[I_-]_{\mathcal{B}_{c\overline 3}}$ is
\begin{eqnarray}
 [I_-]_{\mathcal{B}_{c\overline 3}}= \left( \begin{array}{ccc}
   0   & 0  & 0 \\
     -1 &  0  & 0 \\
    0 & 0 & 0 \\
  \end{array}\right).
\end{eqnarray}
The pseudoscalar meson octet is
\begin{eqnarray}
 [M_8]=  \left( \begin{array}{ccc}
   \frac{1}{\sqrt 2} \pi^0+  \frac{1}{\sqrt 6} \eta_8    & \pi^+  & K^+ \\
    \pi^- &   - \frac{1}{\sqrt 2} \pi^0+ \frac{1}{\sqrt 6} \eta_8   & K^0 \\
    K^- & \overline K^0 & -\sqrt{2/3}\eta_8 \\
  \end{array}\right).
\end{eqnarray}
The isospin lowering operator $I_-$ acting on a pseudoscalar meson octet is a commutator,
\begin{align}\label{m}
 I_- \langle [M_8]_\alpha|&=[I_-, \langle [M_8]_\alpha|] =I_-\cdot \langle[M_8]_\alpha| - \langle[M_8]_\alpha| \cdot I_-\nonumber\\&~~~= \sum_\beta{\rm Tr}\{[\,I_-,[M_8]_\alpha\,]\cdot [M_8]_\beta^T\}\langle [M_8]_\beta | = \sum_\beta([I_-]_{M_8})^\beta_\alpha \langle [M_8]_\beta |.
\end{align}
If we define the pseudoscalar meson octet bases as
\begin{align}
 \langle [M_8]_\beta| = ( \langle \pi^+|,\,\,\langle \pi^0|,\,\,\langle \pi^-|,\,\,\langle K^+|,\,\,\langle K^0|,\,\,\langle \overline K^0|,\,\,\langle K^-|,\,\,\langle \eta_8|    ),
\end{align}
$[I_-]_{M_8}$ is derived to be
\begin{eqnarray}
 [I_-]_{M_8}= \left( \begin{array}{cccccccc}
  0 & 0& 0& 0& 0& 0& 0& 0 \\
  -\sqrt{2}& 0& 0& 0& 0& 0& 0& 0 \\
 0& \sqrt{2}& 0& 0& 0& 0& 0& 0 \\
  0& 0& 0& 0& 0& 0& 0& 0 \\
  0& 0& 0& 1& 0& 0& 0& 0\\
 0& 0& 0& 0& 0& 0& 0& 0\\
 0& 0& 0& 0& 0& -1& 0& 0 \\
 0& 0& 0& 0 & 0&0& 0& 0 \\
  \end{array}\right).
\end{eqnarray}
The light baryon octet is
\begin{eqnarray}
 [\mathcal{B}_8]=  \left( \begin{array}{ccc}
   \frac{1}{\sqrt 2} \Sigma^0+  \frac{1}{\sqrt 6} \Lambda^0    & \Sigma^+  & p \\
    \Sigma^- &   - \frac{1}{\sqrt 2} \Sigma^0+ \frac{1}{\sqrt 6} \Lambda^0   & n \\
    \Xi^- & \Xi^0 & -\sqrt{2/3}\Lambda^0 \\
  \end{array}\right).
\end{eqnarray}
Matrix $ [I_-]_{\mathcal{B}_8}$ is the same as $[I_-]_{M_8}$ if the light baryon octet bases are defined by
\begin{align}
 \langle [\mathcal{B}_8]_\beta| = ( \langle \Sigma^+|,\,\,\langle \Sigma^0|,\,\,\langle \Sigma^-|,\,\,\langle p|,\,\,\langle n|,\,\,\langle \Xi^0|,\,\,\langle \Xi^-|,\,\,\langle \Lambda^0|).
\end{align}

With the matrix $[I_-]_{M_8}$, $[I_-]_{\mathcal{B}_8}$ and $[I_-]_{\mathcal{B}_{c\overline 3}}$, the sum of decay amplitudes generated by the isospin lowering operator $I_-$ is
\begin{align}\label{rule1}
{ SumI_-}\,[\gamma, \alpha,\beta]=   \sum_\mu\left[([I_-]_{M_8})_\alpha^\mu \mathcal{A}_{ \gamma \to \mu \beta} +  ([I_-]_{\mathcal{B}_8})_\beta^\mu \mathcal{A}_{\gamma\to \alpha\mu } + ([I_-]_{\mathcal{B}_{ c\overline 3}})_\gamma^\mu \mathcal{A}_{\mu\to \alpha \beta }\right].
\end{align}
Isospin sum rules of the $\mathcal{B}_{c\overline 3}\to M_8 \mathcal{B}_8$ decays can be derived by Eq.~\eqref{rule1} if appropriate $\alpha$, $\beta$, $\gamma$ and the number of operation $n$ are selected.
The isospin sum rules for the $\mathcal{B}_{c\overline 3}\to M_8 \mathcal{B}_8$ modes are shown in Appendix \ref{singly1}.

The light baryon decuplet is given by
{\small \begin{align}\label{b10}
[\mathcal{B}_{10}] = \left(\left( \begin{array}{ccc}
      \Delta^{++} &  \frac{1}{\sqrt{3}}\Delta^{+}  & \frac{1}{\sqrt{3}}\Sigma^{*+} \\
   \frac{1}{\sqrt{3}}\Delta^{+} &  \frac{1}{\sqrt{3}}\Delta^{0}  & \frac{1}{\sqrt{6}}\Sigma^{*0} \\
    \frac{1}{\sqrt{3}}\Sigma^{*+} & \frac{1}{\sqrt{6}}\Sigma^{*0} & \frac{1}{\sqrt{3}}\Xi^{*0} \\
  \end{array}\right)\left( \begin{array}{ccc}\frac{1}{\sqrt{3}}\Delta^{+} &  \frac{1}{\sqrt{3}}\Delta^{0}  & \frac{1}{\sqrt{6}}\Sigma^{*0} \\
   \frac{1}{\sqrt{3}}\Delta^{0} & \Delta^{-}  & \frac{1}{\sqrt{3}}\Sigma^{*-} \\
    \frac{1}{\sqrt{6}}\Sigma^{*0} & \frac{1}{\sqrt{3}}\Sigma^{*-} & \frac{1}{\sqrt{3}}\Xi^{*-}\\\end{array}\right)
    \left(\begin{array}{ccc}\frac{1}{\sqrt{3}}\Sigma^{*+} &  \frac{1}{\sqrt{6}}\Sigma^{*0}  & \frac{1}{\sqrt{3}}\Xi^{*0} \\
   \frac{1}{\sqrt{6}}\Sigma^{*0} & \frac{1}{\sqrt{3}}\Sigma^{*-}  & \frac{1}{\sqrt{3}}\Xi^{*-} \\
    \frac{1}{\sqrt{3}}\Xi^{*0} & \frac{1}{\sqrt{3}}\Xi^{*-} & \Omega^{-}\\\end{array}\right)\right).
\end{align}}
There are three matrices in Eq.~\eqref{b10}, which are labeled as $X_1$, $X_2$ and $X_3$.
The light baryon decuplet can be written as
\begin{align}\label{cb10}
|[X_a]_\alpha\rangle = ([X_a]_\alpha)^{i}_j |[X_a]^{i}_j\rangle.
\end{align}
Both the indices $i$ and $j$ in Eq.~\eqref{cb10} serve as the bases of basic representation of $SU(3)$ group.
The charmed baryon decuplet is transformed under $I_-$ as
\begin{align}
 I_- \langle [X_a]_\alpha|=I_-\cdot\langle [X_a]_\alpha| = \sum_\beta([I_-]_{X_a})^\beta_\alpha \langle [X_a]_\beta |.
\end{align}
If we define the light baryon decuplet bases as
\begin{align}
 \langle [X_1]_\beta| & = ( \langle \Delta^{++}|,\,\,\langle \Delta^0|,\,\,\langle \Xi^{*0}|,\,\,\langle \Delta^+|,\,\,\langle \Sigma^{*+}|,\,\,\langle \Sigma^{*0}| ),\\
  \langle [X_2]_\beta| & = ( \langle \Delta^{+}|,\,\,\langle \Delta^-|,\,\,\langle \Xi^{*-}|,\,\,\langle \Delta^0|,\,\,\langle \Sigma^{*0}|,\,\,\langle \Sigma^{*-}| ),\\
  \langle [X_3]_\beta| & = ( \langle \Sigma^{*+}|,\,\,\langle \Sigma^{*-}|,\,\,\langle \Omega^{-}|,\,\,\langle \Sigma^{*0}|,\,\,\langle \Xi^{*0}|,\,\,\langle \Xi^{*-}| ),
\end{align}
the coefficient matrices $[I_-]_{X_a}$ are derived to be
\begin{eqnarray}\label{x1}
 [I_-]_{X_1}= \left( \begin{array}{cccccc}
  0 & 0& 0& 0& 0& 0 \\
  0 & 0& 0& \frac{1}{3}& 0& 0 \\
  0 & 0& 0& 0& 0& 0 \\
  \frac{1}{\sqrt{3}} & 0& 0& 0& 0& 0 \\
  0 & 0& 0& 0& 0& 0 \\
  0 & 0& 0& 0& \frac{1}{3\sqrt{2}}& 0 \\
  \end{array}\right),
\end{eqnarray}
\begin{eqnarray}\label{x2}
 [I_-]_{X_2}= \left( \begin{array}{cccccc}
  0 & 0& 0& 0& 0& 0 \\
  0 & 0& 0& \frac{1}{\sqrt{3}}& 0& 0 \\
  0 & 0& 0& 0& 0& 0 \\
  \frac{1}{3} & 0& 0& 0& 0& 0 \\
  0 & 0& 0& 0& 0& 0 \\
  0 & 0& 0&0& \frac{1}{3\sqrt{2}}& 0 \\
  \end{array}\right),
\end{eqnarray}
\begin{eqnarray}\label{x3}
 [I_-]_{X_3}= \left( \begin{array}{cccccc}
  0 & 0& 0& 0& 0& 0 \\
  0 & 0& 0& \frac{1}{3\sqrt{2}}& 0& 0 \\
  0 & 0& 0& 0& 0& 0 \\
  \frac{1}{3\sqrt{2}} & 0& 0& 0& 0& 0 \\
  0 & 0& 0& 0& 0& 0 \\
  0 & 0& 0& 0& \frac{1}{3}& 0 \\
  \end{array}\right).
\end{eqnarray}
If the light baryon decuplet bases are defined as
\begin{align}
 \langle [\mathcal{B}_{10}]_\beta| = ( \langle \Delta^{++}|,\,\,\langle \Delta^+|,\,\,\langle \Delta^0|,\,\,\langle \Delta^-|,\,\,\langle \Sigma^{*+}|,\,\,\langle \Sigma^{*0}|,\,\,\langle \Sigma^{*-}|,\,\,\langle \Xi^{*0}|,\,\,\langle \Xi^{*-}|,\,\,\langle \Omega^{-}|),
\end{align}
the coefficient matrix $[I_-]_{\mathcal{B}_{10}}$, which can be obtained by summing Eqs.~\eqref{x1}$\sim$\eqref{x3}, is
\begin{eqnarray}
 [I_-]_{\mathcal{B}_{10}}= \left( \begin{array}{cccccccccc}
  0 & 0& 0& 0& 0& 0 &0 &0 &0&0\\
   \frac{1}{\sqrt{3}}& 0& 0& 0& 0& 0& 0&0&0&0\\
  0 & \frac{2}{3}& 0&0& 0& 0& 0&0&0&0 \\
 0&0& \frac{1}{\sqrt{3}} & 0&0& 0& 0& 0& 0& 0 \\
 0& 0& 0& 0&0 & 0& 0& 0& 0& 0 \\
  0 & 0& 0& 0& \frac{\sqrt{2}}{3} & 0& 0&0& 0&0 \\
  0& 0& 0& 0&0 & \frac{\sqrt{2}}{3}&0&  0& 0&0\\0& 0& 0&0& 0& 0&0& 0& 0&0\\0& 0& 0&0& 0& 0& 0&\frac{1}{3}&0& 0\\ 0 & 0& 0& 0& 0& 0 &0 &0 &0&0
  \end{array}\right).
\end{eqnarray}
The sum of decay amplitudes generated by the isospin lowering operator $I_-$ for the $\mathcal{B}_{c\overline 3}\to M_8\mathcal{B}_{10}$ modes is
\begin{align}\label{rule2}
{ SumI_-}\,[\gamma, \alpha,\beta]=  \sum_\mu\left[([I_-]_{M_8})_\alpha^\mu \mathcal{A}_{\gamma\to \mu \beta } +  3\,([I_-]_{\mathcal{B}_{10}})_\beta^\mu \mathcal{A}_{\gamma\to \alpha\mu }  + ([I_-]_{\mathcal{B}_{ c\overline 3}})_\gamma^\mu \mathcal{A}_{\mu\to \alpha \beta }\right].
\end{align}
The factor $3$ in the second term is arisen from the fact that $I_-$ acts on the first, second and third quarks in the light baryon decuplet getting the same coefficient matrices.
The isospin sum rules for the $\mathcal{B}_{c\overline 3}\to M_8 \mathcal{B}_{10}$ modes are shown in Appendix \ref{singly2}.

Above discussions can be extended to the three-body decays.
The sum of decay amplitudes generated by the isospin lowering operator $I_-$ for the $\mathcal{B}_{c\overline{3}} \to \mathcal{B}_{8}{M_8}{M_8}$ is
\begin{align}\label{rule11}\nonumber
{ SumI_-}\,[\gamma,\alpha,\beta, \delta]= \sum_\mu\left[([I_-]_{\mathcal{B}_8})_\alpha^\mu \mathcal{A}_{\gamma\to \mu \beta\delta } + ([I_-]_{M_8})_\delta^\mu \mathcal{A}_{\gamma\to \alpha \beta\mu } \right.\\\left.~~+([I_-]_{M_8})_\beta^\mu \mathcal{A}_{\gamma\to \alpha \mu\delta }+([I_-]_{\mathcal{B}_{c\overline{3}}})_\gamma^\mu \mathcal{A}_{\mu\to\alpha \beta\delta }\right].
\end{align}
The isospin sum rules for the $\mathcal{B}_{c\overline{3}} \to \mathcal{B}_{8}{M_8}{M_8}$ modes are shown in Appendix \ref{singly3}.
The sum of decay amplitudes generated by $I_-$ for the $\mathcal{B}_{c\overline{3}} \to \mathcal{B}_{10}{M_8}{M_8}$ is
\begin{align}\label{rule12}\nonumber
{ SumI_-}\,[\gamma,\alpha,\beta, \delta]= \sum_\mu\left[3\,([I_-]_{\mathcal{B}_{10}})_\alpha^\mu \mathcal{A}_{\gamma\to \mu \beta\delta } + ([I_-]_{M_8})_\delta^\mu \mathcal{A}_{\gamma\to \alpha \beta\mu }\right.\\\left.~~+([I_-]_{M_8})_\beta^\mu \mathcal{A}_{\gamma\to \alpha \mu\delta } +([I_-]_{\mathcal{B}_{c\overline{3}}})_\gamma^\mu \mathcal{A}_{\mu\to\alpha \beta\delta }\right].
\end{align}
The isospin sum rules for the $\mathcal{B}_{c\overline{3}} \to \mathcal{B}_{10}{M_8}{M_8}$ modes are shown in Appendix \ref{singly4}.

\subsection{Doubly charmed baryon decays}\label{doubly}

The doubly charmed baryon triplet can be written as
$|[\mathcal{B}_{cc}]_\alpha\rangle = ([\mathcal{B}_{cc}]_\alpha)^i |[\mathcal{B}_{cc}]^{i}\rangle$, in which the light quarks serve bases of basic representations of $SU(3)$ group.
A basic representation in the final state is equivalent to a conjugate representation in the initial state.
So the isospin lowering operator $I_-$ acting on $|[\mathcal{B}_{cc}]_\alpha\rangle$ is written as
\begin{align}
 I_- |[\mathcal{B}_{cc}]_\alpha\rangle
 =|[\mathcal{B}_{cc}]_\alpha\rangle \cdot I_- = \sum_\beta([I_-]_{\mathcal{B}_{cc}})^\beta_\alpha |[\mathcal{B}_{cc}]_\beta\rangle.
\end{align}
If we define the doubly charmed baryon triplet bases as
 $|[\mathcal{B}_{cc}]_\beta\rangle = ( |\Xi_{cc}^{++}\rangle,\,\, |\Xi_{cc}^{+}\rangle ,\,\, |\Omega_{cc}^{+}\rangle )$,
the coefficient matrix $[I_-]_{\mathcal{B}_{cc}}$ is
\begin{eqnarray}
 [I_-]_{\mathcal{B}_{cc}}= \left( \begin{array}{ccc}
   0   & 1  & 0 \\
     0 &  0  & 0 \\
    0 & 0 & 0 \\
  \end{array}\right).
\end{eqnarray}
The sum of decay amplitudes generated by the isospin lowering operator $I_-$ for the $\mathcal{B}_{cc}\to M_8\mathcal{B}_{\overline c3}$ modes is
\begin{align}\label{rule3}
{ SumI_-}\,[\gamma, \alpha,\beta]=  \sum_\mu\left[([I_-]_{M_8})_\alpha^\mu \mathcal{A}_{\gamma\to \mu \beta } -  ([I_-]^T_{\mathcal{B}_{ c\overline 3}})_\beta^\mu \mathcal{A}_{\gamma\to \alpha\mu } -([I_-]_{\mathcal{B}_{cc}})_\gamma^\mu \mathcal{A}_{\mu\to\alpha \beta }\right].
\end{align}
The transposition in the second term is arisen from the initial-final transformation for the charmed baryon anti-triplet.
The minus signs in the last two terms are used to match the minus sign of commutator in Eq.~\eqref{m}.
The isospin sum rules for the $\mathcal{B}_{cc}\to M_8\mathcal{B}_{\overline c3}$  modes are shown in Appendix \ref{doubly1}.

The charmed baryon sextet is
\begin{eqnarray}
 [\mathcal{B}_{c6}]=  \left( \begin{array}{ccc}
   \Sigma_c^{++}   &  \frac{1}{\sqrt{2}}\Sigma_c^{+}  & \frac{1}{\sqrt{2}}\Xi_c^{*+} \\
   \frac{1}{\sqrt{2}}\Sigma_c^{+} &   \Sigma_c^{0}   & \frac{1}{\sqrt{2}}\Xi_c^{*0} \\
    \frac{1}{\sqrt{2}}\Xi_c^{*+} & \frac{1}{\sqrt{2}}\Xi_c^{*0} & \Omega_c^0 \\
  \end{array}\right).
\end{eqnarray}
The isospin lowering operator $I_-$ acting on the charmed baryon sextet is written as
\begin{align}
 I_- \langle [\mathcal{B}_{c6}]_\alpha|=I_-\cdot\langle [\mathcal{B}_{c6}]_\alpha| = \sum_\beta([I_-]_{\mathcal{B}_{c6}})^\beta_\alpha \langle [\mathcal{B}_{c6}]_\beta |.
\end{align}
If we define the charmed baryon sextet bases as
 $\langle[\mathcal{B}_{c6}]_\beta |= ( \langle\Sigma_{c}^{++}|,\,\, \langle\Sigma_{c}^{0}|,\,\, \langle\Omega_{c}^{0}|, \,\, \langle\Sigma_{c}^{+}|,\,\, \langle\Xi_{c}^{*+}|,\,\, \langle\Xi_{c}^{*0}| )$,
the coefficient matrix $[I_-]_{\mathcal{B}_{c6}}$ is derived to be
\begin{eqnarray}
 [I_-]_{\mathcal{B}_{c6}}= \left( \begin{array}{cccccc}
  0 & 0& 0& 0& 0& 0 \\
  0 & 0& 0& \frac{1}{\sqrt{2}}& 0& 0 \\
  0 & 0& 0& 0& 0& 0 \\
  \frac{1}{\sqrt{2}} & 0& 0& 0& 0& 0 \\
  0 & 0& 0& 0& 0& 0 \\
  0 & 0& 0& 0& \frac{1}{2}& 0 \\
  \end{array}\right).
\end{eqnarray}
The sum of decay amplitudes generated by $I_-$ for the $\mathcal{B}_{cc}\to M_8\mathcal{B}_{c6}$ modes is
\begin{align}\label{rule4}
{ SumI_-}\,[\gamma, \alpha,\beta]=  \sum_\mu\left[([I_-]_{M_8})_\alpha^\mu \mathcal{A}_{\gamma\to \mu \beta } +  2\,([I_-]_{\mathcal{B}_{c6}})_\beta^\mu \mathcal{A}_{\gamma\to \alpha\mu } - ([I_-]_{\mathcal{B}_{cc}})_\gamma^\mu \mathcal{A}_{ \mu\to \alpha \beta}\right].
\end{align}
The factor $2$ in the second term is arisen from the fact that $I_-$ acts on the first and second quarks in the charmed baryon sextet getting the same results.
The only one isospin sum rule for the $\mathcal{B}_{cc}\to M_8\mathcal{B}_{c6}$ modes is shown in Appendix \ref{doubly1}.

The $D$ meson anti-triplet can be written as $\langle D_\alpha | = (D_\alpha)^i \langle D^{i}|$.
Under the isospin lowering operator $I_-$, we have
\begin{align}
&I_-\langle D_\alpha| = \langle D_\alpha|\cdot I_- = \sum_\beta([I_-]_{D})^{\beta}_{\alpha}\langle D_\beta|.
\end{align}
If the charmed meson anti-triplet bases are defined as
 $\langle D_\beta| = (\langle D^0|, \langle D^+|,\langle D^+_s| )$,
the coefficient matrix $[I_-]_{D}$ is derived to be
\begin{eqnarray}
 [I_-]_{D}= \left( \begin{array}{ccc}
   0   & 1  & 0 \\
     0 &  0  & 0 \\
    0 & 0 & 0 \\
  \end{array}\right).
\end{eqnarray}
The sum of decay amplitudes generated by the isospin lowering operator $I_-$ for the $\mathcal{B}_{cc}\to D\mathcal{B}_{8}$ modes is
\begin{align}\label{rule5}
{ SumI_-}\,[\gamma, \alpha,\beta]=  \sum_\mu\left[-([I_-]_{D})_\alpha^\mu \mathcal{A}_{\gamma\to \mu \beta } +  ([I_-]_{\mathcal{B}_{8}})_\beta^\mu \mathcal{A}_{\gamma\to \alpha\mu } - ([I_-]_{\mathcal{B}_{cc}})_\gamma^\mu \mathcal{A}_{ \mu\to \alpha \beta}\right].
\end{align}
The sum of decay amplitudes generated by $I_-$ for the $\mathcal{B}_{cc}\to D\mathcal{B}_{10}$ modes is
\begin{align}\label{rule6}
{ SumI_-}\,[\gamma, \alpha,\beta]=  \sum_\mu\left[-([I_-]_{D})_\alpha^\mu \mathcal{A}_{\gamma\to \mu \beta } +  3\,([I_-]_{\mathcal{B}_{10}})_\beta^\mu \mathcal{A}_{\gamma\to \alpha\mu }  - ([I_-]_{\mathcal{B}_{cc}})_\gamma^\mu \mathcal{A}_{ \mu\to \alpha \beta}\right].
\end{align}
The isospin sum rules for the $\mathcal{B}_{cc}\to D\mathcal{B}_{8}$ and $\mathcal{B}_{cc}\to D\mathcal{B}_{10}$ modes are shown in Appendix \ref{doubly2}.

For three body decays, the sum of decay amplitudes generated by the isospin lowering operation $I_-$ for the $\mathcal{B}_{cc} \to \mathcal{B}_{c\overline{3}}{M_8}{M_8}$ is
\begin{align}\label{rule7}\nonumber
{ SumI_-}\,[\gamma, \alpha,\beta,\delta]=  \sum_\mu\left[([I_-]_{M_8})_\beta^\mu \mathcal{A}_{\gamma\to \alpha \mu\delta } + ([I_-]_{M_8})_\delta^\mu \mathcal{A}_{\gamma\to \alpha \beta\mu } \right.\\\left.~~-  ([I_-]^T_{\mathcal{B}_{ c\overline 3}})_\alpha^\mu \mathcal{A}_{\gamma\to \mu\beta\delta } -([I_-]_{\mathcal{B}_{cc}})_\gamma^\mu \mathcal{A}_{\mu\to\alpha \beta\delta }\right].
\end{align}
The sum of decay amplitudes generated by $I_-$ for the $\mathcal{B}_{cc} \to \mathcal{B}_{c6}{M_8}{M_8}$ is
\begin{align}\label{rule8}\nonumber
{ SumI_-}\,[\gamma,\alpha,\beta, \delta]= \sum_\mu\left[2\,([I_-]_{\mathcal{B}_{c6}})_\alpha^\mu \mathcal{A}_{\gamma\to \mu \beta\delta } + ([I_-]_{M_8})_\beta^\mu \mathcal{A}_{\gamma\to \alpha \mu\delta }\right.\\\left.~~+([I_-]_{M_8})_\delta^\mu \mathcal{A}_{\gamma\to \alpha \beta\mu } -([I_-]_{\mathcal{B}_{cc}})_\gamma^\mu \mathcal{A}_{\mu\to\alpha \beta\delta }\right].
\end{align}
The sum of decay amplitudes generated by $I_-$ for the $\mathcal{B}_{cc} \to \mathcal{B}_{8}{D}{M_8}$ is
\begin{align}\label{rule9}\nonumber
{ SumI_-}\,[\gamma,\alpha,\beta, \delta]= \sum_\mu\left[([I_-]_{\mathcal{B}_{8}})_\alpha^\mu \mathcal{A}_{\gamma\to \mu \beta\delta } + ([I_-]_{M_8})_\delta^\mu \mathcal{A}_{\gamma\to \alpha \beta\mu }\right.\\\left.~~-([I_-]_D)_\beta^\mu \mathcal{A}_{\gamma\to \alpha \mu\delta } -([I_-]_{\mathcal{B}_{cc}})_\gamma^\mu \mathcal{A}_{\mu\to\alpha \beta\delta }\right].
\end{align}
The sum of decay amplitudes generated by $I_-$ for the $\mathcal{B}_{cc} \to \mathcal{B}_{10}{D}{M_8}$ is
\begin{align}\label{rule10}\nonumber
{ SumI_-}\,[\gamma,\alpha,\beta, \delta]= \sum_\mu\left[3\,([I_-]_{\mathcal{B}_{10}})_\alpha^\mu \mathcal{A}_{\gamma\to \mu \beta\delta } + ([I_-]_{M_8})_\delta^\mu \mathcal{A}_{\gamma\to \alpha \beta\mu }\right.\\\left.~~-([I_-]_D)_\beta^\mu \mathcal{A}_{\gamma\to \alpha \mu\delta } -([I_-]_{\mathcal{B}_{cc}})_\gamma^\mu \mathcal{A}_{\mu\to\alpha \beta\delta }\right].
\end{align}
The isospin sum rules for three-body decays of doubly charmed baryons are shown in Appendices \ref{doubly3} and  \ref{doubly4}.

In the isospin sum rules listed Appendices \ref{singlyx} and \ref{doublyx}, some of them have been derived by using the Wigner-Eckart invariants in literature.
For example, the isospin sum rules \eqref{test1}, \eqref{test7}, \eqref{test8}, \eqref{test9}, \eqref{test10}, \eqref{test11} and \eqref{test12} were  first derived in Ref.~\cite{Savage:1989qr} and are consistent with our results.
Eq.~\eqref{test5} is verified by Refs.~\cite{Lu:2016ogy,Gronau:2018vei}.
The different sign in Eq.~\eqref{test5} compared to Refs.~\cite{Lu:2016ogy,Gronau:2018vei} raises from the conventions of $|\pi^+\rangle$ and $|\overline K^0\rangle$.
And Eqs.~\eqref{test2} and \eqref{test3} are verified by Ref.~\cite{Gronau:2018vei}.
However, most of the isospin sum rules listed Appendices \ref{singlyx} and \ref{doublyx} are derived first in this work  due to the use of a programmatic approach.

\section{Phenomenological analysis}\label{pa}
The decay amplitude for the $\mathcal{B}_{c\overline 3}\to\mathcal{B}_8M_8$ mode is given by
\begin{eqnarray}\label{eq:A&B}
\mathcal{A}(\mathcal{B}_{c\overline 3}\to\mathcal{B}_8M_8)=i\overline u_\mathcal{B}(A-B\gamma_5)u_{\mathcal{B}_c}
\end{eqnarray}
where $A$ and $B$ are the parity-violating $S$-wave and parity-conserving $P$-wave amplitudes with strong phases $\delta_S$ and $\delta_P$, respectively.
The decay width $\Gamma$ and Lee-Yang parameters $\alpha^\prime$, $\beta^\prime$ and $\gamma^\prime$ are computed by\footnote{Superscript "$\prime$" is used to differentiate the Greek indices in Eq.~\eqref{amp}.}
\begin{align}
&\Gamma = \frac{p_c}{8\pi}\frac{(m_{\mathcal{B}_c}+m_\mathcal{B})^2-m_M^2}
{m_{\mathcal{B}_c}^2}\left(|A|^2
+ \kappa^2|B|^2\right),\nonumber\\
& \alpha^\prime=\frac{2\kappa |A^*B|\cos(\delta_P-\delta_S)}{|A|^2+\kappa^2 |B|^2},~~
\beta^\prime=\frac{2\kappa |A^*B|\sin(\delta_P-\delta_S)}{|A|^2+\kappa^2 |B|^2},~~
\gamma^\prime=\frac{|A|^2-\kappa^2 |B|^2}{|A|^2+\kappa^2 |B|^2},
\end{align}
where $p_c$ is the c.m. three-momentum in the rest
frame of initial baryon and $\kappa$ is defined as $\kappa=p_c/(E_\mathcal{B}+m_\mathcal{B})=\sqrt{(E_\mathcal{B}-m_\mathcal{B})
/(E_\mathcal{B}+m_\mathcal{B})}$.
The Lee-Yang parameters $\alpha^\prime$, $\beta^\prime$ and $\gamma^\prime$ describe the interference between different partial waves.
The isospin sum rules work for all partial waves.
Thereby, if two decay channels form an isospin sum rule, they have the same decay
asymmetries $\alpha^\prime$, $\beta^\prime$ and $\gamma^\prime$.
According to the isospin sum rule \eqref{test1}, the branching fraction and $\alpha^\prime$ parameter of $\Lambda_c^+\to\Sigma^+\pi^0$ and $\Lambda_c^+\to\Sigma^0\pi^+$ modes are equal,
\begin{equation}
  \mathcal{B}r(\Lambda_c^+\to\Sigma^+\pi^0)=\mathcal{B}r(\Lambda_c^+\to\Sigma^0\pi^+), \qquad \alpha^\prime(\Lambda_c^+\to\Sigma^+\pi^0)=\alpha^\prime(\Lambda_c^+\to\Sigma^0\pi^+).
\end{equation}
Experimental data imply that \cite{PDG}
\begin{align}\label{ib}
 \mathcal{B}r(\Lambda_c^+\to\Sigma^+\pi^0)/\mathcal{B}r(\Lambda_c^+\to\Sigma^0\pi^+)&=0.98\pm0.12,\\
  \alpha^\prime(\Lambda_c^+\to\Sigma^+\pi^0)/\alpha^\prime(\Lambda_c^+\to\Sigma^0\pi^+)&=0.75\pm0.24.
\end{align}
One can find the ratio of branching fraction is well consistent with the prediction of isospin symmetry.
And a precise measurement of $\alpha^\prime$ parameter is required to test the isospin symmetry.
According to the isospin sum rule \eqref{test7}, we have following relations under the isospin symmetry,
\begin{align}
  \mathcal{B}r(\Lambda_c^+\to\Sigma^{*+}\pi^0)&
  =\mathcal{B}r(\Lambda_c^+\to\Sigma^{*0}\pi^+), \nonumber\\ \alpha^\prime(\Lambda_c^+\to\Sigma^{*+}\pi^0)&=\alpha^\prime(\Lambda_c^+\to\Sigma^{*0}\pi^+).
\end{align}
The ratios of the branching fraction and $\alpha^\prime$ parameter for the $\Lambda_c^+\to\Sigma^{*+}\pi^0$ and $\Lambda_c^+\to\Sigma^{*0}\pi^+$ modes are given by \cite{PDG}
\begin{align}
 \mathcal{B}r(\Lambda_c^+\to\Sigma^{*+}\pi^0)/\mathcal{B}r(\Lambda_c^+\to\Sigma^{*0}\pi^+)&=0.89\pm0.18,\\
 \alpha^\prime(\Lambda_c^+\to\Sigma^{*+}\pi^0)/\alpha^\prime(\Lambda_c^+\to\Sigma^{*0}\pi^+)&=1.16\pm0.20,
\end{align}
which are consistent with the predictions of isospin symmetry.

If three decay channels form an isospin sum rule, the decay amplitudes form a triangle in the complex plane.
The sum of any two sides of a triangle is greater than the third side, and the difference between any two sides is less than the third side.
The branching fractions for the $\Xi^0_c\to \Sigma^0K^0_S$ and $\Xi^0_c\to \Sigma^+K^-$ modes are given by \cite{PDG}
\begin{align}
 \mathcal{B}r(\Xi^0_c\to \Sigma^0K^0_S) = (5.4\pm 1.6)\times 10^{-4},\qquad \mathcal{B}r(\Xi^0_c\to \Sigma^+K^-) = (1.8\pm 0.4)\times 10^{-3}.
\end{align}
If the contribution of DCS decay $\Xi^0_c\to \Sigma^0K^0$ is neglected, we have
\begin{align}
\mathcal{B}r(\Xi^0_c\to \Sigma^0\overline K^0) = 2\,\mathcal{B}r(\Xi^0_c\to \Sigma^0K^0_S).
\end{align}
Then we can estimate the branching fraction of the $\Xi^+_c\to \Sigma^+\overline K^0$ mode according to the isospin sum rule \eqref{testa},
\begin{align}
(0.05\pm 0.21)\times 10^{-3}<\mathcal{B}r(\Xi^+_c\to \Sigma^+\overline K^0)<(23.81\pm 4.47)\times 10^{-3}.
\end{align}
The branching fractions of the $\Xi^0_c\to \Xi^-\pi^+$ and $\Xi^+_c\to \Xi^0\pi^+$ modes are given by \cite{PDG}
\begin{align}
 \mathcal{B}r(\Xi^0_c\to \Xi^-\pi^+) = (1.43\pm 0.32)\%,\qquad \mathcal{B}r(\Xi^+_c\to \Xi^0\pi^+) = (1.6\pm 0.8)\%.
\end{align}
The branching fraction of the $\Xi^+_c\to \Sigma^+\overline K^0$ mode is estimated according to the isospin sum rule \eqref{testb} as
\begin{align}
(0.11\pm 0.11)\%<\mathcal{B}r(\Xi^0_c\to \Xi^0\pi^0)<(1.85\pm 0.44)\%.
\end{align}
The isospin sum rule is valid for the excited states. For example, the $K$ meson in the isospin sum rule \eqref{testa} can be replaced by a $K^*$ meson.
Branching fractions of the $\Xi^0_c\to \Sigma^+K^{*-}$, $\Xi^0_c\to \Sigma^0\overline K^{*0}$ and $\Xi^+_c\to \Sigma^+\overline K^{*0}$ modes are given by \cite{PDG}
\begin{align}
 \mathcal{B}r(\Xi^0_c\to \Sigma^+K^{*-}) & = (4.9\pm 1.4)\times 10^{-3},\qquad \mathcal{B}r(\Xi^0_c\to \Sigma^0\overline K^{*0}) = (9.8\pm 2.3)\times 10^{-3},\nonumber\\
\mathcal{B}r(\Xi^+_c\to \Sigma^+\overline K^{*0}) & = (23\pm 11)\times 10^{-3}.
\end{align}
One can check the decay amplitudes form a triangle indeed.
The isospin sum rule \eqref{test5} can be tested by the total branching fractions of the $\Lambda_c^+\to pK^-\pi^+$, $\Lambda_c^+\to pK^0_S\pi^0$ and $\Lambda_c^+\to nK^0_S\pi^+$ modes \cite{PDG},
\begin{align}
\mathcal{B}r(\Lambda_c^+\to pK^-\pi^+)&=(6.26\pm 0.29)\%,\qquad
\mathcal{B}r(\Lambda_c^+\to pK^0_S\pi^0)=(1.96\pm 0.12)\%,\nonumber\\
\mathcal{B}r(\Lambda_c^+\to nK^0_S\pi^+)&=(1.82\pm 0.25)\%.
\end{align}
One can check their decay amplitudes form an isospin triangle as well.

For multi-body decays, the intermediate resonances contribute a lot to the branch fractions.
Our approach of generating isospin rules works when intermediate resonances are considered.
The isospin sum rules are derived by applying $I_-^n$ to the initial/final states and expanding the results with the initial/final states.
If there is a resonance $R_I$ serves as an intermediate state, we have
\begin{align}
 I_- (| R_I \rangle \langle R_I |) =  (I_-|R_I\rangle)\langle R_I |+|R_I\rangle(I_-\langle R_I |)=|R_{I+1}\rangle\langle R_I |+|R_I\rangle\langle R_{I-1} | =0.
\end{align}
Therefore, intermediate resonances do not generate new terms in the isospin sum rules.
For example, the isospin sum rule \eqref{test5} is derived by $I_-$ acting on the $|\Lambda^+_c\rangle$, $\langle p|$, $\langle \pi^+|$ and $\langle \overline{K}^0|$ states,
\begin{align}
 I_-|\Lambda^+_c\rangle = 0, \qquad I_-\langle p| = \langle n|, \qquad I_-\langle \pi^+| = -\sqrt{2}\,\langle\pi^0|, \qquad I_-\langle \overline{K}^0| = -\langle K^-|,
\end{align}
and then
\begin{align}\label{tes3}
    {SumI_-}\,[\Lambda^+_c,p,\pi^+,\overline{K}^0]&= \langle\, p\, \pi^+\, \overline{K}^0 |I_-\,\mathcal{H}_{\rm eff}| \Lambda^+_c\rangle\nonumber\\&=\langle\, p\, \pi^+\, \overline{K}^0 |\mathcal{H}_{\rm eff}| I_-\,\Lambda^+_c\rangle+\langle \,(I_-\,p)\, \pi^+ \,\overline{K}^0 |\mathcal{H}_{\rm eff}| \Lambda^+_c\rangle\nonumber\\&~~~~~~+\langle \,p\, (I_-\,\pi^+)\, \overline{K}^0 |\mathcal{H}_{\rm eff}| \Lambda^+_c\rangle+\langle \,p \, \pi^+ \,(I_-\,\overline{K}^0) |\mathcal{H}_{\rm eff}| \Lambda^+_c\rangle\nonumber\\&=0+\langle \,n\, \pi^+ \,\overline{K}^0 |\mathcal{H}_{\rm eff}| \Lambda^+_c\rangle-\sqrt{2}\,\langle \,p\,\pi^0\, \overline{K}^0 |\mathcal{H}_{\rm eff}| \Lambda^+_c\rangle-\langle \,p \, \pi^+ \,K^-) |\mathcal{H}_{\rm eff}| \Lambda^+_c\rangle\nonumber\\&=\mathcal{A}(\Lambda^+_c\to n\pi^+\overline{K}^0)-\sqrt{2}\,\mathcal{A}(\Lambda^+_c\to p\pi^0\overline{K}^0)-\mathcal{A}(\Lambda^+_c\to p\pi^+K^-)=0.
\end{align}
With the inclusion of the intermediate resonance $\overline K^{*0}$, which decays into $\pi^0\overline K^0$ and $\pi^+ K^-$,
Eq.~\eqref{tes3} takes the following form,
\begin{align}\label{x11}
   \langle\, p\, \pi^+\, \overline{K}^0 |I_-\,\mathcal{H}_{\rm eff}| \Lambda^+_c\rangle= \langle\, p\, \pi^+\, \overline{K}^0 |I_-\,\mathcal{H}_{\rm eff}| \Lambda^+_c\rangle_{\rm non-res}+\langle\, \pi^+\, \overline{K}^0|\overline K^{*0} \rangle\langle\, p\,\overline K^{*0}|I_-\,\mathcal{H}_{\rm eff}| \Lambda^+_c\rangle.
\end{align}
The first term of Eq.~\eqref{x11} is
\begin{align}\label{tes1}
 \langle\, p\, \pi^+\, \overline{K}^0 |I_-\,\mathcal{H}_{\rm eff}| \Lambda^+_c\rangle_{\rm non-res} &= \mathcal{A}(\Lambda^+_c\to n\pi^+\overline{K}^0)_{\rm non-res}-\sqrt{2}\mathcal{A}(\Lambda^+_c\to p\pi^0\overline{K}^0)_{\rm non-res}\nonumber\\&~~~~~~~-\mathcal{A}(\Lambda^+_c\to p\pi^+K^-)_{\rm non-res}=0.
\end{align}
And the second term of Eq.~\eqref{x11} is
\begin{align}\label{tes2}
&~~~~~~\langle\, \pi^+\, \overline{K}^0|\overline K^{*0} \rangle\langle\, p\,\overline K^{*0}|I_-\,\mathcal{H}_{\rm eff}| \Lambda^+_c\rangle\nonumber\\& =\langle\, \pi^+\, \overline{K}^0|\overline K^{*0} \rangle\langle\, p\,\overline K^{*0}|\mathcal{H}_{\rm eff}| I_-\,\Lambda^+_c\rangle+\langle\, \pi^+\, \overline{K}^0|\overline K^{*0} \rangle\langle\, (I_-\,p)\,\overline K^{*0}|\mathcal{H}_{\rm eff}| \Lambda^+_c\rangle\nonumber\\&~~~~+\langle\, (I_-\,\pi^+)\, \overline{K}^0|\overline K^{*0} \rangle\langle\, p\,\overline K^{*0}|\mathcal{H}_{\rm eff}| \Lambda^+_c\rangle+\langle\, \pi^+\, (I_-\,\overline{K}^0)|\overline K^{*0} \rangle\langle\, p\,\overline K^{*0}|\mathcal{H}_{\rm eff}| \Lambda^+_c\rangle\nonumber\\&~~~~+\langle\, \pi^+\, \overline{K}^0(I_-\,|\overline K^{*0} \rangle)\langle\, p\,\overline K^{*0}|\mathcal{H}_{\rm eff}| \Lambda^+_c\rangle+\langle\, \pi^+\, \overline{K}^0|\overline K^{*0} \rangle\langle\, p\,(I_-\,\overline K^{*0})|\mathcal{H}_{\rm eff}| \Lambda^+_c\rangle
\nonumber\\& =0+0-\sqrt{2}\,\langle\, \pi^0\, \overline{K}^0|\overline K^{*0} \rangle\langle\, p\,\overline K^{*0}|\mathcal{H}_{\rm eff}| \Lambda^+_c\rangle-\langle\, \pi^+\,K^-|\overline K^{*0} \rangle\langle\, p\,\overline K^{*0}|\mathcal{H}_{\rm eff}| \Lambda^+_c\rangle+0+0\nonumber\\
&= -\sqrt{2}\,\mathcal{A}(\Lambda^+_c\to p\overline K^{*0}[\to \pi^0\overline{K}^0])-\mathcal{A}(\Lambda^+_c\to p\overline K^{*0}[\to\pi^+K^-])=0,
\end{align}
in which $\langle\, \pi^+\, \overline{K}^0|\overline K^{*0} \rangle=0$ and  $I_-\,|\overline K^{*0} \rangle=0$ are used.
$\mathcal{A}(\Lambda^+_c\to p\pi^0\overline{K}^0)$ and $\mathcal{A}(\Lambda^+_c\to p\pi^+K^-)$ receive contributions without and with resonance $\overline K^{*0}$,
\begin{align}
 \mathcal{A}(\Lambda^+_c\to p\pi^0\overline{K}^0) &= \mathcal{A}(\Lambda^+_c\to p\pi^0\overline{K}^0)_{\rm non-res}+\mathcal{A}(\Lambda^+_c\to p\overline K^{*0}[\to \pi^0\overline{K}^0]),\nonumber \\
\mathcal{A}(\Lambda^+_c\to p\pi^+K^-) &= \mathcal{A}(\Lambda^+_c\to p\pi^+K^-)_{\rm non-res}+\mathcal{A}(\Lambda^+_c\to p\overline K^{*0}[\to\pi^+K^-]).
\end{align}
Combining Eq.~\eqref{tes1} and Eq.~\eqref{tes2}, it is found Eq.~\eqref{tes3} remains valid.
As resonance states do not violate isospin symmetry in three-body decays, isospin sum rules can be utilized to investigate the isospin multiplets of exotic hadrons in heavy hadron weak decays.

If two mesons within one isospin multiplet lie in an isospin sum rule of three-body decays, the order of them cannot be exchanged arbitrarily in the derivation of the isospin sum rule.
For example, the isospin sum rule \eqref{test4} is derived via following procedure,
\begin{align}\label{x4}
    {SumI_-}\,[\Lambda^+_c,\Lambda^0,\pi^+,\pi^+]&= \langle\, \Lambda^0\, \pi^+\, \pi^+ |\,I_-\,\mathcal{H}_{\rm eff}\,|\, \Lambda^+_c\rangle\nonumber\\&= \langle\, \Lambda^0\, \pi^+\, \pi^+ |\,\mathcal{H}_{\rm eff}\,|\, I_-\, \Lambda^+_c\rangle+ \langle\, (I_-\, \Lambda^0)\, \pi^+\, \pi^+ |\,\mathcal{H}_{\rm eff}\,|\, \Lambda^+_c\rangle\nonumber\\&~~~~+ \langle\, \Lambda^0\,(I_-\, \pi^+)\, \pi^+ |\,\mathcal{H}_{\rm eff}\,|\, \Lambda^+_c\rangle+ \langle\,  \Lambda^0\, \pi^+\, (I_-\,\pi^+) |\,\mathcal{H}_{\rm eff}\,|\, \Lambda^+_c\rangle\nonumber\\&=-\sqrt{2}\mathcal{A}(\Lambda^+_c\to \Lambda^0\pi^0\pi^+)-\sqrt{2}\mathcal{A}(\Lambda^+_c\to\Lambda^0\pi^+\pi^0)
    =0.
\end{align}
It is not correct to assume that
\begin{align}
 \langle\, \Lambda^0\,(I_-\, \pi^+)\, \pi^+ |\,\mathcal{H}_{\rm eff}\,|\, \Lambda^+_c\rangle+ \langle\,  \Lambda^0\, \pi^+\, (I_-\,\pi^+) |\,\mathcal{H}_{\rm eff}\,|\, \Lambda^+_c\rangle = 2\langle\, \Lambda^0\,\pi^+ (I_-\, \pi^+)\, |\,\mathcal{H}_{\rm eff}\,|\, \Lambda^+_c\rangle,
\end{align}
and then Eq.~\eqref{x4} is changed to be
\begin{align}
    {SumI_-}\,[\Lambda^+_c,\Lambda^0,\pi^+,\pi^+]&=-2\sqrt{2}\mathcal{A}(\Lambda^+_c\to\Lambda^0\pi^+\pi^0)=0.
\end{align}
The reason is that if there is a resonance, for instance $\Sigma^{*0}$, contributes to the first term of isospin sum rule \eqref{x4} with $\Lambda^+_c\to\Sigma^{*0}[\,\to\Lambda^0\pi^0]\pi^+$, there must be an isospin parter $\Sigma^{*+}$, which contributes to the second term of isospin sum rule \eqref{x4} with $\Lambda^+_c\to\Sigma^{*+}[\,\to\Lambda^0\pi^+]\pi^0$.
Analogous to Eq.~\eqref{x11}, it is equivalent to inserting $|\Sigma^{*0}\rangle\langle \Sigma^{*0}|+|\Sigma^{*+}\rangle\langle \Sigma^{*+}|$ between two $\pi^+$ states of $\langle\, \Lambda^0\, \pi^+\, \pi^+ |\,I_-\,\mathcal{H}_{\rm eff}\,|\, \Lambda^+_c\rangle$.
Under the narrow width approximation, we have
\begin{align}
\mathcal{A}(\Lambda^+_c\to\Sigma^{*+,0}[\,\to\Lambda^0\pi^{+,0}]\pi^{0,+})
=\mathcal{A}(\Lambda^+_c\to\Sigma^{*+,0}\pi^{0,+})\times
\mathcal{A}(\Sigma^{*+,0}\to\Lambda^0\pi^{+,0})
\end{align}
Under the isospin symmetry, we have
\begin{align}\label{x5}
I_-\langle \Lambda^0\pi^+|\Sigma^{*0}\rangle &=\langle \Lambda^0\pi^+|I_-\,\Sigma^{*0}\rangle+\langle (I _-\,\Lambda^0)\pi^+|\Sigma^{*0}\rangle+\langle \Lambda^0 (I _-\,\pi^+)|\Sigma^{*0}\rangle\nonumber \\
&=-\sqrt{2}\langle \Lambda^0\pi^+|I_-\,\Sigma^{*+}\rangle+0-\sqrt{2}\langle \Lambda^0\pi^0|\Sigma^{*0}\rangle
\nonumber \\
&=-\sqrt{2}[\mathcal{A}(\Sigma^{*+}\to\Lambda^0\pi^+)+ \mathcal{A}(\Sigma^{*0}\to\Lambda^0\pi^0)]=0.
\end{align}
Here we have used the fact that $I_-\,H_{\rm strong}=0$ since the strong interaction is flavor conservative.
Under the narrow width approximation, isospin sum rule \eqref{x4} for three-body decays degrades into isospin sum rule \eqref{test7} for two-body decays,
\begin{align}\label{x9}
-\sqrt{2}\,\mathcal{A}(\Lambda^+_c\to \Sigma^{*+}\pi^0)+ \sqrt{2}\, \mathcal{A}(\Lambda^+_c\to \Sigma^{*0}\pi^+)=0.
\end{align}
Exchanging $\pi^+$ and $\pi^0$ in $\mathcal{A}(\Lambda^+_c\to\Lambda^0\pi^0\pi^+)$ within Eq.~\eqref{x4} results in an interchange of $\pi^+$ and $\pi^0$ in $\Lambda^+_c\to\Sigma^{*0}[\,\to\Lambda^0\pi^0]\pi^+$, violating isospin sum rule \eqref{x5} and \eqref{x9}.
Moreover, not only the first two particles of the $\Lambda^+_c\to \Lambda^0\pi^0\pi^+$ can generate form a resonance particle, but also the first and third particles, and the last two particles can generate form resonance states.
If the first and third particles of $\Lambda^+_c\to \Lambda^0\pi^0\pi^+$ decay, $\Lambda^0$ and $\pi^+$, forming a resonance $\Sigma^{*+}$, the first and third particles of $\Lambda^+_c\to \Lambda^0\pi^+\pi^0$ decay, $\Lambda^0$ and $\pi^0$, will form a resonance $\Sigma^{*0}$.
The analysis of this case is similar to the last case.
If the last two particles of the $\Lambda^+_c\to \Lambda^0\pi^0\pi^+$ decay receive contributions from decay chains such as $\Lambda^+_c\to\Lambda^0\rho^+[\,\to\pi^0\pi^+]$,
the last two particles of the $\Lambda^+_c\to \Lambda^0\pi^+\pi^0$ decay will receive contribution from decay chain
$\Lambda^+_c\to\Lambda^0\rho^+[\,\to\pi^+\pi^0]$.
According to the chiral Lagrangian
\begin{align}\label{x7}
\mathcal{L}_{VPP} = \frac{i}{\sqrt{2}}g_{VPP}Tr[V^{\mu}[P,\partial_\mu P]],
\end{align}
the amplitudes $\mathcal{A}(\rho^+\to\pi^+\pi^0)$ and $\mathcal{A}(\rho^+\to\pi^0\pi^+)$ are opposite because of the commutator in Eq.~\eqref{x7},
\begin{align}\label{x6}
 \mathcal{A}(\rho^+\to\pi^+\pi^0) = -\mathcal{A}(\rho^+\to\pi^0\pi^+).
\end{align}
Under the narrow width approximation, isospin sum rule \eqref{x4} degrades into
\begin{align}
 \sqrt{2}\,\mathcal{A}(\Lambda^+_c\to\Lambda^0\rho^+) - \sqrt{2}\, \mathcal{A}(\Lambda^+_c\to\Lambda^0\rho^+) =0.
\end{align}
This equation is trivial without physical result.
Exchanging $\pi^+$ and $\pi^0$ in $\mathcal{A}(\Lambda^+_c\to\Lambda^0\pi^0\pi^+)$ in Eq.~\eqref{x4} is equivalent to setting $\mathcal{A}(\rho^+\to\pi^+\pi^0) = \mathcal{A}(\rho^+\to\pi^0\pi^+)$ and then $\mathcal{A}(\Lambda^+_c\to\Lambda^0\rho^+)=0$.
It is in conflict with the recent BESIII measurements in which the contribution of $\mathcal{A}(\Lambda^+_c\to\Lambda^0\rho^+)$ is dominated in the $\Lambda^+_c\to\Lambda^0\pi^+\pi^0$ decay \cite{BESIII:2022udq}.
According to the above analysis, an isospin sum rule for three-body decays involves several isospin sum rules for two-body decays in the narrow width approximation in which the three-body decay can be factorized into several underlying two-body decays.

The isospin sum rules for three-body decays with two particles within one isospin multiplet can not be tested by data of total branching fractions without performing a partial wave analysis.
For example, the isospin sum rule \eqref{test2} given
\begin{align}\label{x8}
{SumI_-}\,[\Lambda^+_c,\Sigma^0,\pi^+,\pi^+]&=
-\sqrt{2}\big[\mathcal{A}(\Lambda^+_c\to\Sigma^0\pi^0\pi^+)+
\mathcal{A}(\Lambda^+_c\to\Sigma^0\pi^+\pi^0)\nonumber\\&~~~~~
-\mathcal{A}(\Lambda^+_c\to\Sigma^-\pi^+\pi^+)\big]=0
\end{align}
cannot be tested by measuring the total branching fractions of $\mathcal{B}r(\Lambda^+_c\to\Sigma^0\pi^+\pi^0)$ and $\mathcal{B}r(\Lambda^+_c\to\Sigma^-\pi^+\pi^+)$.
however, Eq.~\eqref{x8} can be tested through partial wave analysis.
The Hamiltonian is symmetric and anti-symmetric if the two mesons are in relatively even and odd angular momentum $L$.
In the case of $L=0,2,...$, amplitudes $\mathcal{A}(\Lambda^+_c\to\Sigma^0\pi^+\pi^0)$ and $\mathcal{A}(\Lambda^+_c\to\Sigma^0\pi^0\pi^+)$ are the same.
Then Eq.~\eqref{x8} becomes
\begin{align}\label{x10}
2\,\mathcal{A}(\Lambda^+_c\to\Sigma^0\pi^+\pi^0)_{L=0,2,...}
-\mathcal{A}(\Lambda^+_c\to\Sigma^-\pi^+\pi^+)_{L=0,2,...}=0.
\end{align}
In the case of $L=1,3,...$, amplitudes $\mathcal{A}(\Lambda^+_c\to\Sigma^0\pi^+\pi^0)$ in sign, and $\mathcal{A}(\Lambda^+_c\to\Sigma^0\pi^0\pi^+)$ are opposite and $\mathcal{A}(\Lambda^+_c\to\Sigma^-\pi^+\pi^+)=0$.
Eq.~\eqref{x8} reduces to a trivial equation of $0=0$.
Above discussion can be applied to the other isospin sum rules listed in Appendices \ref{singlyx} and \ref{doublyx}.
The isospin equations given by Ref.~\cite{Savage:1989qr}, such as
\begin{align}
\mathcal{A}(\Lambda^+_c\to\Sigma^0\pi^+\pi^0)_{L=1,3,...}
-\mathcal{A}(\Lambda^+_c\to\Sigma^+\pi^+\pi^-)_{L=1,3,...}=0,
\end{align}
are re-derived in different manner.
In fact, the discussion about $\rho^+$ resonance in the isospin sum rule \eqref{x4} is a practical example for $L=1$.
As a systematic approach, many new isospin sum rules for cases of even and odd
$L$ can be given according to Appendices \ref{singlyx} and \ref{doublyx}.
These could be tested in future experiments and provide useful information in the search for new decay modes.

\section{Summary}\label{summary}

Flavor symmetry is a powerful tool to analyze the charmed baryon weak decays.
In this work, we derive the isospin sum rules for the two- and three-body non-leptonic decays of singly and doubly charmed baryons in a systematic approach.
Hundreds of isospin sum rules are derived to test the isospin symmetry and study the isospin partners of intermediate resonances in the charmed baryon decays.

\begin{acknowledgements}

This work was supported in part by the National Natural Science Foundation of China under Grants No. 12105099.

\end{acknowledgements}

\begin{appendix}
\section{Isospin sum rules for singly charmed baryon decays}\label{singlyx}

\subsection{$\mathcal{B}_{c\overline 3}\to M_8 \mathcal{B}_8$ modes}\label{singly1}

The isospin sum rules for the $\mathcal{B}_{c\overline 3}\to M_8 \mathcal{B}_8$ modes are derived as
\begin{align}\label{test1}
{ SumI_-}\,[\Lambda^+_c, \pi^+,\Sigma^+]&=-\sqrt{2}\,\big[ \,\mathcal{A}(\Lambda^+_c\to \pi^0\Sigma^+)+\mathcal{A}(\Lambda^+_c\to \pi^+\Sigma^0)\big]=0,
\end{align}
\begin{align}\label{testa}
{ SumI_-}\,[\Xi^0_c, \overline K^0,\Sigma^+]&=- \mathcal{A}(\Xi^0_c\to K^-\Sigma^+)-\sqrt{2}\,\mathcal{A}(\Xi^0_c\to \overline K^0\Sigma^0)-\mathcal{A}(\Xi^+_c\to \overline K^0\Sigma^+)=0,
\end{align}
\begin{align}\label{testb}
{ SumI_-}\,[\Xi^0_c, \pi^+,\Xi^0]&=- \sqrt{2}\,\mathcal{A}(\Xi^0_c\to \pi^0\Xi^0)-\mathcal{A}(\Xi^0_c\to \pi^+\Xi^-)-\mathcal{A}(\Xi^+_c\to \pi^+\Xi^0)=0,
\end{align}
\begin{align}
{ SumI_-^2}\,[\Xi^0_c, \pi^+,\Sigma^+]&=2\,\big[\, 2\,\mathcal{A}( \Xi^0_c\to\pi^0\Sigma^0)-\mathcal{A}(\Xi^0_c\to\pi^-\Sigma^+)
-\mathcal{A}(\Xi^0_c\to\pi^+\Sigma^-)\nonumber\\&~~~~~
+\sqrt{2}\,\big(\mathcal{A}(\Xi^+_c\to\pi^0\Sigma^+)+
\mathcal{A}(\Xi^+_c\to\pi^+\Sigma^0)\big)\big]=0,
\end{align}
\begin{align}
{ SumI_-^2}\,[\Xi^0_c, K^+,\Sigma^+]&=-2\,\big[ \sqrt{2}\,\mathcal{A}( \Xi^0_c\to K^0\Sigma^0)+\mathcal{A}(\Xi^0_c\to K^+\Sigma^-)
\nonumber\\&~~~~~
+\mathcal{A}(\Xi^+_c\to K^0\Sigma^+)-\sqrt{2}\,\mathcal{A}(\Xi^+_c\to K^+\Sigma^0)\big]=0,
\end{align}
\begin{align}
{ SumI_-^2}\,[\Xi^0_c, \pi^+,p]&=-2\,\big[ \sqrt{2}\,\mathcal{A}( \Xi^0_c\to \pi^0n)+\mathcal{A}(\Xi^0_c\to \pi^-p)
\nonumber\\&~~~~~
-\sqrt{2}\,\mathcal{A}(\Xi^+_c\to \pi^0p)+\mathcal{A}(\Xi^+_c\to \pi^+n)\big]=0.
\end{align}

\subsection{$\mathcal{B}_{c\overline 3}\to M_8 \mathcal{B}_{10}$ modes}\label{singly2}

The isospin sum rules for the $\mathcal{B}_{c\overline 3}\to M_8 \mathcal{B}_{10}$ modes are derived as
\begin{align}\label{test7}
{ SumI_-}\,[\Lambda^+_c, \pi^+,\Sigma^{*+}]&= \sqrt{2}\,\big[-\mathcal{A}(\Lambda^+_c\to \pi^0\Sigma^{*+})+ \mathcal{A}(\Lambda^+_c\to \pi^+\Sigma^{*0})\big]=0,
\end{align}
\begin{align}\label{test8}
{ SumI_-}\,[\Lambda^+_c, \overline K^0,\Delta^{++}]&=- \mathcal{A}(\Lambda^+_c\to K^-\Delta^{++})+\sqrt{3}\,\mathcal{A}(\Lambda^+_c\to \overline K^0\Delta^{+})=0,
\end{align}
\begin{align}
{ SumI_-}\,[\Xi^0_c, \overline K^0,\Sigma^{*+}]&= -\mathcal{A}(\Xi^0_c\to K^-\Sigma^{*+})+ \sqrt{2}\,\mathcal{A}(\Xi^0_c\to \overline K^0\Sigma^{*0})-\mathcal{A}(\Xi^+_c\to \overline K^0\Sigma^{*+})=0,
\end{align}
\begin{align}
{ SumI_-}\,[\Xi^0_c, \pi^+,\Xi^{*0}]&= -\sqrt{2}\,\mathcal{A}(\Xi^0_c\to \pi^0\Xi^{*0})+ \mathcal{A}(\Xi^0_c\to \pi^+\Xi^{*-})-\mathcal{A}(\Xi^+_c\to \pi^+\Xi^{*0})=0,
\end{align}
\begin{align}
{ SumI_-^2}\,[\Lambda^+_c, \pi^+,\Delta^{++}]&=-2\,\big[\sqrt{6}\,\mathcal{A}( \Lambda^+_c\to\pi^0\Delta^+)+\mathcal{A}(\Lambda^+_c\to\pi^-\Delta^{++})
\nonumber\\&~~~~~-\sqrt{3}\,\mathcal{A}(\Lambda^+_c\to\pi^+\Delta^0)\big]=0,
\end{align}
\begin{align}
{ SumI_-^2}\,[\Xi^0_c, \overline K^0,\Delta^{++}]&=2\,\big[-\sqrt{3}\,\mathcal{A}( \Xi^0_c\to K^-\Delta^+)+\sqrt{3}\,\mathcal{A}(\Xi^0_c\to\overline K^0\Delta^{0})\nonumber\\&~~~~~
+\mathcal{A}(\Xi^+_c\to K^-\Delta^{++})-\sqrt{3}\,\mathcal{A}( \Xi^+_c\to \overline K^0\Delta^+)\big]=0,
\end{align}
\begin{align}
{ SumI_-^2}\,[\Xi^0_c, \pi^+,\Sigma^{*+}]&=-2\,\big[ 2\,\mathcal{A}( \Xi^0_c\to\pi^0\Sigma^{*0})+\mathcal{A}(\Xi^0_c\to\pi^-\Sigma^{*+})\nonumber\\&~~~~~
-\mathcal{A}(\Xi^0_c\to\pi^+\Sigma^{*-})
-\sqrt{2}\,\mathcal{A}(\Xi^+_c\to\pi^0\Sigma^{*+})\nonumber\\&~~~~~+
\sqrt{2}\,\mathcal{A}(\Xi^+_c\to\pi^+\Sigma^{*0})\big]=0,
\end{align}
\begin{align}\label{test9}
{ SumI_-^2}\,[\Lambda^+_c,  K^+,\Delta^{++}]&=2\sqrt{3}\,\big[\mathcal{A}(\Lambda^+_c\to K^0\Delta^{+})+\mathcal{A}(\Lambda^+_c\to  K^+\Delta^{0})\big]
=0,
\end{align}
\begin{align}
{ SumI_-^2}\,[\Xi^0_c,  \eta_8,\Delta^{++}]&=2\sqrt{3}\,\big[\mathcal{A}(\Xi^0_c\to \eta_8\Delta^{0})-\mathcal{A}(\Xi^+_c\to  \eta_8\Delta^{+})\big]
=0,
\end{align}
\begin{align}
{ SumI_-^2}\,[\Xi^0_c,  K^+,\Sigma^{*+}]&=2\,\big[\sqrt{2}\,\mathcal{A}(\Xi^0_c\to K^0\Sigma^{*0})+\mathcal{A}(\Xi^0_c\to  K^+\Sigma^{*-})\nonumber\\&~~~~~
-\mathcal{A}(\Xi^+_c\to  K^0\Sigma^{*+})-\sqrt{2}\,\mathcal{A}(\Xi^+_c\to  K^+\Sigma^{*0})\big]=0,
\end{align}
\begin{align}
{ SumI_-^3}\,[\Xi^0_c,  \pi^+,\Delta^{++}]&=6\,\big[-\sqrt{6}\,\mathcal{A}(\Xi^0_c\to \pi^0\Delta^{0})-\sqrt{3}\,\mathcal{A}(\Xi^0_c\to  \pi^-\Delta^{+})\nonumber\\&~~~~~+\mathcal{A}(\Xi^0_c\to  \pi^+\Delta^{-})
+\sqrt{6}\,\mathcal{A}(\Xi^+_c\to \pi^0\Delta^{+})\nonumber\\&~~~~~+\mathcal{A}(\Xi^+_c\to  \pi^-\Delta^{++})-\sqrt{3}\,\mathcal{A}(\Xi^+_c\to  \pi^+\Delta^{0})\big]=0,
\end{align}
\begin{align}
{ SumI_-^2}\,[\Xi^0_c,  \pi^+,\Delta^{+}]&=-2\,\big[\,2\sqrt{2}\,\mathcal{A}(\Xi^0_c\to \pi^0\Delta^{0})+\mathcal{A}(\Xi^0_c\to  \pi^-\Delta^{+})\nonumber\\&~~~~~-\sqrt{3}\,\mathcal{A}(\Xi^0_c\to  \pi^+\Delta^{-})
-\sqrt{2}\,\mathcal{A}(\Xi^+_c\to \pi^0\Delta^{+})\nonumber\\&~~~~~+2\,\mathcal{A}(\Xi^+_c\to  \pi^+\Delta^{0})\big]=0,
\end{align}
\begin{align}
{ SumI_-^2}\,[\Xi^0_c,  \pi^0,\Delta^{++}]&=2\,\big[\,\sqrt{3}\,\mathcal{A}(\Xi^0_c\to \pi^0\Delta^{0})+\sqrt{6}\,\mathcal{A}(\Xi^0_c\to  \pi^-\Delta^{+})
\nonumber\\&~~~~~-\sqrt{3}\,\mathcal{A}(\Xi^+_c\to \pi^0\Delta^{+})-\sqrt{2}\,\mathcal{A}(\Xi^+_c\to  \pi^-\Delta^{++})\big]=0,
\end{align}
\begin{align}
{ SumI_-^2}\,[\Xi^+_c,  \pi^+,\Delta^{++}]&-2\,\big[\sqrt{6}\,\mathcal{A}(\Xi^+_c\to \pi^0\Delta^{+})+\mathcal{A}(\Xi^+_c\to  \pi^-\Delta^{++})\nonumber\\&~~~~~
-\sqrt{3}\,\mathcal{A}(\Xi^+_c\to \pi^+\Delta^{0})\big]=0.
\end{align}

\subsection{$\mathcal{B}_{c\overline 3} \to \mathcal{B}_{8}{M_8}{M_8}$ modes}\label{singly3}

The isospin rules for the $\mathcal{B}_{c\overline 3} \to \mathcal{B}_{8}{M_8}{M_8}$  modes derived to be
\begin{align}\label{test5}
    {SumI_-}\,[\Lambda^+_c,p,\pi^+,\overline{K}^0]&=\mathcal{A}(\Lambda^+_c\to n\pi^+\overline{K}^0)-\sqrt{2}\mathcal{A}(\Lambda^+_c\to p\pi^0\overline{K}^0)\nonumber\\&~~~~~-\mathcal{A}(\Lambda^+_c\to p\pi^+K^-)=0,
\end{align}
\begin{align}\label{test4}
    {SumI_-}\,[\Lambda^+_c,\Lambda^0,\pi^+,\pi^+]&=-\sqrt{2}\mathcal{A}(\Lambda^+_c\to\Lambda^0\pi^+\pi^0)-\sqrt{2}\mathcal{A}(\Lambda^+_c\to \Lambda^0\pi^0\pi^+)=0,
\end{align}
\begin{align}\label{test2}
     {SumI_-}\,[\Lambda^+_c,\Sigma^0,\pi^+,\pi^+]&=-\sqrt{2}\big[\mathcal{A}(\Lambda^+_c\to\Sigma^0\pi^0\pi^+)+\mathcal{A}(\Lambda^+_c\to\Sigma^0\pi^+\pi^0)\nonumber\\&~~~~~-\mathcal{A}(\Lambda^+_c\to\Sigma^-\pi^+\pi^+)\big]=0,
\end{align}
\begin{align}\label{test3}
    {SumI_-}\,[\Lambda^+_c,\Sigma^+,\pi^+,\pi^0]&=-\sqrt{2}\big[\mathcal{A}(\Lambda^+_c\to\Sigma^0\pi^+\pi^0)+\mathcal{A}(\Lambda^+_c\to\Sigma^+\pi^0\pi^0)\nonumber\\&~~~~~-\mathcal{A}(\Lambda^+_c\to\Sigma^+\pi^+\pi^-)\big]=0,
\end{align}
\begin{align}
    {SumI_-^2}\,[\Lambda^+_c,\Sigma^+,\pi^+,\pi^+]&=4\big[\mathcal{A}(\Lambda^+_c\to\Sigma^0\pi^0\pi^+)+\mathcal{A}(\Lambda^+_c\to\Sigma^0\pi^+\pi^0)\nonumber\\&~~~~~+\mathcal{A}(\Lambda^+_c\to\Sigma^+\pi^0\pi^0)\big]-2\big[\mathcal{A}(\Lambda^+_c\to\Sigma^-\pi^+\pi^+)\nonumber\\&~~~~~+\mathcal{A}(\Lambda^+_c\to\Sigma^+\pi^-\pi^+)+\mathcal{A}(\Lambda^+_c\to\Sigma^+\pi^+\pi^-)\big]=0,
\end{align}
\begin{align}\label{test13}
    {SumI_-^2}\,[\Xi^0_c,\Xi^0,\pi^+,\pi^+]&
    =2\big[-\mathcal{A}(\Xi^0_c\to\Xi^0\pi^-\pi^+)
    -\mathcal{A}(\Xi^0_c\to\Xi^0\pi^+\pi^-)\nonumber\\&~~~~~
    +\sqrt{2}\mathcal{A}(\Xi^0_c\to\Xi^-\pi^0\pi^+)
    +\sqrt{2}\mathcal{A}(\Xi^0_c\to\Xi^-\pi^+\pi^0)\nonumber\\&~~~~~
    +\sqrt{2}\mathcal{A}(\Xi^+_c\to\Xi^0\pi^0\pi^+)
    +\sqrt{2}\mathcal{A}(\Xi^+_c\to\Xi^0\pi^+\pi^0)\nonumber\\&~~~~~
    +\mathcal{A}(\Xi^+_c\to\Xi^-\pi^+\pi^+)
    +2\mathcal{A}(\Xi^0_c\to\Xi^0\pi^0\pi^0)\big]=0,
\end{align}
\begin{align}
    {SumI_-}\,[\Xi^+_c,\Xi^0,\pi^+,\pi^+]&=-\sqrt{2}\mathcal{A}(\Xi^+_c\to\Xi^0\pi^0\pi^+)-\sqrt{2}\mathcal{A}(\Xi^+_c\to\Xi^0\pi^+\pi^0)\nonumber\\&~~~~~-\mathcal{A}(\Xi^+_c\to\Xi^-\pi^+\pi^+)=0,
\end{align}
\begin{align}\label{test14}
     {SumI_-}\,[\Xi^0_c,\Xi^0,\pi^+,\pi^0]&=\sqrt{2}\mathcal{A}(\Xi^0_c\to\Xi^0\pi^+\pi^-)-\mathcal{A}(\Xi^0_c\to\Xi^-\pi^+\pi^0)\nonumber\\&~~~~~-\mathcal{A}(\Xi^+_c\to\Xi^0\pi^+\pi^0)-\sqrt{2}\mathcal{A}(\Xi^0_c\to\Xi^0\pi^0\pi^0)=0,
\end{align}
\begin{align}\label{test10}
    {SumI_-}\,[\Lambda^+_c,\Sigma^+,\pi^+,\eta_8]&=-\sqrt{2}\mathcal{A}(\Lambda^+_c\to\Sigma^+\pi^0\eta_8)-\sqrt{2}\mathcal{A}(\Lambda^+_c\to\Sigma^0\pi^+\eta_8)=0,
\end{align}
\begin{align}
    {SumI_-}\,[\Lambda^+_c,\Sigma^+,K^+,\overline{K}^0]&=-\sqrt{2}\mathcal{A}(\Lambda^+_c\to\Sigma^0K^+\overline{K}^0)+\mathcal{A}(\Lambda^+_c\to\Sigma^+K^0\overline{K}^0)\nonumber\\&~~~~~-\mathcal{A}(\Lambda^+_c\to\Sigma^+K^+K^-)=0,
\end{align}
\begin{align}
    {SumI_-}\,[\Lambda^+_c,\Xi^0,\pi^+,K^+]&=-\sqrt{2}\mathcal{A}(\Lambda^+_c\to\Xi^0\pi^0K^+)+\mathcal{A}(\Lambda^+_c\to \Xi^0\pi^+{K}^0)\nonumber\\&~~~~~
    -\mathcal{A}(\Lambda^+_c\to\Xi^-\pi^+K^+)=0,
\end{align}
\begin{align}
    {SumI_-}\,[\Xi^0_c,\Sigma^+,\overline{K}^0,\eta_8]&=-\sqrt{2}\mathcal{A}(\Xi^0_c\to\Sigma^0\overline{K}^0\eta_8)-\mathcal{A}(\Xi^0_c\to\Sigma^+{K}^-\eta_8)\nonumber\\&~~~~~-\mathcal{A}(\Xi^+_c\to\Sigma^+\overline{K}^0\eta_8)=0,
\end{align}
\begin{align}
     {SumI_-}\,[\Xi^0_c,\Sigma^+,\pi^0,\overline{K}^0]&=-\sqrt{2}\mathcal{A}(\Xi^0_c\to\Sigma^0\pi^0\overline{K}^0)-\mathcal{A}(\Xi^0_c\to\Sigma^+\pi^0{K}^-)\nonumber\\&~~~~~+\sqrt{2}\mathcal{A}(\Xi^0_c\to\Sigma^+\pi^-\overline{K}^0)-\mathcal{A}(\Xi^+_c\to\Sigma^+\pi^0\overline{K}^0)=0,
\end{align}
\begin{align}
    {SumI_-}\,[\Xi^0_c,p,\overline{K}^0,\overline{K}^0]&=\mathcal{A}(\Xi^0_c\to n\overline{K}^0\overline{K}^0)-\mathcal{A}(\Xi^0_c\to p{K}^-\overline{K}^0)\nonumber\\&~~~~~-\mathcal{A}(\Xi^0_c\to p\overline{K}^0{K}^-)-\mathcal{A}(\Xi^+_c\to p\overline{K}^0\overline{K}^0)=0,
\end{align}
\begin{align}
     {SumI_-}\,[\Xi^0_c,\Sigma^0,\pi^+,\overline{K}^0]&=-\sqrt{2}\mathcal{A}(\Xi^0_c\to\Sigma^0\pi^0\overline{K}^0)-\mathcal{A}(\Xi^0_c\to\Sigma^0\pi^+{K}^-)\nonumber\\&~~~~~+\sqrt{2}\mathcal{A}(\Xi^0_c\to\Sigma^-\pi^+\overline{K}^0)-\mathcal{A}(\Xi^+_c\to\Sigma^0\pi^+\overline{K}^0)=0,
\end{align}
\begin{align}
   {SumI_-}\,[\Xi^0_c,\Xi^0,\pi^+,\eta_8]&=-\sqrt{2}\mathcal{A}(\Xi^0_c\to\Xi^0\pi^0\eta_8)-\mathcal{A}(\Xi^0_c\to\Xi^-\pi^+\eta_8)\nonumber\\&~~~~~-\mathcal{A}(\Xi^+_c\to\Xi^0\pi^+\eta_8)=0,
\end{align}
\begin{align}
     {SumI_-}\,[\Xi^0_c,\Lambda^0,\pi^+,\overline{K}^0]&=-\sqrt{2}\mathcal{A}(\Xi^0_c\to\Lambda^0\pi^0\overline{K}^0)-\mathcal{A}(\Xi^0_c\to\Lambda^0\pi^+{K}^-)\nonumber\\&~~~~~-\mathcal{A}(\Xi^+_c\to\Lambda^0\pi^+\overline{K}^0)=0,
\end{align}
\begin{align}
    {SumI_-}\,[\Xi^0_c,\Sigma^-,\pi^+,\pi^+]&=-\sqrt{2}\mathcal{A}(\Xi^0_c\to\Sigma^-\pi^0\pi^+)-\sqrt{2}\mathcal{A}(\Xi^0_c\to\Sigma^-\pi^+\pi^0)\nonumber\\&~~~~~-\mathcal{A}(\Xi^+_c\to\Sigma^-\pi^+\pi^+)=0,
\end{align}
\begin{align}
   {SumI_-}\,[\Xi^+_c,\Sigma^+,\pi^+,\overline{K}^0]&=-\sqrt{2}\mathcal{A}(\Xi^+_c\to\Sigma^0\pi^+\overline{K}^0)-\sqrt{2}\mathcal{A}(\Xi^+_c\to\Sigma^+\pi^0\overline{K}^0)\nonumber\\&~~~~~-\mathcal{A}(\Xi^+_c\to\Sigma^+\pi^+K^-)=0,
\end{align}
\begin{align}
     {SumI_-^2}\,[\Xi^0_c,\Sigma^+,\pi^+,\overline{K}^0]&
     =2\big[2\mathcal{A}(\Xi^0_c\to\Sigma^0\pi^0\overline{K}^0)
     +\sqrt{2}\mathcal{A}(\Xi^0_c\to\Sigma^0\pi^+{K}^-)\nonumber\\&~~~~~
     +\sqrt{2}\mathcal{A}(\Xi^0_c\to\Sigma^+\pi^0{K}^-)
     -\mathcal{A}(\Xi^0_c\to\Sigma^-\pi^+\overline{K}^0)\nonumber\\&~~~~~
     -\mathcal{A}(\Xi^0_c\to\Sigma^+\pi^-\overline{K}^0)
     +\sqrt{2}\mathcal{A}(\Xi^+_c\to\Sigma^0\pi^+\overline{K}^0)\nonumber\\&~~~~~
     +\sqrt{2}\mathcal{A}(\Xi^+_c\to\Sigma^+\pi^0\overline{K}^0)
     +\mathcal{A}(\Xi^+_c\to\Sigma^+\pi^+{K}^-)\big]=0,
\end{align}
\begin{align}
     {SumI_-^2}\,[\Xi^0_c,\Sigma^+,\pi^+,\pi^0]&=-2\big[-2\mathcal{A}(\Xi^0_c\to\Sigma^0\pi^0\pi^0)+2\mathcal{A}(\Xi^0_c\to\Sigma^0\pi^+\pi^-)\nonumber\\&~~~~~+2\mathcal{A}(\Xi^0_c\to\Sigma^+\pi^0\pi^-)+\mathcal{A}(\Xi^0_c\to\Sigma^-\pi^+\pi^0)\nonumber\\&~~~~~+\mathcal{A}(\Xi^0_c\to\Sigma^+\pi^-\pi^0)-\sqrt{2}\big(\mathcal{A}(\Xi^+_c\to\Sigma^0\pi^0\pi^+)\nonumber\\&~~~~~+\mathcal{A}(\Xi^+_c\to\Sigma^+\pi^0\pi^0)-\mathcal{A}(\Xi^+_c\to\Sigma^+\pi^+\pi^-)\big)\big]= 0,
\end{align}
\begin{align}
    {SumI_-^2}\,[\Xi^0_c,\Sigma^+,\pi^+,\eta_8]&=4\mathcal{A}(\Xi^0_c\to\Sigma^0\pi^0\eta_8)-2\big[\mathcal{A}(\Xi^0_c\to\Sigma^-\pi^+\eta_8)\nonumber\\&~~~~~+\mathcal{A}(\Xi^0_c\to\Sigma^+\pi^-\eta_8)-\sqrt{2}\mathcal{A}(\Xi^+_c\to\Sigma^0\pi^+\eta_8)\nonumber\\&~~~~~-\sqrt{2}\mathcal{A}(\Xi^+_c\to\Sigma^+\pi^0\eta_8)
     \big]=0,
\end{align}
\begin{align}
    {SumI_-^2}\,[\Xi^0_c,\Sigma^+,K^+,\overline{K}^0]&=-2\big[\sqrt{2}\mathcal{A}(\Xi^0_c\to\Sigma^0K^0\overline{K}^0)-\sqrt{2}  \mathcal{A}(\Xi^0_c\to\Sigma^0K^+{K}^-)\nonumber\\&~~~~~+ \mathcal{A}(\Xi^0_c\to\Sigma^+K^-{K}^0)+ \mathcal{A}(\Xi^0_c\to\Sigma^-K^+\overline{K}^0)\nonumber\\&~~~~~-\sqrt{2} \mathcal{A}(\Xi^+_c\to\Sigma^0K^+\overline{K}^0)- \mathcal{A}(\Xi^+_c\to\Sigma^+K^+{K}^-)\nonumber\\&~~~~~+ \mathcal{A}(\Xi^+_c\to\Sigma^+K^0\overline{K}^0)\big]=0,
\end{align}
\begin{align}
     {SumI_-^2}\,[\Xi^0_c,p,\pi^+,\overline{K}^0]&=-2\big[\sqrt{2}\mathcal{A}(\Xi^0_c\to p\pi^0K^-)+\sqrt{2}\mathcal{A}(\Xi^0_c\to n\pi^0\overline{K}^0)\nonumber\\&~~~~~ +\mathcal{A}(\Xi^0_c\to n\pi^+K^-)+\mathcal{A}(\Xi^0_c\to p\pi^-\overline{K}^0)\nonumber\\&~~~~~ +\mathcal{A}(\Xi^+_c\to n\pi^+\overline{K}^0)-\sqrt{2}\mathcal{A}(\Xi^+_c\to p\pi^0\overline{K}^0)\nonumber\\&~~~~~ -\mathcal{A}(\Xi^+_c\to p\pi^+{K}^-)\big]=0,
\end{align}
\begin{align}
     {SumI_-^2}\,[\Xi^0_c,\Sigma^0,\pi^+,\pi^+]&=-2\big[-2\mathcal{A}(\Xi^0_c\to\Sigma^0\pi^0\pi^0)+\mathcal{A}(\Xi^0_c\to\Sigma^0\pi^-\pi^+)\nonumber\\&~~~~~+\mathcal{A}(\Xi^0_c\to\Sigma^0\pi^+\pi^-)+2\mathcal{A}(\Xi^0_c\to\Sigma^-\pi^0\pi^+)\nonumber\\&~~~~~-\sqrt{2}\mathcal{A}(\Xi^+_c\to\Sigma^0\pi^0\pi^+)-\sqrt{2}\mathcal{A}(\Xi^+_c\to\Sigma^0\pi^+\pi^0)\nonumber\\&~~~~~-\sqrt{2}\mathcal{A}(\Xi^+_c\to\Sigma^-\pi^+\pi^+)+2\mathcal{A}(\Xi^0_c\to\Sigma^-\pi^+\pi^0)\big]=0,
\end{align}
\begin{align}
     {SumI_-^2}\,[\Xi^0_c,\Xi^0,\pi^+,K^+]&=-2\big[\mathcal{A}(\Xi^0_c\to\Xi^0\pi^-K^+)-\sqrt{2}\mathcal{A}(\Xi^0_c\to\Xi^-\pi^0K^+)\nonumber\\&~~~~~+\mathcal{A}(\Xi^0_c\to\Xi^-\pi^+K^0)-\sqrt{2}\mathcal{A}(\Xi^+_c\to\Xi^0\pi^0K^+)\nonumber\\&~~~~~+\mathcal{A}(\Xi^+_c\to\Xi^0\pi^+K^0)-\mathcal{A}(\Xi^+_c\to\Xi^-\pi^+K^+)\nonumber\\&~~~~~+\sqrt{2}\mathcal{A}(\Xi^0_c\to\Xi^0\pi^0K^0)\big]=0,
\end{align}
\begin{align}
    {SumI_-^2}\,[\Xi^0_c,\Lambda^0,\pi^+,\pi^+]&=4 \mathcal{A}(\Xi^0_c\to\Lambda^0\pi^0\pi^0)-2\mathcal{A}(\Xi^0_c\to\Lambda^0\pi^-\pi^+)\nonumber\\&~~~~~-2\mathcal{A}(\Xi^0_c\to\Lambda^0\pi^+\pi^-)+2\sqrt{2}\mathcal{A}(\Xi^+_c\to\Lambda^0\pi^+\pi^0)\nonumber\\&~~~~~+2\sqrt{2}\mathcal{A}(\Xi^+_c\to\Lambda^0\pi^0\pi^+)=0,
\end{align}
\begin{align}
    {SumI_-^2}\,[\Xi^+_c,\Sigma^+,\pi^+,\pi^+]&= 4\mathcal{A}(\Xi^+_c\to\Sigma^0\pi^0\pi^+)+4\mathcal{A}(\Xi^+_c\to\Sigma^0\pi^+\pi^0)\nonumber\\&~~~~~+4\mathcal{A}(\Xi^+_c\to\Sigma^+\pi^0\pi^0)-2\mathcal{A}(\Xi^+_c\to\Sigma^-\pi^+\pi^+)\nonumber\\&~~~~~-2\mathcal{A}(\Xi^+_c\to\Sigma^+\pi^-\pi^+)-2\mathcal{A}(\Xi^+_c\to\Sigma^+\pi^+\pi^-)=0,
\end{align}
\begin{align}
   {SumI_-^2}\,[\Lambda^+_c,\Sigma^+,\pi^+,K^+]&= -2\big[ \sqrt{2}\mathcal{A} (\Lambda^+_c\to\Sigma^0\pi^+K^0)+\mathcal{A} (\Lambda^+_c\to\Sigma^+\pi^-K^+)\nonumber\\&~~~~~+\sqrt{2}\mathcal{A} (\Lambda^+_c\to\Sigma^+\pi^0K^0)-2\mathcal{A} (\Lambda^+_c\to\Sigma^0\pi^0K^+)\nonumber\\&~~~~~+\mathcal{A} (\Lambda^+_c\to\Sigma^-\pi^+K^+)
   \big] =0,
\end{align}
\begin{align}
   {SumI_-^2}\,[\Lambda^+_c,p,\pi^+,\pi^+]&= -2\big[-2\mathcal{A} (\Lambda^+_c\to p\pi^0\pi^0)+\mathcal{A} (\Lambda^+_c\to p\pi^-\pi^+)\nonumber\\&~~~~~+ \mathcal{A} (\Lambda^+_c\to p\pi^+\pi^-)+\sqrt{2}\mathcal{A} (\Lambda^+_c\to n\pi^+\pi^0)\nonumber\\&~~~~~+\sqrt{2}\mathcal{A} (\Lambda^+_c\to n\pi^0\pi^+)=0,
\end{align}
\begin{align}
   {SumI_-^2}\,[\Xi^0_c,\Sigma^+,\pi^+,K^0]&= 4\mathcal{A} (\Xi^0_c\to\Sigma^0\pi^0K^0)-2\big[\mathcal{A} (\Xi^0_c\to\Sigma^-\pi^+K^0)\nonumber\\&~~~~~+\mathcal{A} (\Xi^0_c\to\Sigma^+\pi^-K^0)-\sqrt{2}\mathcal{A} (\Xi^+_c\to\Sigma^0\pi^+K^0)\nonumber\\&~~~~~-\sqrt{2}\mathcal{A} (\Xi^+_c\to\Sigma^+\pi^0K^0)
   \big] =0,
\end{align}
\begin{align}
     {SumI_-^2}\,[\Xi^0_c,\Sigma^+,\pi^0,K^+]&=-2\big[\sqrt{2}\mathcal{A} (\Xi^0_c\to\Sigma^0\pi^0K^0)+2\mathcal{A} (\Xi^0_c\to\Sigma^0\pi^-K^+)\nonumber\\&~~~~~+\mathcal{A} (\Xi^0_c\to\Sigma^-\pi^0K^+)-\sqrt{2}\mathcal{A} (\Xi^0_c\to\Sigma^+\pi^-K^0)\nonumber\\&~~~~~-\sqrt{2}\mathcal{A} (\Xi^+_c\to\Sigma^0\pi^0K^+)+\mathcal{A} (\Xi^+_c\to\Sigma^+\pi^0K^0)\nonumber\\&~~~~~+\sqrt{2}\mathcal{A} (\Xi^+_c\to\Sigma^+\pi^-K^+) \big]=0,
\end{align}
\begin{align}
    {SumI_-^2}\,[\Xi^0_c,\Sigma^+,K^+,\eta_8]&=-2\big[\sqrt{2}\mathcal{A}(\Xi^0_c\to\Sigma^0K^0\eta_8)+\mathcal{A}(\Xi^0_c\to\Sigma^-K^+\eta_8)\nonumber\\&~~~~~-\sqrt{2}\mathcal{A}(\Xi^+_c\to\Sigma^0K^+\eta_8)+\mathcal{A}(\Xi^+_c\to\Sigma^+K^0\eta_8)\big] =0,
\end{align}
\begin{align}
   {SumI_-^2}\,[\Xi^0_c,p,\pi^+,\pi^0]&=-2\big[\sqrt{2} \mathcal{A}(\Xi^0_c\to n\pi^0\pi^0)-\sqrt{2} \mathcal{A}(\Xi^0_c\to n\pi^+\pi^-)\nonumber\\&~~~~~-\sqrt{2} \mathcal{A}(\Xi^+_c\to p\pi^0\pi^0 )+\sqrt{2} \mathcal{A}(\Xi^+_c\to p\pi^+\pi^-)\nonumber\\&~~~~~+2\mathcal{A}(\Xi^0_c\to p\pi^0\pi^-)+\mathcal{A}(\Xi^+_c\to n\pi^+\pi^0)\nonumber\\&~~~~~+\mathcal{A}(\Xi^0_c\to p\pi^-\pi^0)\big]=0,
\end{align}
\begin{align}
   {SumI_-^2}\,[\Xi^0_c,p,\pi^+,\eta_8]&=-2\big[\sqrt{2} \mathcal{A}(\Xi^0_c\to n\pi^0\eta_8)+\mathcal{A}(\Xi^0_c\to p\pi^-\eta_8)\nonumber\\&~~~~~+ \mathcal{A}(\Xi^+_c\to n\pi^+\eta_8 )-\sqrt{2} \mathcal{A}(\Xi^+_c\to p\pi^0\eta_8)\big]=0,
\end{align}
\begin{align}
    {SumI_-^2}\,[\Xi^0_c,p,K^+,\overline{K}^0]&=2\big[\mathcal{A}(\Xi^0_c\to n\overline{K}^0{K}^0)-\mathcal{A}(\Xi^0_c\to pK^0K^-)\nonumber\\&~~~~~-\mathcal{A}(\Xi^0_c\to nK^+K^-)-\mathcal{A}(\Xi^0_c\to n\overline{K}^0K^+)\nonumber\\&~~~~~-\mathcal{A}(\Xi^0_c\to pK^0\overline{K}^0)+\mathcal{A}(\Xi^0_c\to pK^+K^-) \big]=0,
\end{align}
\begin{align}
     {SumI_-^2}\,[\Xi^0_c,\Sigma^0,\pi^+,K^+]&=-2\big[\sqrt{2}\mathcal{A} (\Xi^0_c\to\Sigma^0\pi^0K^0)+\mathcal{A} (\Xi^0_c\to\Sigma^0\pi^-K^+)\nonumber\\&~~~~~+2\mathcal{A} (\Xi^0_c\to\Sigma^-\pi^0K^+)-\sqrt{2}\mathcal{A} (\Xi^0_c\to\Sigma^-\pi^+K^0)\nonumber\\&~~~~~-\sqrt{2}\mathcal{A} (\Xi^+_c\to\Sigma^0\pi^0K^+)+\mathcal{A} (\Xi^+_c\to\Sigma^0\pi^+K^0)\nonumber\\&~~~~~+\sqrt{2}\mathcal{A} (\Xi^+_c\to\Sigma^-\pi^+K^+) \big]=0,
\end{align}
\begin{align}
    {SumI_-^2}\,[\Xi^0_c,n,\pi^+,\pi^+]&=4\mathcal{A}(\Xi^0_c\to n\pi^0\pi^0)-2\big[\mathcal{A}(\Xi^0_c\to n\pi^-\pi^+)\nonumber\\&~~~~~+\mathcal{A}(\Xi^0_c\to n\pi^+\pi^-)-\sqrt{2}\mathcal{A}(\Xi^+_c\to n\pi^+\pi^0)\nonumber\\&~~~~~-\sqrt{2}\mathcal{A}(\Xi^+_c\to n\pi^0\pi^+) =0,
\end{align}
\begin{align}
     {SumI_-^2}\,[\Xi^0_c,\Xi^0,K^+,K^+]&=2\big[\mathcal{A}(\Xi^0_c\to\Xi^0K^0K^0)-\mathcal{A}(\Xi^0_c\to\Xi^-K^0K^+)\nonumber\\&~~~~~-\mathcal{A}(\Xi^0_c\to\Xi^-K^+K^0)-\mathcal{A}(\Xi^+_c\to\Xi^0K^0K^+)\nonumber\\&~~~~~-\mathcal{A}(\Xi^+_c\to\Xi^0K^+K^0)+\mathcal{A}(\Xi^+_c\to\Xi^-K^+K^+)\big]=0,
\end{align}
\begin{align}
     {SumI_-^2}\,[\Xi^0_c,\Lambda^0,\pi^+,K^+]&=-2\big[\sqrt{2}\mathcal{A}(\Xi^0_c\to\Lambda^0\pi^0K^0)+\mathcal{A}(\Xi^0_c\to\Lambda^0\pi^-K^+)\nonumber\\&~~~~~-\sqrt{2}\mathcal{A}(\Xi^+_c\to\Lambda^0\pi^0K^+)+\mathcal{A}(\Xi^+_c\to\Lambda^0\pi^+K^0)\big] =0,
\end{align}
\begin{align}
      {SumI_-^2}\,[\Xi^+_c,\Sigma^+,\pi^+,K^+]&=-2\big[-2\mathcal{A}(\Xi^+_c\to\Sigma^0\pi^0K^+)+\sqrt{2}\mathcal{A}(\Xi^+_c\to\Sigma^0\pi^+K^0)\nonumber\\&~~~~~+\sqrt{2}\mathcal{A}(\Xi^+_c\to\Sigma^+\pi^0K^0)+\mathcal{A}(\Xi^+_c\to\Sigma^-\pi^+K^+)\nonumber\\&~~~~~+\mathcal{A}(\Xi^+_c\to\Sigma^+\pi^-K^+)\big]=0,
\end{align}
\begin{align}
    {SumI_-^2}\,[\Xi^+_c,p,\pi^+,\pi^+]&=4\mathcal{A}(\Xi^+\to p\pi^0\pi^0)-2\mathcal{A}(\Xi^+_c\to p\pi^+\pi^-)\nonumber\\&~~~~~-2\mathcal{A}(\Xi^+_c\to p\pi^-\pi^+)-2\sqrt{2}\mathcal{A}(\Xi^+_c\to n\pi^+\pi^0) \nonumber\\&~~~~~-2\sqrt{2}\mathcal{A}(\Xi^+_c\to n\pi^0\pi^+) =0,
\end{align}
\begin{align}
    {SumI_-^2}\,[\Lambda^+_c,\Sigma^+,K^+,K^+]&=-2\big[\sqrt{2}\mathcal{A}(\Lambda^+_c\to\Sigma^0K^0K^+)+\sqrt{2}\mathcal{A}(\Lambda^+_c\to\Sigma^0K^+K^0)\nonumber\\&~~~~~-\mathcal{A}(\Lambda^+_c\to\Sigma^+K^0K^0)+\mathcal{A}(\Lambda^+_c\to\Sigma^-K^+K^+)\big]=0,
\end{align}
\begin{align}
   {SumI_-^2}\,[\Lambda^+_c,p,\pi^+,K^+]&=-2\big[\sqrt{2}\mathcal{A}(\Lambda^+\to n\pi^0K^+)-\mathcal{A}(\Lambda^+_c\to n\pi^+K^0)\nonumber\\&~~~~~+\sqrt{2}\mathcal{A}(\Lambda^+_c\to p\pi^0K^0) +\mathcal{A}(\Lambda^+_c\to p\pi^-K^+)\big]=0,
\end{align}
\begin{align}
     {SumI_-^3}\,[\Xi^0_c,\Sigma^+,\pi^+,\pi^+]&=
     6\big[-2\sqrt{2}\mathcal{A}(\Xi^0_c\to\Sigma^0\pi^0\pi^0)
     +\sqrt{2}\mathcal{A}(\Xi^0_c\to\Sigma^0\pi^-\pi^+)\nonumber\\&~~~~~
     +\sqrt{2}\mathcal{A}(\Xi^0_c\to\Sigma^0\pi^+\pi^-)
     +\sqrt{2}\mathcal{A}(\Xi^0_c\to\Sigma^-\pi^0\pi^+)\nonumber\\&~~~~~
     +\sqrt{2}\mathcal{A}(\Xi^0_c\to\Sigma^+\pi^0\pi^-)
     +\sqrt{2}\mathcal{A}(\Xi^0_c\to\Sigma^-\pi^+\pi^0)\nonumber\\&~~~~~
     +\sqrt{2}\mathcal{A}(\Xi^0_c\to\Sigma^+\pi^-\pi^0)
     -2\mathcal{A}(\Xi^+_c\to\Sigma^0\pi^0\pi^+)\nonumber\\&~~~~~
     -2\mathcal{A}(\Xi^+_c\to\Sigma^0\pi^+\pi^0)
     -2\mathcal{A}(\Xi^+_c\to\Sigma^+\pi^0\pi^0)\nonumber\\&~~~~~
     +\mathcal{A}(\Xi^+_c\to\Sigma^-\pi^+\pi^+) +\mathcal{A}(\Xi^+_c\to\Sigma^+\pi^-\pi^+)  \nonumber\\&~~~~~+\mathcal{A}(\Xi^+_c\to\Sigma^+\pi^+\pi^-)  \big]=0,
\end{align}
\begin{align}
     {SumI_-^3}\,[\Xi^0_c,\Sigma^+,\pi^+,K^+]&=6\big[2\mathcal{A}(\Xi^0_c\to\Sigma^0\pi^0K^0)+\sqrt{2}\mathcal{A}(\Xi^0_c\to\Sigma^0\pi^-K^+)\nonumber\\&~~~~~+\sqrt{2}\mathcal{A}(\Xi^0_c\to\Sigma^-\pi^0K^+)-\mathcal{A}(\Xi^0_c\to\Sigma^-\pi^+K^0)\nonumber\\&~~~~~-\mathcal{A}(\Xi^0_c\to\Sigma^+\pi^-K^0)-2\mathcal{A}(\Xi^+_c\to\Sigma^0\pi^0K^+)\nonumber\\&~~~~~+\sqrt{2}\mathcal{A}(\Xi^+_c\to\Sigma^0\pi^+K^0)+\sqrt{2}\mathcal{A}(\Xi^+_c\to\Sigma^+\pi^0K^0)\nonumber\\&~~~~~+\mathcal{A}(\Xi^+_c\to\Sigma^-\pi^+K^+) + \mathcal{A}(\Xi^+_c\to\Sigma^+\pi^-K^+) \big]=0,
\end{align}
\begin{align}
     {SumI_-^3}\,[\Xi^0_c,p,\pi^+,\pi^+]&=6\big[\,2\mathcal{A}(\Xi^0_c\to n\pi^0\pi^0)-\mathcal{A}(\Xi^0_c\to n\pi^+\pi^-)\nonumber\\&~~~~~-\mathcal{A}(\Xi^0_c\to n\pi^-\pi^+)-2\mathcal{A}(\Xi^+_c\to p\pi^0\pi^0)\nonumber\\&~~~~~+\mathcal{A}(\Xi^+_c\to p\pi^-\pi^+)+\sqrt{2}\mathcal{A}(\Xi^0_c\to p\pi^0\pi^-) \nonumber\\&~~~~~+\sqrt{2}\mathcal{A}(\Xi^+_c\to n\pi^0\pi^+) +\sqrt{2}\mathcal{A}(\Xi^0_c\to p\pi^-\pi^0)\nonumber\\&~~~~~+\mathcal{A}(\Xi^+_c\to p\pi^+\pi^-)+\sqrt{2}\mathcal{A}(\Xi^+_c\to n\pi^+\pi^0) \big]=0.
\end{align}
\subsection{$\mathcal{B}_{c\overline 3} \to \mathcal{B}_{10}{M_8}{M_8}$ modes}\label{singly4}
The isospin rules for the $\mathcal{B}_{c\overline{3}} \to \mathcal{B}_{10}{M_8}{M_8}$  modes derived via Eq.~\eqref{rule11} are
\begin{align}\label{test11}
     { SumI_-}\,[\Lambda^{+}_{c},\Delta^{++},\pi^+,K^-]&
     =-\sqrt{2}\mathcal{A}(\Lambda^{+}_{c}\to \Delta^{++}\pi^0K^-)+\sqrt{3}\mathcal{A}(\Lambda^{+}_{c}\to \Delta^{+}\pi^+K^-)=0,
\end{align}
\begin{align}\label{test12}
     { SumI_-}\,[\Lambda^{+}_{c},\Sigma^{*+},\pi^+,\eta_8]&=\sqrt{2}\mathcal{A}(\Lambda^{+}_{c}\to \Sigma^{*0}\pi^+\eta_8)-\sqrt{2}\mathcal{A}(\Lambda^{+}_{c}\to \Sigma^{*+}\pi^0\eta_8)=0,
\end{align}
\begin{align}
    { SumI_-}\,[\Xi^{0}_{c},\Delta^{++},\overline{K}^0,K^-]&=\sqrt{3}\mathcal{A}(\Xi^{0}_{c}\to \Delta^+\overline{K}^0K^-)-\mathcal{A}(\Xi^{+}_{c}\to \Delta^{++}\overline{K}^0K^-)\nonumber\\&~~~~~-\mathcal{A}(\Xi^{0}_{c}\to \Delta^{++}{K}^-K^-)= 0,
\end{align}
\begin{align}
    { SumI_-}\,[\Xi^{0}_{c},\Sigma^{*0},\pi^+,\overline{K}^0]&=-\sqrt{2}\mathcal{A}(\Xi^{0}_{c}\to \Sigma^{*0}\pi^0\overline{K}^0)-\mathcal{A}(\Xi^{0}_{c}\to \Sigma^{*0}\pi^+K^-)\nonumber\\&~~~~~-\mathcal{A}(\Xi^{+}_{c}\to\Sigma^{*0} \pi^+\overline{K}^0)+\sqrt{2}\mathcal{A}(\Xi^0_c\to\Sigma^{*-}\pi^+\overline{K}^0)= 0,
\end{align}
\begin{align}
    { SumI_-}\,[\Xi^{0}_{c},\Xi^{*0},\pi^+,\pi^0]&=\mathcal{A}(\Xi^{0}_{c}\to \Xi^{*-}\pi^+\pi^0)-\mathcal{A}(\Xi^{+}_{c}\to \Xi^{*0}\pi^+\pi^0)\nonumber\\&~~~~~-\sqrt{2}\mathcal{A}(\Xi^{0}_{c}\to \Xi^{*0}\pi^0\pi^0)+\sqrt{2}\mathcal{A}(\Xi^{0}_{c}\to \Xi^{*0}\pi^+\pi^-)= 0,
\end{align}
\begin{align}
    { SumI_-}\,[\Xi^{0}_{c},\Xi^{*0},\pi^+,\eta_8]&=\mathcal{A}(\Xi^{0}_{c}\to \Xi^{*-}\pi^+\eta_8)-\mathcal{A}(\Xi^{+}_{c}\to \Xi^{*0}\pi^+\eta_8)\nonumber\\&~~~~~-\sqrt{2}\mathcal{A}(\Xi^{0}_{c}\to \Xi^{*0}\pi^0\eta_8)= 0,
\end{align}
\begin{align}
    { SumI_-}\,[\Xi^{0}_{c},\Xi^{*0},K^+,\overline{K}^0]&=\mathcal{A}(\Xi^{0}_{c}\to \Xi^{*-}K^+\overline{K}^0)-\mathcal{A}(\Xi^{+}_{c}\to \Xi^{*0}K^+\overline{K}^0)\nonumber\\&~~~~~+\mathcal{A}(\Xi^{0}_{c}\to\Xi^{*0} K^0\overline{K}^0)-\mathcal{A}(\Xi^0_c\to\Xi^{*0}K^+{K}^-)= 0,
\end{align}
\begin{align}
    { SumI_-}\,[\Xi^{0}_{c},\Xi^{*-},\pi^+,\pi^+]&=-\sqrt{2}\mathcal{A}(\Xi^{0}_{c}\to \Xi^{*-}\pi^+\pi^0)-\sqrt{2}\mathcal{A}(\Xi^{0}_{c}\to \Xi^{*-}\pi^0\pi^+)\nonumber\\&~~~~~-\mathcal{A}(\Xi^{+}_{c}\to \Xi^{*-}\pi^+\pi^+)=0,
\end{align}
\begin{align}
   { SumI_-}\,[\Xi^{0}_{c},\Omega^{-},\pi^+,K^+]&=-\sqrt{2}\mathcal{A}(\Xi^{0}_{c}\to \Omega^{-}\pi^0K^+)+\mathcal{A}(\Xi^{0}_{c}\to \Omega^{-}\pi^+K^0) \nonumber\\&~~~~~+\mathcal{A}(\Xi^{+}_{c}\to\Omega^{-} \pi^+{K}^+)=0,
\end{align}
\begin{align}
    { SumI_-}\,[\Xi^{+}_{c},\Delta^{++},\overline{K}^0,\overline{K}^0]&=-\mathcal{A}(\Xi^{+}_{c}\to \Delta^{++}\overline{K}^0K^-)-\mathcal{A}(\Xi^{+}_{c}\to \Delta^{++}K^-\overline{K}^0)\nonumber\\&~~~~~+\sqrt{3}\mathcal{A}(\Xi^{+}_{c}\to \Delta^{+}\overline{K}^0\overline{K}^0)= 0,
\end{align}
\begin{align}
    { SumI_-}\,[\Xi^{+}_{c},\Sigma^{*+},\pi^+,\overline{K}^0]&=\sqrt{2}\mathcal{A}(\Xi^{+}_{c}\to \Sigma^{*0}\pi^+\overline{K}^0)-\sqrt{2}\mathcal{A}(\Xi^{+}_{c}\to \Sigma^{*+}\pi^0\overline{K}^0)\nonumber\\&~~~~~-\mathcal{A}(\Xi^{+}_{c}\to\Sigma^{*+} \pi^+{K}^-)= 0,
\end{align}
\begin{align}
    { SumI_-}\,[\Xi^{+}_{c},\Xi^{*0},\pi^+,\pi^+]&=-\sqrt{2}\mathcal{A}(\Xi^{+}_{c}\to \Xi^{*0}\pi^+\pi^0)-\sqrt{2}\mathcal{A}(\Xi^{+}_{c}\to \Xi^{*0}\pi^0\pi^+)\nonumber\\&~~~~~+\mathcal{A}(\Xi^{+}_{c}\to \Xi^{*-}\pi^+\pi^+)=0,
\end{align}
\begin{align}
    { SumI_-}\,[\Lambda^{+}_{c},\Delta^{++},\pi^0,\overline{K}^0]&=-\mathcal{A}(\Lambda^{+}_{c}\to \Delta^{++}\pi^0{K}^-)+\sqrt{3}\mathcal{A}(\Lambda^{+}_{c}\to \Delta^{+}\pi^0\overline{K}^0)\nonumber\\&~~~~~+\sqrt{2}\mathcal{A}(\Lambda^{+}_{c}\to\Delta^{++} \pi^-\overline{K}^0)= 0,
\end{align}
\begin{align}
     { SumI_-}\,[\Lambda^{+}_{c},\Delta^{++},\overline{K}^0,\eta_8]&=-\mathcal{A}(\Lambda^{+}_{c}\to \Delta^{++}{K}^-\eta_8)+\sqrt{3}\mathcal{A}(\Lambda^{+}_{c}\to \Delta^{+}\overline{K}^0\eta_8)=0,
\end{align}
\begin{align}
     { SumI_-}\,[\Lambda^{+}_{c},\Delta^{+},\pi^+,\overline{K}^0]&=2\mathcal{A}(\Lambda^{+}_{c}\to \Delta^{0}\pi^+\overline{K}^0)-\sqrt{2}\mathcal{A}(\Lambda^{+}_{c}\to \Delta^{+}\pi^0\overline{K}^0)\nonumber\\&~~~~~-\mathcal{A}(\Lambda^+_c\to\Delta^{+}\pi^+K^-)=0,
\end{align}
\begin{align}
    { SumI_-}\,[\Lambda^{+}_{c},\Sigma^{*+},\pi^+,\pi^0]&=\sqrt{2}\big[\mathcal{A}(\Lambda^{+}_{c}\to \Sigma^{*0}\pi^+\pi^0)-\mathcal{A}(\Lambda^{+}_{c}\to \Sigma^{*+}\pi^0\pi^0)\nonumber\\&~~~~~+\mathcal{A}(\Lambda^{+}_{c}\to\Sigma^{*+} \pi^+\pi^-)\big]=0,
\end{align}
\begin{align}
    { SumI_-}\,[\Lambda^{+}_{c},\Sigma^{*+},K^+,\overline{K}^0]&=\sqrt{2}\mathcal{A}(\Lambda^{+}_{c}\to \Sigma^{*0}K^+\overline{K}^0)+\mathcal{A}(\Lambda^{+}_{c}\to \Sigma^{*+}K^0\overline{K}^0)\nonumber\\&~~~~~-\mathcal{A}(\Lambda^{+}_{c}\to\Sigma^{*+} K^+{K}^-)=0,
 \end{align}
\begin{align}
     { SumI_-}\,[\Lambda^{+}_{c},\Sigma^{*0},\pi^+,\pi^+]&=-\sqrt{2}\big[\mathcal{A}(\Lambda^{+}_{c}\to \Sigma^{*0}\pi^0\pi^+)+\mathcal{A}(\Lambda^{+}_{c}\to \Sigma^{*0}\pi^+\pi^0)\nonumber\\&~~~~~-\mathcal{A}(\Lambda^{+}_{c}\to \Sigma^{*-}\pi^+\pi^+)\big]=0,
\end{align}
\begin{align}
    { SumI_-}\,[\Lambda^{+}_{c},\Xi^{*0},\pi^+,K^+]&=-\sqrt{2}\mathcal{A}(\Lambda^{+}_{c}\to \Xi^{*0}\pi^0K^+)+\mathcal{A}(\Lambda^{+}_{c}\to \Xi^{*0}\pi^+K^0)\nonumber\\&~~~~~+\mathcal{A}(\Lambda^{+}_{c}\to\Xi^{*-} \pi^+K^+)=0,
\end{align}
\begin{align}
    { SumI_-^2}\,[\Xi^{0}_{c},\Delta^{++},\overline{K}^0,\overline{K}^0]&=2\big[\sqrt{3}\mathcal{A}(\Xi^0_c\to\Delta^{0}\overline{K}^0\overline{K}^0)-\sqrt{3}\mathcal{A}(\Xi^0_c\to\Delta^{+}K^-\overline{K}^0)\nonumber\\&~~~~~-\sqrt{3}\mathcal{A}(\Xi^0_c\to\Delta^{+}\overline{K}^0K^-)+\mathcal{A}(\Xi^0_c\to\Delta^{++}K^-\overline{K}^0)\nonumber\\&~~~~~+\mathcal{A}(\Xi^0_c\to\Delta^{++}\overline{K}^0K^-)-\sqrt{3}\mathcal{A}(\Xi^+_c\to\Delta^{+}\overline{K}^0\overline{K}^0)\nonumber\\&~~~~~+\mathcal{A}(\Xi^0_c\to\Delta^{++}K^-K^-)\big]=0,
\end{align}
\begin{align}
     {SumI_-^2}\,[\Xi^0_c,\Sigma^{*+},\pi^+,\overline{K}^0]&=-2\big[2\mathcal{A}(\Xi^0_c\to\Sigma^{*0}\pi^0\overline{K}^0)+\sqrt{2}\mathcal{A}(\Xi^0_c\to\Sigma^{*0}\pi^+{K}^-)\nonumber\\&~~~~~-\sqrt{2}\mathcal{A}(\Xi^0_c\to\Sigma^{*+}\pi^0{K}^-)+\mathcal{A}(\Xi^0_c\to\Sigma^{*+}\pi^-\overline{K}^0)\nonumber\\&~~~~~+\sqrt{2}\mathcal{A}(\Xi^+_c\to\Sigma^{*0}\pi^+\overline{K}^0)-\sqrt{2}\mathcal{A}(\Xi^+_c\to\Sigma^{*+}\pi^0\overline{K}^0)\nonumber\\&~~~~~-\mathcal{A}(\Xi^+_c\to\Sigma^{*+}\pi^+{K}^-)-\mathcal{A}(\Xi^0_c\to\Sigma^{*-}\pi^+\overline{K}^0)\big]=0,
\end{align}
\begin{align}
      {SumI_-^2}\,[\Xi^0_c,\Xi^{*0},\pi^+,\pi^+]&=-2\big[\sqrt{2}\mathcal{A}(\Xi^0_c\to\Xi^{*-}\pi^0\pi^+)+\sqrt{2}\mathcal{A}(\Xi^0_c\to\Xi^{*-}\pi^+\pi^0)\nonumber\\&~~~~~+\mathcal{A}(\Xi^+_c\to\Xi^{*-}\pi^+\pi^+)-\sqrt{2}\mathcal{A}(\Xi^+_c\to\Xi^{*0}\pi^0\pi^+)\nonumber\\&~~~~~-\sqrt{2}\mathcal{A}(\Xi^+_c\to\Xi^{*0}\pi^+\pi^0)-2\mathcal{A}(\Xi^0_c\to\Xi^{*0}\pi^0\pi^0)\nonumber\\&~~~~~+\mathcal{A}(\Xi^0_c\to\Xi^{*0}\pi^-\pi^+)+\mathcal{A}(\Xi^0_c\to\Xi^{*0}\pi^+\pi^-)\big]=0,
\end{align}
\begin{align}
     { SumI_-^2}\,[\Lambda ^{+}_{c},\Delta^{++},\pi^+,\overline{K}^0]&=2\big[\sqrt{3}\mathcal{A}(\Lambda^+_c\to\Delta^0\pi^+\overline{K}^0)+\sqrt{2}\mathcal{A}(\Lambda^+_c\to\Delta^{++}\pi^0{K}^-)\nonumber\\&~~~~~-\sqrt{6}\mathcal{A}(\Lambda^+_c\to\Delta^+\pi^0\overline{K}^0)-\sqrt{3}\mathcal{A}(\Lambda^+_c\to\Delta^+\pi^+{K}^-)\nonumber\\&~~~~~-\mathcal{A}(\Lambda^+_c\to\Delta^{++}\pi^-\overline{K}^0)]=0,
\end{align}
\begin{align}
    {SumI_-^2}\,[\Lambda^+_c,\Sigma^{*+},\pi^+,\pi^+]&=-2\big[2\mathcal{A}(\Lambda^+_c\to\Sigma^{*0}\pi^0\pi^+)+2\mathcal{A}(\Lambda^+_c\to\Sigma^{*0}\pi^+\pi^0)\nonumber\\&~~~~~~-2\mathcal{A}(\Lambda^+_c\to\Sigma^{*+}\pi^0\pi^0)-\mathcal{A}(\Lambda^+_c\to\Sigma^{*-}\pi^+\pi^+)\nonumber\\&~~~~~~+\mathcal{A}(\Lambda^+_c\to\Sigma^{*+}\pi^-\pi^+)+\mathcal{A}(\Lambda^+_c\to\Sigma^{*+}\pi^+\pi^-)\big]=0,
\end{align}
\begin{align}
     { SumI_-^2}\,[\Xi^{0}_{c},\Delta^{++},\pi^0,\overline{K}^0]&= -2\big[\sqrt{3}\mathcal{A}(\Xi^{0}_{c}\to \Delta^+\pi^0K^-)+\sqrt{2}\mathcal{A}(\Xi^{0}_{c}\to \Delta^{++}\pi^-K^-)\nonumber\\&~~~~~-\sqrt{6}\mathcal{A}(\Xi^{0}_{c}\to\Delta^+\pi^-\overline{K}^0)-\mathcal{A}(\Xi^{+}_{c}\to\Delta^{++}\pi^0{K}^-)\nonumber\\&~~~~~+\sqrt{3}\mathcal{A}(\Xi^{+}_{c}\to \Delta^+\pi^0\overline{K}^0)+\sqrt{2}\mathcal{A}(\Xi^{+}_{c}\to \Delta^{++}\pi^-\overline{K}^0)\nonumber\\&~~~~~-\sqrt{3}\mathcal{A}(\Xi^{0}_{c}\to\Delta^0\pi^0\overline{K}^0)\big]= 0,
\end{align}
\begin{align}
    { SumI_-^2}\,[\Xi^{0}_{c},\Delta^{++},\overline{K}^0,\eta_8]&=2\big[-\sqrt{3}\mathcal{A}(\Xi^0_c\to\Delta^+K^-\eta_8)+\mathcal{A}(\Xi^+_c\to\Delta^{++}K^-\eta_8)\nonumber\\&~~~~~-\sqrt{3}\mathcal{A}(\Xi^+_c\to\Delta^+\overline{K}^0\eta_8)+\sqrt{3}\mathcal{A}(\Xi^0_c\to\Delta^0\overline{K}^0\eta_8)\big]=0,
\end{align}
\begin{align}
     { SumI_-^2}\,[\Xi^{0}_{c},\Delta^{+},\pi^+,\overline{K}^0]&=2\big[\sqrt{2}\mathcal{A}(\Xi^{0}_{c}\to\Delta^{+}\pi^0{K}^-)-\mathcal{A}(\Xi^{0}_{c}\to\Delta^{+}\pi^-\overline{K}^0)\nonumber\\&~~~~~+\sqrt{2}\mathcal{A}(\Xi^{+}_{c}\to\Delta^{+}\pi^0\overline{K}^0)+\mathcal{A}(\Xi^{+}_{c}\to\Delta^{+}\pi^+{K}^-)\nonumber\\&~~~~~-2\mathcal{A}(\Xi^{+}_{c}\to\Delta^{0}\pi^+\overline{K}^0)-2\sqrt{2}\mathcal{A}(\Xi^{0}_{c}\to\Delta^{0}\pi^0\overline{K}^0)\nonumber\\&~~~~~-2\mathcal{A}(\Xi^{0}_{c}\to\Delta^{0}\pi^+{K}^-)+\sqrt{3}\mathcal{A}(\Xi^{0}_{c}\to\Delta^{-}\pi^+\overline{K}^0)\big]=0,
\end{align}
\begin{align}
     { SumI_-^2}\,[\Xi^{0}_{c},\Sigma^{*+},\pi^+,\pi^0]&=-2\big[2\mathcal{A}(\Xi^{0}_{c}\to \Sigma^{*+}\pi^0\pi^-)+\mathcal{A}(\Xi^{+}_{c}\to \Sigma^{*+}\pi^-\pi^0)\nonumber\\&~~~~~+\sqrt{2}\mathcal{A}(\Xi^{+}_{c}\to \Sigma^{*0}\pi^+\pi^0)-\sqrt{2}\mathcal{A}(\Xi^{+}_{c}\to \Sigma^{*+}\pi^0\pi^0)\nonumber\\&~~~~~+\sqrt{2}\mathcal{A}(\Xi^{+}_{c}\to \Sigma^{*+}\pi^+\pi^-)+2\mathcal{A}(\Xi^{0}_{c}\to \Sigma^{*0}\pi^0\pi^0)\nonumber\\&~~~~~-2\mathcal{A}(\Xi^{0}_{c}\to \Sigma^{*0}\pi^+\pi^-)-\mathcal{A}(\Xi^{0}_{c}\to \Sigma^{*-}\pi^+\pi^0)\big]=0,
\end{align}
\begin{align}
     { SumI_-^2}\,[\Xi^{0}_{c},\Sigma^{*+},\pi^+,\eta_8]&=-2\big[\mathcal{A}(\Xi^0_c\to\Sigma^{*+}\pi^-\eta_8)+\sqrt{2}\mathcal{A}(\Xi^+_c\to\Sigma^{*0}\pi^+\eta_8)\nonumber\\&~~~~~-\sqrt{2}\mathcal{A}(\Xi^+_c\to\Sigma^{*+}\pi^0\eta_8)+2\mathcal{A}(\Xi^0_c\to\Sigma^{*0}\pi^0\eta_8)\nonumber\\&~~~~~-\mathcal{A}(\Xi^0_c\to\Sigma^{*-}\pi^+\eta_8)\big]=0,
\end{align}
\begin{align}
    { SumI_-^2}\,[\Xi^{0}_{c},\Sigma^{*+},K^+,\overline{K}^0]&=-2\big[\mathcal{A}(\Xi^0_c\to\Sigma^{*+}K^0K^-)+\sqrt{2}\mathcal{A}(\Xi^+_c\to\Sigma^{*0}K^+\overline{K}^0)\nonumber\\&~~~~~+\mathcal{A}(\Xi^+_c\to\Sigma^{*+}K^0\overline{K}^0)-\mathcal{A}(\Xi^+_c\to\Sigma^{*+}K^+{K}^-)\nonumber\\&~~~~~-\sqrt{2}\mathcal{A}(\Xi^0_c\to\Sigma^{*0}K^0\overline{K}^0)+\sqrt{2}\mathcal{A}(\Xi^0_c\to\Sigma^{*0}K^+{K}^-)\nonumber\\&~~~~~-\mathcal{A}(\Xi^0_c\to\Sigma^{*-}K^+\overline{K}^0)=0,
\end{align}
\begin{align}
   { SumI_-^2}\,[\Xi^{0}_{c},\Delta^{0},\pi^+,\pi^+]& =-2\big[-2\mathcal{A}(\Xi^0_c\to\Delta^0\pi^0\pi^0)+\mathcal{A}(\Xi^0_c\to\Delta^0\pi^+\pi^-)\nonumber\\&~~~~~+\mathcal{A}(\Xi^0_c\to\Delta^0\pi^-\pi^+)+\sqrt{6}\mathcal{A}(\Xi^0_c\to\Delta^-\pi^0\pi^+)\nonumber\\&~~~~~+\sqrt{6}\mathcal{A}(\Xi^0_c\to\Delta^-\pi^+\pi^0)-\sqrt{2}\mathcal{A}(\Xi^+_c\to\Delta^0\pi^0\pi^+)\nonumber\\&~~~~~-\sqrt{2}\mathcal{A}(\Xi^0_c\to\Delta^0\pi^+\pi^0)+\sqrt{3}\mathcal{A}(\Xi^+_c\to\Delta^-\pi^+\pi^+)\big]=0,
\end{align}
\begin{align}
    { SumI_-^2}\,[\Xi^{0}_{c},\Xi^{*0},\pi^+,K^+]&=-2\big[\sqrt{2}\mathcal{A}(\Xi^0_c\to\Xi^{*-}\pi^0K^+)-\mathcal{A}(\Xi^0_c\to\Xi^{*-}\pi^+K^0)\nonumber\\&~~~~~+\mathcal{A}(\Xi^+_c\to\Xi^{*-}\pi^+K^+)-\sqrt{2}\mathcal{A}(\Xi^+_c\to\Xi^{*0}\pi^0K^+)\nonumber\\&~~~~~+\mathcal{A}(\Xi^+_c\to\Xi^{*0}\pi^+K^0)+\sqrt{2}\mathcal{A}(\Xi^0_c\to\Xi^{*0}\pi^0K^0)\nonumber\\&~~~~~+\mathcal{A}(\Xi^0_c\to\Xi^{*0}\pi^-K^+)\big]=0,
\end{align}
\begin{align}
     { SumI_-^2}\,[\Xi^{+}_{c},\Delta^{++},\pi^+,\overline{K}^0]&=2\sqrt{2}\mathcal{A}(\Xi^+_c\to\Delta^{++}\pi^0K^-)-2\big[\sqrt{6}\mathcal{A}(\Xi^+_c\to\Delta^+\pi^0\overline{K}^0)\nonumber\\&~~~~~+\sqrt{3}\mathcal{A}(\Xi^+_c\to\Delta^{+}\pi^+K^-)+\mathcal{A}(\Xi^+_c\to\Delta^{++}\pi^-\overline{K}^0)\nonumber\\&~~~~~-\sqrt{3}\mathcal{A}(\Xi^+_c\to\Delta^0\pi^+\overline{K}^0)\big]=0,
\end{align}
\begin{align}
    { SumI_-^2}\,[\Xi^{+}_{c},\Sigma^{*+},\pi^+,\pi^+]&=-4\mathcal{A}(\Xi^+_c\to\Sigma^{*+}\pi^0\pi^0)-2\big[\mathcal{A}(\Xi^+_c\to\Sigma^{*-}\pi^+\pi^+)\nonumber\\&~~~~~+\mathcal{A}(\Xi^+_c\to\Sigma^{*+}\pi^-\pi^+)+\mathcal{A}(\Xi^+_c\to\Sigma^{*+}\pi^+\pi^-)\big]=0,
\end{align}
\begin{align}
    { SumI_-^2}\,[\Lambda^{+}_{c},\Delta^{++},\pi^+,\pi^0]&=2\sqrt{3}\mathcal{A}(\Lambda^{+}_{c}\to\Delta^{0}\pi^+\pi^0)-2\big[2\mathcal{A}(\Lambda^{+}_{c}\to\Delta^{++}\pi^0\pi^-)\nonumber\\&~~~~~+\sqrt{6}\mathcal{A}(\Lambda^{+}_{c}\to\Delta^{+}\pi^0\pi^0)-\sqrt{6}\mathcal{A}(\Lambda^{+}_{c}\to\Delta^{+}\pi^+\pi^-)\nonumber\\&~~~~~+\mathcal{A}(\Lambda^{+}_{c}\to\Delta^{++}\pi^-\pi^0)\big]=0,
\end{align}
\begin{align}
    { SumI_-^2}\,[\Lambda^{+}_{c},\Delta^{++},\pi^+,\eta_8]&=2\sqrt{3}\mathcal{A}(\Lambda^+_c\to\Delta^0\pi^+\eta_8)-2\big[\sqrt{6}\mathcal{A}(\Lambda^+_c\to\Delta^+\pi^0\eta_8)\nonumber\\&~~~~~+\mathcal{A}(\Lambda^+_c\to\Delta^{++}\pi^-\eta_8)\big]=0,
\end{align}
\begin{align}
  { SumI_-^2}\,[\Lambda^{+}_{c},\Delta^{++},K^+,\overline{K}^0] & =2\sqrt{3}\mathcal{A}(\Lambda^+_c\to\Delta^0K^+\overline{K}^0)-2\mathcal{A}(\Lambda^+_c\to\Delta^{++}K^0{K}^-)\nonumber\\&~~~~~+2\sqrt{3}\mathcal{A}(\Lambda^+_c\to\Delta^+K^0\overline{K}^0)-2\sqrt{3}\mathcal{A}(\Lambda^+_c\to\Delta^+K^+{K}^-)=0,
\end{align}
\begin{align}
    { SumI_-^2}\,[\Lambda^{+}_{c},\Delta^{++},\pi^+,\pi^+]&=-2\big[2\sqrt{2}\mathcal{A}(\Lambda^{+}_{c}\to\Delta^{0}\pi^0\pi^+)+2\sqrt{2}\mathcal{A}(\Lambda^{0}_{c}\to\Delta^{0}\pi^+\pi^0)\nonumber\\&~~~~~-2\mathcal{A}(\Lambda^{+}_{c}\to\Delta^{+}\pi^0\pi^0)-\sqrt{3}\mathcal{A}(\Lambda^{+}_{c}\to\Delta^{-}\pi^+\pi^+)\nonumber\\&~~~~~+\mathcal{A}(\Lambda^{+}_{c}\to\Delta^{+}\pi^-\pi^+)+\mathcal{A}(\Lambda^{+}_{c}\to\Delta^{+}\pi^+\pi^-)\big]=0,
\end{align}
\begin{align}
    { SumI_-^2}\,[\Lambda^{+}_{c},\Sigma^{*+},\pi^+,K^+]&=2\big[\sqrt{2}\mathcal{A}(\Lambda^+_c\to\Sigma^{*0}\pi^+K^0)-2\mathcal{A}(\Lambda^+_c\to\Sigma^{*0}\pi^0K^+)\nonumber\\&~~~~~-\sqrt{2}\mathcal{A}(\Lambda^+_c\to\Sigma^{*+}\pi^0K^0)+\mathcal{A}(\Lambda^+_c\to\Sigma^{*-}\pi^+K^+)\nonumber\\&~~~~~-\mathcal{A}(\Lambda^+_c\to\Sigma^{*+}\pi^-K^+)\big]=0,
\end{align}
\begin{align}
     { SumI_-^2}\,[\Xi^{0}_{c},\Delta^{++},\pi^0,\pi^0]&=2\big[\sqrt{3}\mathcal{A}(\Xi^0_c\to\Delta^0\pi^0\pi^0)+2\mathcal{A}(\Xi^0_c\to\Delta^{++}\pi^-\pi^-)\nonumber\\&~~~~~+\sqrt{6}\mathcal{A}(\Xi^0_c\to\Delta^+\pi^0\pi^-)+\sqrt{6}\mathcal{A}(\Xi^0_c\to\Delta^+\pi^-\pi^0)\nonumber\\&~~~~~-\sqrt{2}\mathcal{A}(\Xi^+_c\to\Delta^{++}\pi^0\pi^-)-\sqrt{2}\mathcal{A}(\Xi^+_c\to\Delta^{++}\pi^-\pi^0)\nonumber\\&~~~~~-\sqrt{3}\mathcal{A}(\Xi^+_c\to\Delta^+\pi^0\pi^0)\big]=0,
\end{align}
\begin{align}
    { SumI_-^2}\,[\Xi^{0}_{c},\Delta^{++},\pi^0,\eta_8]&=2\big[\sqrt{3}\mathcal{A}(\Xi^0_c\to\Delta^0\pi^0\eta_8)+\sqrt{6}\mathcal{A}(\Xi^0_c\to\Delta^+\pi^-\eta_8)\nonumber\\&~~~~~-\sqrt{3}\mathcal{A}(\Xi^+_c\to\Delta^+\pi^0\eta_8)-\sqrt{2}\mathcal{A}(\Xi^+_c\to\Delta^{++}\pi^-\eta_8)\big]=0,
\end{align}
\begin{align}
    { SumI_-^2}\,[\Xi^0_{c},\Delta^{++},K^0,\overline{K}^0] & =2\big[\sqrt{3}\mathcal{A}(\Xi^0_c\to\Delta^0K^0\overline{K}^0)-\sqrt{3}\mathcal{A}(\Xi^0_c\to\Delta^{+}K^0{K}^-)\nonumber\\&~~~~~+\mathcal{A}(\Xi^+_c\to\Delta^{++}K^0{K}^-)+\sqrt{3}\mathcal{A}(\Xi^+_c\to\Delta^+K^0\overline{K}^0)\big]=0,
\end{align}
\begin{align}
    { SumI_-^2}\,[\Xi^0_{c},\Delta^{++},\eta_8,\eta_8]&=2\sqrt{3}\mathcal{A}(\Xi^0_c\to\Delta^0\eta_8\eta_8)-2\sqrt{3}\mathcal{A}(\Xi^+_c\to\Delta^+\eta_8\eta_8)=0,
\end{align}
\begin{align}
     { SumI_-^2}\,[\Xi^{0}_{c},\Delta^{+},\pi^+,\pi^0]&=-2\big[2\sqrt{2}\mathcal{A}(\Xi^0_c\to\Delta^0\pi^0\pi^0)-2\sqrt{2}\mathcal{A}(\Xi^0_c\to\Delta^0\pi^+\pi^-)\nonumber\\&~~~~~+2\mathcal{A}(\Xi^0_c\to\Delta^+\pi^0\pi^-)-\sqrt{3}\mathcal{A}(\Xi^0_c\to\Delta^-\pi^+\pi^0)\nonumber\\&~~~~~+\mathcal{A}(\Xi^0_c\to\Delta^+\pi^-\pi^0)+2\mathcal{A}(\Xi^+_c\to\Delta^0\pi^+\pi^0)\nonumber\\&~~~~~-\sqrt{2}\mathcal{A}(\Xi^+_c\to\Delta^+\pi^0\pi^0)+\sqrt{2}\mathcal{A}(\Xi^+_c\to\Delta^+\pi^+\pi^-)\big]=0,
\end{align}
\begin{align}
   { SumI_-^2}\,[\Xi^{0}_{c},\Delta^{+},\pi^+,\eta_8]&=-2\big[2\sqrt{2}\mathcal{A}(\Xi^0_c\to\Delta^0\pi^0\eta_8)-\sqrt{3}\mathcal{A}(\Xi^0_c\to\Delta^-\pi^+\eta_8)\nonumber\\&~~~~~ +\mathcal{A}(\Xi^0_c\to\Delta^+\pi^-\eta_8) +2\mathcal{A}(\Xi^+_c\to\Delta^0\pi^+\eta_8) \nonumber\\&~~~~~-\sqrt{2}\mathcal{A}(\Xi^+_c\to\Delta^+\pi^0\eta_8)\big] =0,
\end{align}
\begin{align}
  { SumI_-^2}\,[\Xi^0_{c},\Delta^{+},K^+,\overline{K}^0] &= -2\big[-2\mathcal{A}(\Xi^0_c\to\Delta^0K^0\overline{K}^0) +2\mathcal{A}(\Xi^0_c\to\Delta^0K^+{K}^-) \nonumber\\&~~~~~+\mathcal{A}(\Xi^0_c\to\Delta^+K^0{K}^-) -\sqrt{3}\mathcal{A}(\Xi^0_c\to\Delta^-K^+\overline{K}^0) \nonumber\\&~~~~~+2\mathcal{A}(\Xi^+_c\to\Delta^0K^+\overline{K}^0) +\mathcal{A}(\Xi^+_c\to\Delta^+K^0\overline{K}^0)\nonumber\\&~~~~~-\mathcal{A}(\Xi^+_c\to\Delta^+K^+{K}^-) \big]=0,
\end{align}
\begin{align}
     { SumI_-^2}\,[\Xi^{0}_{c},\Sigma^{*+},\pi^+,K^0]&=-2\big[2\mathcal{A}(\Xi^0_c\to\Sigma^{*0}\pi^0K^0)-\mathcal{A}(\Xi^0_c\to\Sigma^{*-}\pi^+K^0)\nonumber\\&~~~~~+\mathcal{A}(\Xi^0_c\to\Sigma^{*+}\pi^-K^0)+\sqrt{2}\mathcal{A}(\Xi^+_c\to\Sigma^{*0}\pi^+K^0)\nonumber\\&~~~~~-\sqrt{2}\mathcal{A}(\Xi^+_c\to\Sigma^{*+}\pi^0K^0)\big]=0,
\end{align}
\begin{align}
  { SumI_-^2}\,[\Xi^{0}_{c},\Sigma^{*+},\pi^0,K^+]&=2\big[\sqrt{2}\mathcal{A}(\Xi^0_c\to\Sigma^{*0}\pi^0K^0)+2\mathcal{A}(\Xi^0_c\to\Sigma^{*0}\pi^-K^+)\nonumber\\&~~~~~+\mathcal{A}(\Xi^0_c\to\Sigma^{*-}\pi^0K^+)+\sqrt{2}\mathcal{A}(\Xi^0_c\to\Sigma^{*+}\pi^-K^0)\nonumber\\&~~~~~-\sqrt{2}\mathcal{A}(\Xi^+_c\to\Sigma^{*0}\pi^0K^+)-\mathcal{A}(\Xi^+_c\to\Sigma^{*+}\pi^0K^0)\nonumber\\&~~~~~-\sqrt{2}\mathcal{A}(\Xi^+_c\to\Sigma^{*+}\pi^-K^+)\big]=0,
\end{align}
\begin{align}
      { SumI_-^2}\,[\Xi^{0}_{c},\Sigma^{*+},K^+,\eta_8]&=2\big[\sqrt{2}\mathcal{A}(\Xi^0_c\to\Sigma^{*0}K^0\eta_8)+\mathcal{A}(\Xi^0_c\to\Sigma^{*-}K^+\eta_8)\nonumber\\&~~~~~-\sqrt{2}\mathcal{A}(\Xi^+_c\to\Sigma^{*0}K^+\eta_8)-\mathcal{A}(\Xi^+_c\to\Sigma^{*+}K^0\eta_8)\big]=0,
\end{align}
\begin{align}
   { SumI_-^2}\,[\Xi^{0}_{c},\Delta^{0},\pi^+,\pi^+]&=-2\big[-2\mathcal{A}(\Xi^0_c\to\Delta^0\pi^0\pi^0)+\mathcal{A}(\Xi^0_c\to\Delta^0\pi^-\pi^+)\nonumber\\&~~~~~+\mathcal{A}(\Xi^0_c\to\Delta^0\pi^+\pi^-)+\sqrt{6}\mathcal{A}(\Xi^0_c\to\Delta^-\pi^0\pi^+)\nonumber\\&~~~~~+\sqrt{6}\mathcal{A}(\Xi^0_c\to\Delta^-\pi^+\pi^0)-\sqrt{2}\mathcal{A}(\Xi^+_c\to\Delta^0\pi^0\pi^+)\nonumber\\&~~~~~-\sqrt{2}\mathcal{A}(\Xi^+_c\to\Delta^0\pi^+\pi^0)+\sqrt{3}\mathcal{A}(\Xi^+_c\to\Delta^-\pi^+\pi^+)\big]=0,
\end{align}
\begin{align}
    { SumI_-^2}\,[\Xi^{0}_{c},\Sigma^{*0},\pi^+,K^+]&=-2\big[\sqrt{2}\mathcal{A}(\Xi^0_c\to\Sigma^{*0}\pi^0K^0)+\mathcal{A}(\Xi^0_c\to\Sigma^{*0}\pi^-K^+)\nonumber\\&~~~~~+2\mathcal{A}(\Xi^0_c\to\Sigma^{*-}\pi^0K^+)-\sqrt{2}\mathcal{A}(\Xi^0_c\to\Sigma^{*-}\pi^+K^0)\nonumber\\&~~~~~-\sqrt{2}\mathcal{A}(\Xi^+_c\to\Sigma^{*0}\pi^0K^+)+\mathcal{A}(\Xi^+_c\to\Sigma^{*0}\pi^+K^0)\nonumber\\&~~~~~+\sqrt{2}\mathcal{A}(\Xi^+_c\to\Sigma^{*-}\pi^+K^+)\big]=0,
\end{align}
\begin{align}
     { SumI_-^2}\,[\Xi^{0}_{c},\Xi^{*0},K^+,K^+]&=2\big[\mathcal{A}(\Xi^0_c\to\Xi^{*-}K^0K^+)+\mathcal{A}(\Xi^0_c\to\Xi^{*-}K^+K^0)\nonumber\\&~~~~~-\mathcal{A}(\Xi^+_c\to\Xi^{*0}K^0K^+)-\mathcal{A}(\Xi^+_c\to\Xi^{*0}K^+K^0)\nonumber\\&~~~~~-\mathcal{A}(\Xi^+_c\to\Xi^{*-}K^+K^+)+\mathcal{A}(\Xi^0_c\to\Xi^{*0}K^0K^0)\big]=0,
\end{align}
\begin{align}
    { SumI_-^2}\,[\Xi^{+}_{c},\Delta^{++},\pi^+,\pi^0]&=2\sqrt{3}\mathcal{A}(\Xi^+_c\to\Delta^0\pi^+\pi^0)-4\mathcal{A}(\Xi^+_c\to\Delta^{++}\pi^0\pi^-)\nonumber\\&~~~~~-2\sqrt{6}\mathcal{A}(\Xi^+_c\to\Delta^+\pi^0\pi^0)+2\sqrt{6}\mathcal{A}(\Xi^+_c\to\Delta^+\pi^+\pi^-)\nonumber\\&~~~~~-2\mathcal{A}(\Xi^+_c\to\Delta^{++}\pi^-\pi^0)=0,
\end{align}
\begin{align}
     { SumI_-^2}\,[\Xi^{+}_{c},\Delta^{++},\pi^+,\eta_8]&= 2\sqrt{3}\mathcal{A}(\Xi^{+}_{c}\to\Delta^{0}\pi^+\eta_8)-2\sqrt{6}\mathcal{A}(\Xi^{+}_{c}\to\Delta^{+}\pi^0\eta_8)\nonumber\\&~~~~~-2\mathcal{A}(\Xi^{+}_{c}\to\Delta^{++}\pi^-\eta_8)=0,
\end{align}
\begin{align}
    { SumI_-^2}\,[\Xi^+_{c},\Delta^{++},K^+,\overline{K}^0] &=2\big[\sqrt{3}\mathcal{A}(\Xi^+_c\to\Delta^0K^+\overline{K}^0)-\mathcal{A}(\Xi^+_c\to\Delta^{++}K^0{K}^-)\nonumber\\&~~~~~+\sqrt{3}\mathcal{A}(\Xi^+_c\to\Delta^+K^0\overline{K}^0)-\sqrt{3}\mathcal{A}(\Xi^+_c\to\Delta^+K^+{K}^-)\big]=0,
\end{align}
\begin{align}
  { SumI_-^2}\,[\Xi^{+}_{c},\Delta^{+},\pi^+,\pi^+]&=  -2\big[2\sqrt{2}\mathcal{A}(\Xi^{+}_{c}\to\Delta^{0}\pi^0\pi^+)+2\sqrt{2}\mathcal{A}(\Xi^{+}_{c}\to\Delta^{0}\pi^+\pi^0)\nonumber\\&~~~~~-2\mathcal{A}(\Xi^{+}_{c}\to\Delta^{+}\pi^0\pi^0)-\sqrt{3}\mathcal{A}(\Xi^{+}_{c}\to\Delta^{-}\pi^+\pi^+)\nonumber\\&~~~~~+\mathcal{A}(\Xi^{+}_{c}\to\Delta^{+}\pi^-\pi^+)+\mathcal{A}(\Xi^{+}_{c}\to\Delta^{+}\pi^+\pi^-)\big]=0,
\end{align}
\begin{align}
  { SumI_-^2}\,[\Xi^{+}_{c},\Sigma^{*+},\pi^+,K^+]&=-2\big[2\mathcal{A}(\Xi^{+}_{c}\to\Sigma^{*0}\pi^0K^+)-\sqrt{2}\mathcal{A}(\Xi^{+}_{c}\to\Sigma^{*0}\pi^+K^0)\nonumber\\&~~~~~+\sqrt{2}\mathcal{A}(\Xi^{+}_{c}\to\Sigma^{*+}\pi^0K^0)-\mathcal{A}(\Xi^{+}_{c}\to\Sigma^{*-}\pi^+K^+)\nonumber\\&~~~~~+\mathcal{A}(\Xi^{+}_{c}\to\Sigma^{*+}\pi^-K^+)\big]=0,
\end{align}
\begin{align}
   { SumI_-^2}\,[\Lambda^{+}_{c},\Delta^{++},\pi^+,K^0]&=  2\sqrt{3}\mathcal{A}(\Lambda^+_c\to\Delta^0\pi^+K^0)-2\sqrt{6}\mathcal{A}(\Lambda^+_c\to\Delta^+\pi^0K^0)\nonumber\\&~~~~~-2\mathcal{A}(\Lambda^+_c\to\Delta^{++}\pi^-K^0)=0,
\end{align}
\begin{align}
    { SumI_-^2}\,[\Lambda^{+}_{c},\Delta^{++},\pi^0,K^+]&= 2\big[\sqrt{3}\mathcal{A}(\Lambda^{+}_{c}\to\Delta^{0}\pi^0K^+)+\sqrt{3}\mathcal{A}(\Lambda^{+}_{c}\to\Delta^{+}\pi^0K^0)\nonumber\\&~~~~~+\sqrt{6}\mathcal{A}(\Lambda^{+}_{c}\to\Delta^{+}\pi^-K^+)+\sqrt{2}\mathcal{A}(\Lambda^{+}_{c}\to\Delta^{++}\pi^-K^0)\big]=0,
\end{align}
\begin{align}
     { SumI_-^2}\,[\Lambda^{+}_{c},\Delta^{++},K^+,\eta_8]&=2\sqrt{3}\big[\mathcal{A}(\Lambda^+_c\to\Delta^{0}K^+\eta_8)+\mathcal{A}(\Lambda^+_c\to\Delta^{+}K^0\eta_8)\big]=0
\end{align}
\begin{align}
    { SumI_-^2}\,[\Xi^{+}_{c},\Delta^{+},\pi^+,K^+]&=-2\big[2\sqrt{2}\mathcal{A}(\Lambda^+_c\to\Delta^0\pi^0K^+)-2\mathcal{A}(\Lambda^+_c\to\Delta^0\pi^+K^0)\nonumber\\&~~~~~+\sqrt{2}\mathcal{A}(\Lambda^+_c\to\Delta^+\pi^0K^0)-\sqrt{3}\mathcal{A}(\Lambda^+_c\to\Delta^-\pi^+K^+)\nonumber\\&~~~~~+\mathcal{A}(\Lambda^+_c\to\Delta^+\pi^-K^+)\big]=0,
\end{align}
\begin{align}
    { SumI_-^2}\,[\Lambda^{+}_{c},\Sigma^{*+},K^+,K^+]&=2\big[\sqrt{2}\mathcal{A}(\Lambda^+_c\to\Sigma^{*0}K^+K^0)+\mathcal{A}(\Lambda^+_c\to\Sigma^{*+}K^0K^0)\nonumber\\&~~~~~+\mathcal{A}(\Lambda^+_c\to\Sigma^{*-}K^+K^+)+\sqrt{2}\mathcal{A}(\Lambda^+_c\to\Sigma^{*0}K^0K^+)\big]=0
\end{align}
\begin{align}
     { SumI_-^3}\,[\Xi^{0}_{c},\Delta^{++},\pi^+,\overline{K}^0]&=6\big[\sqrt{6}\mathcal{A}(\Xi^0_c\to\Delta^+\pi^0K^-)+\mathcal{A}(\Xi^0_c\to\Delta^{++}\pi^-K^-)\nonumber\\&~~~~~-\sqrt{3}\mathcal{A}(\Xi^0_c\to\Delta^+\pi^-\overline{K}^0)-\sqrt{2}\mathcal{A}(\Xi^+_c\to\Delta^{++}\pi^0K^-)\nonumber\\&~~~~~+\sqrt{6}\mathcal{A}(\Xi^+_c\to\Delta^+\pi^0\overline{K}^0)+\sqrt{3}\mathcal{A}(\Xi^+_c\to\Delta^+\pi^+K^-)\nonumber\\&~~~~~+\mathcal{A}(\Xi^+_c\to\Delta^{++}\pi^-\overline{K}^0)-\sqrt{3}\mathcal{A}(\Xi^+_c\to\Delta^0\pi^+\overline{K}^0)\nonumber\\&~~~~~-\sqrt{6}\mathcal{A}(\Xi^0_c\to\Delta^0\pi^0\overline{K}^0)-\sqrt{3}\mathcal{A}(\Xi^0_c\to\Delta^0\pi^+K^-)\nonumber\\&~~~~~+\mathcal{A}(\Xi^0_c\to\Delta^-\pi^+\overline{K}^0)\big]=0,
\end{align}
\begin{align}
    { SumI_-^3}\,[\Lambda^{+}_{c},\Delta^{++},\pi^+,\pi^+]&=6\big[-\sqrt{6}\mathcal{A}(\Lambda^{+}_{c}\to\Delta^{0}\pi^0\pi^+)-\sqrt{6}\mathcal{A}(\Lambda^{+}_{c}\to\Delta^{0}\pi^+\pi^0)\nonumber\\&~~~~~+2\sqrt{3}\mathcal{A}(\Lambda^{+}_{c}\to\Delta^{+}\pi^0\pi^0)-\sqrt{3}\mathcal{A}(\Lambda^{+}_{c}\to\Delta^{+}\pi^-\pi^+)\nonumber\\&~~~~~+\mathcal{A}(\Lambda^{+}_{c}\to\Delta^{-}\pi^+\pi^+)+\sqrt{2}\mathcal{A}(\Lambda^{+}_{c}\to\Delta^{++}\pi^-\pi^0)\nonumber\\&~~~~~-\sqrt{3}\mathcal{A}(\Lambda^{+}_{c}\to\Delta^{+}\pi^+\pi^-)+\sqrt{2}\mathcal{A}(\Lambda^{+}_{c}\to\Delta^{++}\pi^0\pi^-)=0,
\end{align}
\begin{align}
     { SumI_-^3}\,[\Xi^{0}_{c},\Delta^{++},\pi^+,\pi^0]&=-6\big[\sqrt{6}\mathcal{A}(\Xi^0_c\to\Delta^0\pi^0\pi^0)-\sqrt{6}\mathcal{A}(\Xi^0_c\to\Delta^0\pi^+\pi^-)\nonumber\\&~~~~~+\sqrt{2}\mathcal{A}(\Xi^0_c\to\Delta^{++}\pi^-\pi^-)+2\sqrt{3}\mathcal{A}(\Xi^0_c\to\Delta^+\pi^0\pi^-)\nonumber\\&~~~~~-\mathcal{A}(\Xi^0_c\to\Delta^-\pi^+\pi^0)+\sqrt{3}\mathcal{A}(\Xi^0_c\to\Delta^+\pi^-\pi^0)\nonumber\\&~~~~~-2\mathcal{A}(\Xi^+_c\to\Delta^{++}\pi^0\pi^-)-\sqrt{6}\mathcal{A}(\Xi^+_c\to\Delta^+\pi^0\pi^0)\nonumber\\&~~~~~+\sqrt{6}\mathcal{A}(\Xi^+_c\to\Delta^+\pi^+\pi^-)+\sqrt{3}\mathcal{A}(\Xi^+_c\to\Delta^0\pi^+\pi^0)\nonumber\\&~~~~~-\mathcal{A}(\Xi^+_c\to\Delta^{++}\pi^-\pi^0)\big]=0,
\end{align}
\begin{align}
     { SumI_-^3}\,[\Xi^{0}_{c},\Delta^{++},\pi^+,\eta_8]&=6\big[-\sqrt{6}\mathcal{A}(\Xi^0_c\to\Delta^0\pi^0\eta_8)+\mathcal{A}(\Xi^0_c\to\Delta^-\pi^+\eta_8)\nonumber\\&~~~~~-\sqrt{3}\mathcal{A}(\Xi^0_c\to\Delta^+\pi^-\eta_8)-\sqrt{3}\mathcal{A}(\Xi^+_c\to\Delta^0\pi^+\eta_8)\nonumber\\&~~~~~+\sqrt{6}\mathcal{A}(\Xi^+_c\to\Delta^+\pi^0\eta_8)+\mathcal{A}(\Xi^+_c\to\Delta^{++}\pi^-\eta_8)\big]=0,
\end{align}
\begin{align}
    { SumI_-^3}\,[\Xi^{0}_{c},\Delta^{++},K^+,\overline{K}^0]&=6\big[\sqrt{3}\mathcal{A}(\Xi^0_c\to\Delta^0K^0\overline{K}^0)-\sqrt{3}\mathcal{A}(\Xi^0_c\to\Delta^0K^+K^-)\nonumber\\&~~~~~-\sqrt{3}\mathcal{A}(\Xi^0_c\to\Delta^+K^0K^-)+\mathcal{A}(\Xi^0_c\to\Delta^-K^+\overline{K}^0)\nonumber\\&~~~~~-\sqrt{3}\mathcal{A}(\Xi^+_c\to\Delta^0K^+\overline{K}^0)+\mathcal{A}(\Xi^+_c\to\Delta^{++}K^0{K}^-)\nonumber\\&~~~~~-\sqrt{3}\mathcal{A}(\Xi^+_c\to\Delta^+K^0\overline{K}^0)+\sqrt{3}\mathcal{A}(\Xi^+_c\to\Delta^+K^+{K}^-)\big]=0,
\end{align}
\begin{align}
    { SumI_-^3}\,[\Xi^{+}_{c},\Delta^{++},\pi^+,\pi^+]&=6\big[-\sqrt{6}\mathcal{A}(\Xi^+_c\to\Delta^0\pi^0\pi^+)-\sqrt{6}\mathcal{A}(\Xi^+_c\to\Delta^{0}\pi^+\pi^0)\nonumber\\&~~~~~+\sqrt{2}\mathcal{A}(\Xi^+_c\to\Delta^{++}\pi^0\pi^-)+\mathcal{A}(\Xi^+_c\to\Delta^-\pi^+\pi^+)\nonumber\\&~~~~~+2\sqrt{3}\mathcal{A}(\Xi^+_c\to\Delta^+\pi^0\pi^0)-\sqrt{3}\mathcal{A}(\Xi^+_c\to\Delta^+\pi^+\pi^-)\nonumber\\&~~~~~+\sqrt{2}\mathcal{A}(\Xi^+_c\to\Delta^{++}\pi^-\pi^0)-\sqrt{3}\mathcal{A}(\Xi^+_c\to\Delta^+\pi^-\pi^+)\big]=0,
\end{align}
\begin{align}
      { SumI_-^3}\,[\Lambda^{+}_{c},\Delta^{++},\pi^+,K^+]&=-6\big[\sqrt{6}\mathcal{A}(\Lambda^{+}_{c}\to\Delta^{0}\pi^0K^+)-\sqrt{3}\mathcal{A}(\Lambda^{+}_{c}\to\Delta^{0}\pi^+K^0)\nonumber\\&~~~~~+\sqrt{6}\mathcal{A}(\Lambda^{+}_{c}\to\Delta^{+}\pi^0K^0)-\mathcal{A}(\Lambda^{+}_{c}\to\Delta^{-}\pi^+K^+)\nonumber\\&~~~~~+\sqrt{3}\mathcal{A}(\Lambda^{+}_{c}\to\Delta^{+}\pi^-K^+)+\mathcal{A}(\Lambda^{+}_{c}\to\Delta^{++}\pi^-K^0)\big]=0.
\end{align}

\section{Isospin sum rules for doubly charmed baryon decays}\label{doublyx}

\subsection{$\mathcal{B}_{cc}\to M_8\mathcal{B}_{\overline c3}$ and $\mathcal{B}_{cc}\to M_8\mathcal{B}_{c6}$ modes}\label{doubly1}
The isospin sum rule for the $\mathcal{B}_{cc}\to M_8\mathcal{B}_{\overline c3}$ modes is
\begin{align}
{ SumI_-}\,[\Xi^+_{cc}, \pi^+,\Xi^{+}_c]&= -\sqrt{2}\,\mathcal{A}(\Xi^+_{cc}\to \pi^0\Xi^{+}_c)- \mathcal{A}(\Xi^{++}_{cc}\to \pi^+\Xi^{+}_c)\nonumber\\&~~~~~+\mathcal{A}(\Xi^+_{cc}\to \pi^+\Xi^{0}_c)=0.
\end{align}

The isospin sum rules for the $\mathcal{B}_{cc}\to M_8\mathcal{B}_{c6}$ modes are
\begin{align}
{ SumI_-}\,[\Xi^+_{cc}, \overline K^0,\Sigma^{++}_c]&= -\mathcal{A}(\Xi^+_{cc}\to K^-\Sigma^{++}_c)- \mathcal{A}(\Xi^{++}_{cc}\to \overline K^0\Sigma^{++}_c)\nonumber\\&~~~~~+\sqrt{2}\,\mathcal{A}(\Xi^+_{cc}\to \overline K^0\Sigma^{+}_c)=0,
\end{align}
\begin{align}
{ SumI_-}\,[\Xi^+_{cc}, \pi^+,\Xi^{*+}_c]&= -\sqrt{2}\,\mathcal{A}(\Xi^+_{cc}\to \pi^0\Xi^{*+}_c)- \mathcal{A}(\Xi^{++}_{cc}\to \pi^+\Xi^{*+}_c)\nonumber\\&~~~~~+\mathcal{A}(\Xi^+_{cc}\to \pi^+\Xi^{*0}_c)=0,
\end{align}
\begin{align}
{ SumI_-^2}\,[\Xi^+_{cc}, \pi^+,\Sigma^{++}_c]&=2\,\big[ \sqrt{2}\,\mathcal{A}( \Xi^{++}_{cc}\to\pi^0\Sigma^{++}_c)
-2\,\mathcal{A}(\Xi^{+}_{cc}\to\pi^0\Sigma^{+}_c)\nonumber\\&~~~~~
-\sqrt{2}\,\mathcal{A}(\Xi^{++}_{cc}\to\pi^+\Sigma^{+}_c)
-\mathcal{A}(\Xi^{+}_{cc}\to\pi^-\Sigma^{++}_c)\nonumber\\&~~~~~+
\mathcal{A}(\Xi^{+}_{cc}\to\pi^+\Sigma^{0}_c)\big]=0,
\end{align}
\begin{align}
{ SumI_-^2}\,[\Xi^+_{cc}, K^+,\Sigma^{++}_c]&=2\,\big[ -\mathcal{A}( \Xi^{++}_{cc}\to K^0\Sigma^{++}_c)
+\sqrt{2}\,\mathcal{A}(\Xi^{+}_{cc}\to K^0\Sigma^{+}_c)
\nonumber\\&~~~~~~
-\sqrt{2}\,\mathcal{A}(\Xi^{++}_{cc}\to K^+\Sigma^{+}_c)+
\mathcal{A}(\Xi^{+}_{cc}\to K^+\Sigma^{0}_c)\big]=0,
\end{align}
\begin{align}
{ SumI_-^2}\,[\Omega^+_{cc}, \pi^+,\Sigma^{++}_c]&=-2\,\big[\, 2\,\mathcal{A}( \Omega^{+}_{cc}\to \pi^0\Sigma^{+}_c)
+\mathcal{A}(\Omega^{+}_{cc}\to \pi^-\Sigma^{++}_c)
\nonumber\\&~~~~~-\mathcal{A}(\Omega^{+}_{cc}\to \pi^+\Sigma^{0}_c)\big]=0.
\end{align}

\subsection{$\mathcal{B}_{cc}\to D\mathcal{B}_{8}$ and $\mathcal{B}_{cc}\to D\mathcal{B}_{10}$ modes}\label{doubly2}

The isospin sum rule for the $\mathcal{B}_{cc}\to D\mathcal{B}_{8}$ is
\begin{align}
{ SumI_-}\,[\Xi^+_{cc}, D^+,\Sigma^{+}]&= -\sqrt{2}\,\mathcal{A}(\Xi^+_{cc}\to D^+\Sigma^{0})- \mathcal{A}(\Xi^{++}_{cc}\to D^+\Sigma^{+})\nonumber\\&~~~~~-\mathcal{A}(\Xi^+_{cc}\to D^0\Sigma^{+})=0.
\end{align}

The isospin sum rules for the $\mathcal{B}_{cc}\to D\mathcal{B}_{10}$ modes are
\begin{align}
{ SumI_-}\,[\Xi^+_{cc}, D^+,\Sigma^{*+}]&= \sqrt{2}\,\mathcal{A}(\Xi^+_{cc}\to D^+\Sigma^{*0})- \mathcal{A}(\Xi^{++}_{cc}\to D^+\Sigma^{*+})\nonumber\\&~~~~~-\mathcal{A}(\Xi^+_{cc}\to D^0\Sigma^{*+})=0,
\end{align}
\begin{align}
{ SumI_-^2}\,[\Xi^+_{cc}, D^+,\Delta^{++}]&=-2\sqrt{3}\,\big[-
\mathcal{A}(\Xi^{+}_{cc}\to D^+\Delta^{0})
+\mathcal{A}(\Xi^{++}_{cc}\to D^+\Delta^{+})\nonumber\\&~~~~~
+\mathcal{A}(\Xi^{+}_{cc}\to D^0\Delta^{+})\big]=0,
\end{align}
\begin{align}
{ SumI_-^2}\,[\Xi^+_{cc}, D^+_s,\Delta^{++}]&=2\sqrt{3}\,\big[\,
\mathcal{A}(\Xi^{+}_{cc}\to D^+_s\Delta^{0})
-\mathcal{A}(\Xi^{++}_{cc}\to D^+_s\Delta^{+})\big]=0,
\end{align}
\begin{align}
{ SumI_-^2}\,[\Omega^+_{cc}, D^+,\Delta^{++}]&=2\sqrt{3}\,\big[\,
\mathcal{A}(\Omega^{+}_{cc}\to D^+\Delta^{0})
-\mathcal{A}(\Omega^{+}_{cc}\to D^0\Delta^{+})\big]=0.
\end{align}

\subsection{$\mathcal{B}_{cc} \to \mathcal{B}_{c\overline{3}}{M_8}{M_8}$ and $\mathcal{B}_{cc} \to \mathcal{B}_{c6}{M_8}{M_8}$ modes}\label{doubly3}

The isospin rules for the $\mathcal{B}_{cc} \to \mathcal{B}_{c\overline{3}}{M_8}{M_8}$ modes derived via Eq.~\eqref{rule7} are
\begin{align}
{ SumI_-}\,[\Xi^{++}_{cc}, \Xi^+_{c}, \pi^+,\pi^+]&=\mathcal{A}(\Xi^{++}_{cc}\to \Xi^0_{c}\pi^+\pi^+) -\sqrt{2}\mathcal{A}(\Xi^{++}_{cc}\to\Xi^+_{c} \pi^+\pi^0)\nonumber\\&~~~~~-\sqrt{2}\mathcal{A}(\Xi^{++}_{cc}\to\Xi^+_{c} \pi^0\pi^+)=0,
\end{align}
\begin{align}
 { SumI_-}\,[\Xi^{+}_{cc},\Xi^0_{c},\pi^+,\pi^+]&= -\mathcal{A}(\Xi^{++}_{cc}\to \Xi^0_{c}\pi^+\pi^+) -\sqrt{2}\mathcal{A}(\Xi^+_{cc}\to \Xi^0_{c}\pi^+\pi^0)\nonumber\\&~~~~~ -\sqrt{2}\mathcal{A}(\Xi^+_{cc}\to \Xi^0_{c}\pi^0\pi^+)=0 ,
\end{align}
\begin{align}
 {SumI_-}\,[\Xi^+_{cc},\Xi^+_{c},\pi^+,\pi^0]=&\mathcal{A}(\Xi^+_{cc}\to \Xi^0_{c}\pi^+\pi^0)-\sqrt{2}\mathcal{A}(\Xi^{+}_{cc}\to \Xi^+_{c}\pi^0\pi^0)\nonumber\\&+\sqrt{2}\mathcal{A}(\Xi^+_{cc}\to \Xi^+_c\pi^+\pi^-)-\mathcal{A}(\Xi^{++}_{cc}\to \Xi^+_c\pi^+\pi^0)=0,
\end{align}
\begin{align}
  {SumI_-}\,[\Xi^+_{cc},\Xi^+_{c},\pi^+,\eta_8]=&\mathcal{A}(\Xi^+_{cc}\to \Xi^0_{c}\pi^+\eta_8)-\sqrt{2}\mathcal{A}(\Xi^+_{cc}\to\Xi^+_{c}\pi^0\eta_8)\nonumber\\&~~~~~-\mathcal{A}(\Xi^{++}_{cc}\to\Xi^+_{c}\pi^+\eta_8)=0,
\end{align}
\begin{align}
{SumI_-}\,[\Xi^+_{cc},\Xi^+_{c},K^+,\overline{K}^0]&=\mathcal{A}(\Xi^+_{cc}\to \Xi^0_{c}K^+\overline{K}^0)+\mathcal{A}(\Xi^+_{cc}\to\Xi^+_{c}K^0\overline{K}^0)\nonumber\\&~~~~~-\mathcal{A}(\Xi^{+}_{cc}\to\Xi^+_cK^+K^-)-\mathcal{A}(\Xi^{++}_{cc}\to \Xi^+_cK^+\overline{K}^0)=0,
\end{align}
\begin{align}
 {SumI_-}\,[\Xi^+_{cc},\Lambda^+_{c},\pi^+,\overline K^0]&=-\sqrt{2}\mathcal{A}(\Xi^+_{cc}\to \Lambda^+_{c}\pi^0\overline K^0)-\mathcal{A}(\Xi^{+}_{cc}\to \Lambda^+_{c}\pi^+K^-)\nonumber\\&-\mathcal{A}(\Xi^{++}_{cc}\to \Lambda^+_c\pi^+\overline{K}^0)=0,
\end{align}
\begin{align}
{SumI_-}\,[\Omega^+_{cc},\Xi^+_{c},\pi^+,\overline K^0]=&\mathcal{A}(\Omega^+_{cc}\to \Xi^{0}_{c}\pi^+\overline{K}^0)-\sqrt{2}\mathcal{A}(\Omega^{+}_{cc}\to \Xi^+_{c}\pi^0\overline{K}^0)\nonumber\\&-\mathcal{A}(\Omega^{+}_{cc}\to \Xi^+_{c}\pi^+K^-)=0,
\end{align}
\begin{align}
{ SumI_-^2}\,[\Xi^+_{cc}, \Xi^+_c,\pi^{+},\pi^+]&=-2\big[\mathcal{A}(\Xi^{++}_{cc}\to \Xi^0_c\pi^{+}\pi^+)+\sqrt{2}\mathcal{A}(\Xi^{+}_{cc}\to \Xi^0_c\pi^{0}\pi^+)\nonumber\\&~~~~~-2\mathcal{A}(\Xi^+_{cc}\to\Xi^+_c \pi^0\pi^0)+\mathcal{A}(\Xi^+_{cc}\to\Xi^+_c \pi^+\pi^-)\nonumber\\&~~~~~-\sqrt{2}\mathcal{A}(\Xi^{++}_{cc}\to\Xi^+_c\pi^0\pi^+)+\mathcal{A}(\Xi^+_{cc}\to\Xi^+_c \pi^-\pi^+)\nonumber\\&~~~~~+\sqrt{2}\mathcal{A}(\Xi^{+}_{cc}\to\Xi^0_c\pi^+\pi^0)-\sqrt{2}\mathcal{A}(\Xi^{++}_{cc}\to\Xi^+_c \pi^+\pi^0)\big]=0,
\end{align}
\begin{align}
 { SumI_-^2}\,[\Xi^+_{cc}, \Xi^+_c,\pi^{+},K^+]&=-2\big[\sqrt{2}\mathcal{A}(\Xi^{+}_{cc}\to \Xi^0_c\pi^{0}K^+)+\mathcal{A}(\Xi^{++}_{cc}\to \Xi^0_c\pi^{+}K^+)\nonumber\\&~~~~~-\mathcal{A}(\Xi^+_{cc}\to\Xi^0_c \pi^+K^0)-\sqrt{2}\mathcal{A}(\Xi^{++}_{cc}\to\Xi^+_c\pi^0K^+)\nonumber\\&~~~~~+\sqrt{2}\mathcal{A}(\Xi^{+}_{cc}\to\Xi^+_c\pi^0K^0)+\mathcal{A}(\Xi^{+}_{cc}\to\Xi^+_c\pi^-K^+)\nonumber\\&~~~~~+\mathcal{A}(\Xi^{++}_{cc}\to\Xi^+_c\pi^+K^0)\big]=0,
\end{align}
\begin{align}
{ SumI_-^2}\,[\Xi^+_{cc}, \Lambda^+_c,\pi^{+},\pi^+]&=2\big[\sqrt{2}\mathcal{A}(\Xi^{++}_{cc}\to\Lambda^+_c\pi^0\pi^+)+2\mathcal{A}(\Xi^+_{cc}\to\Lambda^+_c\pi^0\pi^0)\nonumber\\&~~~~~-\mathcal{A}(\Xi^+_{cc}\to\Lambda^+_c\pi^+\pi^-)-\mathcal{A}(\Xi^+_{cc}\to\Lambda^+_c\pi^-\pi^+)\nonumber\\&~~~~~\sqrt{2}\mathcal{A}(\Xi^{++}_{cc}\to\Lambda^+_c\pi^+\pi^0)\big]=0,
\end{align}
\begin{align}
 { SumI_-^2}\,[\Omega^+_{cc}, \Xi^+_c,\pi^{+},\pi^+]&=-2\big[\sqrt{2}\mathcal{A}(\Omega^{+}_{cc}\to\Xi^0_c\pi^0\pi^+)-2\mathcal{A}(\Omega^+_{cc}\to\Xi^+_c\pi^0\pi^0)\nonumber\\&~~~~~+\mathcal{A}(\Omega^+_{cc}\to\Xi^+_c\pi^+\pi^-)+\sqrt{2}\mathcal{A}(\Omega^{+}_{cc}\to\Xi^0_c\pi^+\pi^0)\nonumber\\&~~~~~+\mathcal{A}(\Omega^+_{cc}\to\Xi^+_c\pi^-\pi^+)\big]=0,
\end{align}
\begin{align}
  { SumI_-^2}\,[\Xi^+_{cc}, \Lambda^+_c,\pi^+,K^{+}]&=-2\big[-\sqrt{2}\mathcal{A}(\Xi^{++}_{cc}\to\Lambda^+_c\pi^0K^+)+\sqrt{2}\mathcal{A}(\Xi^{+}_{cc}\to\Lambda^+_c\pi^0K^0)\nonumber\\&~~~~~+\mathcal{A}(\Xi^+_{cc}\to\Lambda^+_c\pi^-K^+)+\mathcal{A}(\Xi^{++}_{cc}\to\Lambda^+_c\pi^+K^0)\big]=0,
\end{align}
\begin{align}
 { SumI_-^2}\,[\Omega^+_{cc}, \Xi^+_c,\pi^{+},K^+]&=-2\big[-\mathcal{A}(\Omega^{+}_{cc}\to\Xi^0_c\pi^+K^0)+\sqrt{2}\mathcal{A}(\Omega^+_{cc}\to\Xi^+_c\pi^0K^0)\nonumber\\&~~~~~+\mathcal{A}(\Omega^+_{cc}\to\Xi^+_c\pi^-K^+)+\sqrt{2}\mathcal{A}(\Omega^+_{cc}\to\Xi^0_c\pi^0K^+)\big]=0,
\end{align}
\begin{align}
{ SumI_-^2}\,[\Omega^+_{cc}, \Lambda^+_c,\pi^{+},\pi^+]&=4\mathcal{A}(\Omega^{+}_{cc}\to\Lambda^+_c\pi^0\pi^0)
-2\mathcal{A}(\Omega^+_{cc}\to\Lambda^+_c\pi^+\pi^-)\nonumber\\&~~~~~
-2\mathcal{A}(\Omega^+_{cc}\to\Lambda^+_c\pi^-\pi^+)=0.
\end{align}

The isospin rules for the $\mathcal{B}_{cc} \to \mathcal{B}_{c6}{M_8}{M_8}$  modes derived via Eq.~\eqref{rule8} are
\begin{align}
  { SumI_-}\,[\Omega^+_{cc}, \Sigma^{++}_c,\overline {K}^0,\overline {K}^0]&= -\mathcal{A}(\Omega^+_{cc}\to\Sigma^{++}_c\overline {K}^0K^-) -\mathcal{A}(\Omega^+_{cc}\to\Sigma^{++}_cK^-\overline {K}^0)\nonumber\\&~~~~~+\sqrt{2} \mathcal{A}(\Omega^{+}_{cc}\to\Sigma^{+}_c\overline {K}^0\overline {K}^0)=0,
\end{align}
\begin{align}
 { SumI_-}\,[\Xi^{+}_{cc},\Sigma^{++}_{c},\overline {K}^0,\eta_8]&= -\mathcal{A}(\Xi^{+}_{cc}\to \Sigma^{++}_{c}K^-\eta_8) +\sqrt{2}\mathcal{A}(\Xi^+_{cc}\to \Sigma^+_{c}\overline {K}^0\eta_8)\nonumber\\&~~~~~-\mathcal{A}(\Xi^{++}_{cc}\to \Sigma^{++}_{c}\overline {K}^0\eta_8)=0 ,
\end{align}
\begin{align}
  { SumI_-}\,[\Xi^+_{cc}, \Sigma^{++}_c,\pi^0,\overline {K}^0]&= \sqrt{2}\mathcal{A}(\Xi^+_{cc}\to\Sigma^{+}_c\pi^0\overline {K}^0) - \mathcal{A}(\Xi^+_{cc}\to\Sigma^{++}_c\pi^0K^-)\nonumber\\&~~~~~- \mathcal{A}(\Xi^{++}_{cc}\to\Sigma^{++}_c\pi^0\overline{K}^0)+\sqrt{2}\mathcal{A}(\Xi^+_{cc}\to\Sigma^{++}_c\pi^-\overline{K}^0)=0,
\end{align}
\begin{align}
 { SumI_-}\,[\Xi^+_{cc}, \Omega^{0}_c,\pi^+,{K}^+]&= \mathcal{A}(\Xi^+_{cc}\to\Omega^{0}_c\pi^+{K}^0) -\sqrt{2} \mathcal{A}(\Xi^+_{cc}\to\Omega^0_c\pi^0K^+)\nonumber\\&~~~~~- \mathcal{A}(\Xi^{++}_{cc}\to\Omega^{0}_c\pi^+{K}^+)=0,
\end{align}
\begin{align}
     { SumI_-}\,[\Omega^+_{cc}, \Omega^{0}_c,\pi^+,\pi^+]&=-\sqrt{2}\mathcal{A}(\Omega^+_{cc}\to\Omega^0_c\pi^+\pi^0)-\sqrt{2}\mathcal{A}(\Omega^+_{cc}\to\Omega^0_c\pi^0\pi^+)=0,
\end{align}
\begin{align}
 { SumI_-}\,[\Xi^+_{cc}, \Sigma^{+}_c,\pi^+,\overline {K}^0]&= \sqrt{2}\mathcal{A}(\Xi^+_{cc}\to\Sigma^{0}_c\pi^+\overline {K}^0) - \mathcal{A}(\Xi^+_{cc}\to\Sigma^{+}_c\pi^+K^-)\nonumber\\&~~~~~-\sqrt{2} \mathcal{A}(\Xi^{+}_{cc}\to\Sigma^{+}_c\pi^0\overline{K}^0)-\mathcal{A}(\Xi^{++}_{cc}\to\Sigma^{+}_c\pi^+\overline{K}^0)=0,
\end{align}
\begin{align}
 { SumI_-}\,[\Xi^+_{cc}, \Xi^{*+}_c,\pi^{+},\pi^0]&=\mathcal{A}  (\Xi^+_{cc}\to\Xi^{*0}_c\pi^{+}\pi^0)-\sqrt{2}\mathcal{A}(\Xi^+_{cc}\to\Xi^{*+}_c\pi^{0}\pi^0)\nonumber\\&~~~~~+\sqrt{2}(\Xi^+_{cc}\to\Xi^{*+}_c\pi^{+}\pi^-)-\mathcal{A}(\Xi^{++}_{cc}\to\Xi^{*+}_c\pi^{+}\pi^0)=0,
\end{align}
\begin{align}
 { SumI_-}\,[\Omega^+_{cc}, \Xi^{*+}_{cc},\pi^+,\overline{K}^0]&= \mathcal{A}(\Omega^+_{cc}\to\Xi^{*0}_c\pi^+\overline{K}^0) -\sqrt{2} \mathcal{A}(\Omega^+_{cc}\to\Xi^{*+}_c\pi^0\overline{K}^0)\nonumber\\&~~~~~- \mathcal{A}(\Omega^{+}_{cc}\to\Xi^{*+}_c\pi^{+}{K}^-)=0,
\end{align}
\begin{align}
 { SumI_-}\,[\Xi^+_{cc}, \Xi^{*+}_{c},K^+,\overline{K}^0]&= \mathcal{A}(\Xi^+_{cc}\to\Xi^{*0}_cK^+\overline{K}^0) + \mathcal{A}(\Xi^+_{cc}\to\Xi^{*+}_cK^0\overline{K}^0)\nonumber\\&~~~~~-\mathcal{A}(\Xi^{+}_{cc}\to\Xi^{*+}_cK^{+}{K}^-)-\mathcal{A}(\Xi^{++}_{cc}\to\Xi^{*+}_cK^{+}\overline{K}^0)=0,
\end{align}
\begin{align}
 { SumI_-}\,[\Xi^+_{cc}, \Xi^{*0}_{c},\pi^+,\pi^+]&= -\sqrt{2}\mathcal{A}(\Xi^+_{cc}\to\Xi^{*0}_c\pi^0\pi^+)  -\sqrt{2}\mathcal{A}(\Xi^+_{cc}\to\Xi^{*0}_c\pi^+\pi^0)\nonumber\\&~~~~~-\mathcal{A}(\Xi^{++}_{cc}\to\Xi^{*0}_c\pi^+\pi^+)=0,
\end{align}
\begin{align}
 { SumI_-}\,[\Xi^{++}_{cc}, \Sigma^{++}_c,\pi^+,\overline {K}^0]&= -\sqrt{2}\mathcal{A}(\Xi^{++}_{cc}\to\Sigma^{++}_c\pi^0\overline {K}^0) - \mathcal{A}(\Xi^{++}_{cc}\to\Sigma^{++}_c\pi^+K^-)\nonumber\\&~~~~~+\sqrt{2} \mathcal{A}(\Xi^{++}_{cc}\to\Sigma^{+}_c\pi^+\overline{K}^0)=0,
\end{align}
\begin{align}
   { SumI_-}\,[\Xi^{++}_{cc}, \Xi^{*+}_c,\pi^+,\pi^+]&=\mathcal{A}(\Xi^{++}_{cc}\to\Xi^{*0}_c\pi^+\pi^+)-\sqrt{2} \mathcal{A}(\Xi^{++}_{cc}\to\Xi^{*+}_c\pi^+\pi^0)\nonumber\\&~~~~~-\sqrt{2} \mathcal{A}(\Xi^{++}_{cc}\to\Xi^{*+}_c\pi^0\pi^+)=0,
\end{align}
\begin{align}
   { SumI_-^2}\,[\Xi^{+}_{cc},\Sigma^{++}_{c},\pi ^+,\overline{K}^0]&= 2\big[\mathcal{A}(\Xi^+_{cc}\to\Sigma^0_c\pi^+\overline{K}^0)+\sqrt{2}\mathcal{A}(\Xi^+_{cc}\to\Sigma^{++}_c\pi^0{K}^-)\nonumber\\&~~~~~+\sqrt{2}\mathcal{A}(\Xi^{++}_{cc}\to\Sigma^{++}_c\pi^0\overline{K}^0)-2\mathcal{A}(\Xi^+_{cc}\to\Sigma^+_c\pi^0\overline{K}^0)\nonumber\\&~~~~~+\mathcal{A}(\Xi^{++}_{cc}\to\Sigma^{++}_c\pi^+{K}^-)-\sqrt{2}\mathcal{A}(\Xi^+_{cc}\to\Sigma^+_c\pi^+{K}^-)\nonumber\\&~~~~~-\sqrt{2}\mathcal{A}(\Xi^{++}_{cc}\to\Sigma^+_c\pi^+\overline{K}^0)-\mathcal{A}(\Xi^+_{cc}\to\Sigma^{++}_c\pi^-\overline{K}^0)\big]=0,
\end{align}
\begin{align}
    { SumI_-^2}\,[\Xi^{+}_{cc},\Xi^{*+}_{c},\pi ^+,\pi^+]&=-2\big[\sqrt{2}\mathcal{A}(\Xi^+_{cc}\to\Xi^{*0}_{c}\pi^0\pi^+)+\mathcal{A}(\Xi^{++}_{cc}\to\Xi^{*0}_{c}\pi^+\pi^+)\nonumber\\&~~~~~+\sqrt{2}\mathcal{A}(\Xi^+_{cc}\to\Xi^{*0}_{c}\pi^+\pi^0)-\sqrt{2}\mathcal{A}(\Xi^{++}_{cc}\to\Xi^{*+}_{c}\pi^0\pi^+)\nonumber\\&~~~~~-2\mathcal{A}(\Xi^+_{cc}\to\Xi^{*+}_{c}\pi^0\pi^0)+\mathcal{A}(\Xi^+_{cc}\to\Xi^{*+}_{c}\pi^-\pi^+)\nonumber\\&~~~~~+\mathcal{A}(\Xi^+_{cc}\to\Xi^{*+}_{c}\pi^+\pi^-)-\sqrt{2}\mathcal{A}(\Xi^{++}_{cc}\to\Xi^{*+}_{c}\pi^+\pi^0)\big]=0,
\end{align}
\begin{align}
 { SumI_-^2}\,[\Xi^{++}_{cc},\Sigma^{++}_{c},\pi ^+,\pi^+]&=2\big[\mathcal{A}(\Xi^{++}_{cc}\to\Sigma^{0}_{c}\pi ^+\pi^+)-2\mathcal{A}(\Xi^{++}_{cc}\to\Sigma^{+}_{c}\pi ^0\pi^+)\nonumber\\&~~~~~+2\mathcal{A}(\Xi^{++}_{cc}\to\Sigma^{++}_{c}\pi ^0\pi^0)-\mathcal{A}(\Xi^{++}_{cc}\to\Sigma^{++}_{c}\pi ^-\pi^+)\nonumber\\&~~~~~-2\mathcal{A}(\Xi^{++}_{cc}\to\Sigma^{+}_{c}\pi
 ^+\pi^0)-\mathcal{A}(\Xi^{++}_{cc}\to\Sigma^{++}_{c}\pi ^+\pi^-)\big]=0 ,
\end{align}
\begin{align}
{ SumI_-^2}\,[\Xi^{+}_{cc},\Sigma^{++}_{c},\pi ^+,\pi^0]&=-2\big[-\mathcal{A}(\Xi^{+}_{cc}\to\Sigma^{0}_{c}\pi ^0\pi^+)+2\mathcal{A}(\Xi^{+}_{cc}\to\Sigma^{++}_{c}\pi ^0\pi^-)\nonumber\\&~~~~~-\sqrt{2}\mathcal{A}(\Xi^{++}_{cc}\to\Sigma^{++}_{c}\pi ^0\pi^0)+2\mathcal{A}(\Xi^{+}_{cc}\to\Sigma^{+}_{c}\pi ^0\pi^0)\nonumber\\&~~~~~+\sqrt{2}\mathcal{A}(\Xi^{++}_{cc}\to\Sigma^{++}_{c}\pi ^+\pi^-)-2\mathcal{A}(\Xi^{+}_{cc}\to\Sigma^{+}_{c}\pi ^+\pi^-)\nonumber\\&~~~~~+\sqrt{2}\mathcal{A}(\Xi^{++}_{cc}\to\Sigma^{+}_{c}\pi^+\pi^0)+\mathcal{A}(\Xi^{+}_{cc}\to\Sigma^{++}_{c}\pi^-\pi^0)\big]=0 ,
\end{align}
\begin{align}
    { SumI_-^2}\,[\Omega^{+}_{cc},\Sigma^{++}_{c},\pi ^+,\overline{K}^0]&=2\big[\mathcal{A}(\Omega^{+}_{cc}\to\Sigma^{0}_{c}\pi ^+\overline{K}^0)+\sqrt{2}\mathcal{A}(\Omega^{+}_{cc}\to\Sigma^{++}_{c}\pi ^0K^-)\nonumber\\&~~~~~-2\mathcal{A}(\Omega^{+}_{cc}\to\Sigma^{+}_{c}\pi ^0\overline{K}^0)-\sqrt{2}\mathcal{A}(\Omega^{+}_{cc}\to\Sigma^{+}_{c}\pi ^+K^-)\nonumber\\&~~~~~-\mathcal{A}(\Omega^{+}_{cc}\to\Sigma^{++}_{c}\pi^-\overline{K}^0)\big]=0,
\end{align}
\begin{align}
   {SumI_-^2}\,[\Xi^{+}_{cc},\Sigma^{++}_{c},\pi ^+,\eta_8]&=2\big[\mathcal{A}(\Xi^+_{cc}\to\Sigma^0_c\pi^+\eta_8)+\sqrt{2}\mathcal{A}(\Xi^{++}_{cc}\to\Sigma^{++}_c\pi^0\eta_8)\nonumber\\&~~~~~-2\mathcal{A}(\Xi^+_{cc}\to\Sigma^+_c\pi^0\eta_8)-\sqrt{2}\mathcal{A}(\Xi^{++}_{cc}\to\Sigma^+_c\pi^+\eta_8)\nonumber\\&~~~~~-\mathcal{A}(\Xi^+_{cc}\to\Sigma^{++}_c\pi^-\eta_8)\big]=0,
\end{align}
\begin{align}
   {SumI_-^2}\,[\Xi^{+}_{cc},\Sigma^{++}_{c},K ^+,\overline{K}^0]&=  2\big[\mathcal{A}(\Xi^{+}_{cc}\to\Sigma^{0}_{c}K^+\overline{K}^0)+\sqrt{2}\mathcal{A}(\Xi^{+}_{cc}\to\Sigma^{+}_{c}\overline{K}^0\overline{K}^0)\nonumber\\&~~~~~+\mathcal{A}(\Xi^{++}_{cc}\to\Sigma^{++}_{c}K^+K^-)-\sqrt{2}\mathcal{A}(\Xi^{+}_{cc}\to\Sigma^{+}_{c}K^+{K}^-)\nonumber\\&~~~~~-\mathcal{A}(\Xi^{+}_{cc}\to\Sigma^{++}_{c}K^-{K}^0)-\sqrt{2}\mathcal{A}(\Xi^{++}_{cc}\to\Sigma^{+}_{c}K^+\overline{K}^0)\nonumber\\&~~~~~-\mathcal{A}(\Xi^{++}_{cc}\to\Sigma^{++}_{c}K^0\overline{K}^0)\big]=0,
\end{align}
\begin{align}
   {SumI_-^2}\,[\Xi^{+}_{cc},\Sigma^{+}_{c},\pi^+,\pi^+]&= -2\big[2\mathcal{A}(\Xi^{+}_{cc}\to\Sigma^{0}_{c}\pi^0\pi^+)+\sqrt{2}\mathcal{A}(\Xi^{++}_{cc}\to\Sigma^{0}_{c}\pi^+\pi^+)\nonumber\\&~~~~~-\sqrt{2}\mathcal{A}(\Xi^{++}_{cc}\to\Sigma^{+}_{c}\pi^0\pi^+)-\sqrt{2}\mathcal{A}(\Xi^{++}_{cc}\to\Sigma^{+}_{c}\pi^+\pi^0)\nonumber\\&~~~~~+\mathcal{A}(\Xi^{+}_{cc}\to\Sigma^{+}_{c}\pi^-\pi^+)+\mathcal{A}(\Xi^{+}_{cc}\to\Sigma^{+}_{c}\pi^+\pi^-)\nonumber\\&~~~~~+2\mathcal{A}(\Xi^{+}_{cc}\to\Sigma^{0}_{c}\pi^+\pi^0)-2\mathcal{A}(\Xi^{+}_{cc}\to\Sigma^{+}_{c}\pi^0\pi^0)\big] =0,
\end{align}
\begin{align}
 {SumI_-^2}\,[\Xi^{+}_{cc},\Xi^{*+}_{c},\pi^+,K^+]&=-2\big[\sqrt{2}\mathcal{A}(\Xi^{+}_{cc}\to\Xi^{*0}_{c}\pi^0{K}^+)-\mathcal{A}(\Xi^{+}_{cc}\to\Xi^{*0}_{c}\pi^+{K}^0)\nonumber\\&~~~~~-\sqrt{2}\mathcal{A}(\Xi^{++}_{cc}\to\Xi^{*+}_{c}\pi^0{K}^+)+\sqrt{2}\mathcal{A}(\Xi^{+}_{cc}\to\Xi^{*+}_{c}\pi^0{K}^0)\nonumber\\&~~~~~+\mathcal{A}(\Xi^{+}_{cc}\to\Xi^{*+}_{c}\pi^-{K}^+)+\mathcal{A}(\Xi^{++}_{cc}\to\Xi^{*+}_{c}\pi^+{K}^0)\nonumber\\&~~~~~+\mathcal{A}(\Xi^{++}_{cc}\to\Xi^{*0}_{c}\pi^+{K}^+)\big]=0,
\end{align}
\begin{align}
     {SumI_-^2}\,[\Omega^{+}_{cc},\Xi^{*+}_{c},\pi^+,\pi^+]&=-2\big[\sqrt{2}\mathcal{A}(\Omega^{+}_{cc}\to\Xi^{*0}_{c}\pi^+\pi^0)-2\mathcal{A}(\Omega^{+}_{cc}\to\Xi^{*+}_{c}\pi^0\pi^0)\nonumber\\&~~~~~+\mathcal{A}(\Omega^{+}_{cc}\to\Xi^{*+}_{c}\pi^+\pi^-)+\sqrt{2}\mathcal{A}(\Omega^{+}_{cc}\to\Xi^{*0}_{c}\pi^0\pi^+)\nonumber\\&~~~~~+\mathcal{A}(\Omega^{+}_{cc}\to\Xi^{*+}_{c}\pi^-\pi^+)\big]=0,
\end{align}
\begin{align}
    { SumI_-^2}\,[\Xi^{++}_{cc},\Sigma^{++}_{c},\pi ^+,K^+]&=2\big[\mathcal{A}(\Xi^{++}_{cc}\to \Sigma^{0}_{c}\pi^+K^+)-2\mathcal{A}(\Xi^{++}_{cc}\to \Sigma^{+}_{c}\pi^0K^+)\nonumber\\&~~~~-\sqrt{2}\mathcal{A}(\Xi^{++}_{cc}\to \Sigma^{++}_{c}\pi^0K^0)-\mathcal{A}(\Xi^{++}_{cc}\to \Sigma^{++}_{c}\pi^-K^+)\nonumber\\&~~~~+\sqrt{2}\mathcal{A}(\Xi^{++}_{cc}\to \Sigma^{+}_{c}\pi^+K^0)\big]=0,
\end{align}
\begin{align}
  { SumI_-^2}\,[\Omega^{+}_{cc},\Sigma^{++}_{c},\pi ^+,\pi^0]&= 2\mathcal{A}(\Omega^{+}_{cc}\to\Sigma^{0}_{c}\pi ^+\pi^0)-4\big[\mathcal{A}(\Omega^{+}_{cc}\to\Sigma^{++}_{c}\pi ^0\pi^-)\nonumber\\&~~~~~+\mathcal{A}(\Omega^{+}_{cc}\to\Sigma^{+}_{c}\pi ^0\pi^0)-\mathcal{A}(\Omega^{+}_{cc}\to\Sigma^{+}_{c}\pi ^+\pi^-)\big]\nonumber\\&~~~~~-2\mathcal{A}(\Omega^{+}_{cc}\to\Sigma^{++}_{c}\pi ^-\pi^0)=0,
\end{align}
\begin{align}
  { SumI_-^2}\,[\Xi^{+}_{cc},\Sigma^{++}_{c},\pi ^+,K^0]&=  2\big[\mathcal{A}(\Xi^{+}_{cc}\to\Sigma^{0}_{c}\pi^+K^0)+\sqrt{2}\mathcal{A}(\Xi^{++}_{cc}\to\Sigma^{++}_{c}\pi^0K^0)\nonumber\\&~~~~~-2\mathcal{A}(\Xi^{+}_{cc}\to\Sigma^{+}_{c}\pi^0K^0)-\sqrt{2}\mathcal{A}(\Xi^{++}_{cc}\to\Sigma^{+}_{c}\pi^+K^0)\nonumber\\&~~~~~-\mathcal{A}(\Xi^{+}_{cc}\to\Sigma^{++}_{c}\pi^-K^0)\big]=0,
\end{align}
\begin{align}
    { SumI_-^2}\,[\Omega^{+}_{cc},\Sigma^{++}_{c},\pi ^+,\eta_8]&=2\big[\mathcal{A}(\Omega^{+}_{cc}\to\Sigma^{0}_{c}\pi ^+\eta_8)-2\mathcal{A}(\Omega^{+}_{cc}\to\Sigma^{+}_{c}\pi ^0\eta_8)\nonumber\\&~~~~~-\mathcal{A}(\Omega^{+}_{cc}\to\Sigma^{++}_{c}\pi ^-\eta_8)\big]=0,
\end{align}
\begin{align}
   { SumI_-^2}\,[\Xi^{+}_{cc},\Sigma^{++}_{c},\pi^0,K^+]&=2\big[\mathcal{A}(\Xi^{+}_{cc}\to\Sigma^{0}_{c}\pi^0K^+)-\sqrt{2}\mathcal{A}(\Xi^{++}_{cc}\to\Sigma^{+}_{c}\pi^0K^+)\nonumber\\&~~~~~-\mathcal{A}(\Xi^{++}_{cc}\to\Sigma^{++}_{c}\pi^0K^0)+\sqrt{2}\mathcal{A}(\Xi^{+}_{cc}\to\Sigma^{+}_{c}\pi^0K^0)\nonumber\\&~~~~~-\sqrt{2}\mathcal{A}(\Xi^{++}_{cc}\to\Sigma^{++}_{c}\pi^-K^+)+\sqrt{2}\mathcal{A}(\Xi^{+}_{cc}\to\Sigma^{++}_{c}\pi^-K^0)\nonumber\\&~~~~~+2\mathcal{A}(\Xi^+_{cc}\to\Sigma^{+}_c\pi^-K^+)\big]=0,
\end{align}
\begin{align}
   {SumI_-^2}\,[\Omega^{+}_{cc},\Sigma^{++}_{c},K ^+,\overline{K}^0]&=  2\big[\mathcal{A}(\Omega^{+}_{cc}\to\Sigma^{0}_{c}K ^+\overline{K}^0)-\mathcal{A}(\Omega^{+}_{cc}\to\Sigma^{++}_{c}K ^0{K}^-)\nonumber\\&~~~~~-\sqrt{2}\mathcal{A}(\Omega^{+}_{cc}\to\Sigma^{+ }_{c}K^+{K}^-)+\sqrt{2}\mathcal{A}(\Omega^{+}_{cc}\to\Sigma^{+ }_{c}K^0\overline{K}^0)\big]=0.
\end{align}
\begin{align}
    { SumI_-^2}\,[\Xi^{+}_{cc},\Sigma^{++}_{c},K ^+,\eta_8]&= 2\big[\mathcal{A}(\Xi^{+}_{cc}\to\Sigma^{0}_{c}K ^+\eta_8)-\sqrt{2}\mathcal{A}(\Xi^{++}_{cc}\to\Sigma^{+}_{c}K ^+\eta_8)\nonumber\\&~~~~~-\mathcal{A}(\Xi^{++}_{cc}\to\Sigma^{++}_{c}K ^0\eta_8)+\sqrt{2}\mathcal{A}(\Xi^{+}_{cc}\to\Sigma^{+}_{c}K ^0\eta_8)\big]=0,
\end{align}
\begin{align}
    { SumI_-^2}\,[\Xi^{+}_{cc},\Sigma^{+}_{c},\pi^+,K^+]&=-2\big[\sqrt{2}\mathcal{A}(\Xi^{++}_{cc}\to\Sigma^{0}_{c}\pi^+K^+)+2\mathcal{A}(\Xi^{+}_{cc}\to\Sigma^{0}_{c}\pi^0K^+)\nonumber\\&~~~~-\sqrt{2}\mathcal{A}(\Xi^{+}_{cc}\to\Sigma^{0}_{c}\pi^+K^0)-\sqrt{2}\mathcal{A}(\Xi^{++}_{cc}\to\Sigma^{+}_{c}\pi^0K^+)\nonumber\\&~~~~+\sqrt{2}\mathcal{A}(\Xi^{+}_{cc}\to\Sigma^{+}_{c}\pi^0K^0)+\mathcal{A}(\Xi^{+}_{cc}\to\Sigma^{+}_{c}\pi^-K^+)\nonumber\\&~~~~+\mathcal{A}(\Xi^{++}_{cc}\to\Sigma^{+}_{c}\pi^+K^0)\big]=0,
\end{align}
\begin{align}
    {SumI_-^2}\,[\Omega^{+}_{cc},\Sigma^{+}_{c},\pi ^+,\pi^+]&= -2\big[2\mathcal{A}(\Omega^+_{cc}\to\Sigma^0_c\pi^+\pi^0)-2\mathcal{A}(\Omega^+_{cc}\to\Sigma^+_c\pi^0\pi^0)\nonumber\\&~~~~~+\mathcal{A}(\Omega^+_{cc}\to\Sigma^+_c\pi^+\pi^-)+2\mathcal{A}(\Omega^+_{cc}\to\Sigma^0_c\pi^0\pi^+)\nonumber\\&~~~~~+\mathcal{A}(\Omega^+_{cc}\to\Sigma^+_c\pi^-\pi^+)\big]=0,
\end{align}
\begin{align}
     {SumI_-^2}\,[\Omega^{+}_{cc},\Xi^{*+}_{c},\pi^+,K^+]&=-2\big[\sqrt{2}\mathcal{A}(\Omega^{+}_{cc}\to\Xi^{*0}_{c}\pi^0K^+)-\mathcal{A}(\Omega^{+}_{cc}\to\Xi^{*0}_{c}\pi^+K^0)\nonumber\\&~~~~~+\sqrt{2}\mathcal{A}(\Omega^{+}_{cc}\to\Xi^{*+}_{c}\pi^0K^0)+\mathcal{A}(\Omega^{+}_{cc}\to\Xi^{*+}_{c}\pi^-K^+)\big]=0,
\end{align}
\begin{align}
    {SumI_-^2}\,[\Xi^{+}_{cc},\Xi^{*+}_{c},K^+,K^+]&=2\big[\mathcal{A}(\Xi^{+}_{cc}\to\Xi^{*0}_{c}K^+{K}^0)-\mathcal{A}(\Xi^{++}_{cc}\to\Xi^{*0}_{c}K^+{K}^+)\nonumber\\&~~~~~+\mathcal{A}(\Xi^{+}_{cc}\to\Xi^{*+}_{c}K^0{K}^0)-\mathcal{A}(\Xi^{++}_{cc}\to\Xi^{*+}_{c}K^+{K}^0)\nonumber\\&~~~~~+\mathcal{A}(\Xi^{+}_{cc}\to\Xi^{*0}_{c}K^0{K}^+)-\mathcal{A}(\Xi^{++}_{cc}\to\Xi^{*+}_{c}K^0{K}^+)\big]=0,
\end{align}
\begin{align}
    {SumI_-^3}\,[\Xi^{+}_{cc},\Sigma^{++}_{c},\pi^+,\pi^+]&=-6\big[\sqrt{2}\mathcal{A}(\Xi^+_{cc}\to\Sigma^0_c\pi^0\pi^+)+\mathcal{A}(\Xi^{++}_{cc}\to\Sigma^0_c\pi^+\pi^+)\nonumber\\&~~~~-\sqrt{2}\mathcal{A}(\Xi^{+}_{cc}\to\Sigma^{++}_c\pi^0\pi^-)+2\mathcal{A}(\Xi^{++}_{cc}\to\Sigma^{++}_c\pi^0\pi^0)\nonumber\\&~~~~-2\sqrt{2}\mathcal{A}(\Xi^{+}_{cc}\to\Sigma^+_c\pi^0\pi^0)-\mathcal{A}(\Xi^{++}_{cc}\to\Sigma^{++}_c\pi^+\pi^-)\nonumber\\&~~~~+\sqrt{2}\mathcal{A}(\Xi^{+}_{cc}\to\Sigma^+_c\pi^+\pi^-)-2\mathcal{A}(\Xi^{++}_{cc}\to\Sigma^+_c\pi^+\pi^0)\nonumber\\&~~~~~+\sqrt{2}\mathcal{A}(\Xi^+_{cc}\to\Sigma^0_c\pi^+\pi^0)-2\mathcal{A}(\Xi^{++}_{cc}\to\Sigma^+_c\pi^0\pi^+)\nonumber\\&~~~~-\mathcal{A}(\Xi^{++}_{cc}\to\Sigma^{++}_c\pi^-\pi^+)+\sqrt{2}\mathcal{A}(\Xi^{+}_{cc}\to\Sigma^+_c\pi^-\pi^+)\nonumber\\&~~~~~-\sqrt{2}\mathcal{A}(\Xi^+_{cc}\to\Sigma^{++}_c\pi^-\pi^0)\big]=0,
\end{align}
\begin{align}
    { SumI_-^3}\,[\Xi^{+}_{cc},\Sigma^{++}_{c},\pi^+,K^+]&=-6\big[\mathcal{A}(\Xi^{++}_{cc}\to\Sigma^{0}_{c}\pi^+K^+)+\sqrt{2}\mathcal{A}(\Xi^{+}_{cc}\to\Sigma^{0}_{c}\pi^0K^+)\nonumber\\&~~~~-\mathcal{A}(\Xi^{+}_{cc}\to\Sigma^{0}_{c}\pi^+K^0)-2\mathcal{A}(\Xi^{++}_{cc}\to\Sigma^{+}_{c}\pi^0K^+)\nonumber\\&~~~~-\sqrt{2}\mathcal{A}(\Xi^{++}_{cc}\to\Sigma^{++}_{c}\pi^0K^0)+2\mathcal{A}(\Xi^{+}_{cc}\to\Sigma^{+}_{c}\pi^0K^0)\nonumber\\&~~~~-\mathcal{A}(\Xi^{++}_{cc}\to\Sigma^{++}_{c}\pi^-K^+)+\sqrt{2}\mathcal{A}(\Xi^{+}_{cc}\to\Sigma^{+}_{c}\pi^-K^+)\nonumber\\&~~~~+\sqrt{2}\mathcal{A}(\Xi^{++}_{cc}\to\Sigma^{+}_{c}\pi^+K^0)+\mathcal{A}(\Xi^{+}_{cc}\to\Sigma^{++}_{c}\pi^-K^0)\big]=0,
\end{align}
\begin{align}
    {SumI_-^3}\,[\Omega^{+}_{cc},\Sigma^{++}_{c},\pi ^+,\pi^+]&= -6\sqrt{2}\big[\mathcal{A}(\Omega^+_{cc}\to\Sigma^0_c\pi^+\pi^0)-\mathcal{A}(\Omega^+_{cc}\to\Sigma^{++}_c\pi^0\pi^-)\nonumber\\&~~~~~-2\mathcal{A}(\Omega^+_{cc}\to\Sigma^+_c\pi^0\pi^0)+\mathcal{A}(\Omega^+_{cc}\to\Sigma^+_c\pi^+\pi^-)\nonumber\\&~~~~~+\mathcal{A}(\Omega^+_{cc}\to\Sigma^0_c\pi^0\pi^+)+\mathcal{A}(\Omega^+_{cc}\to\Sigma^+_c\pi^-\pi^+)\nonumber\\&~~~~~+\mathcal{A}(\Omega^+_{cc}\to\Sigma^{++}_c\pi^-\pi^0)\big]=0.
\end{align}

\subsection{$\mathcal{B}_{cc} \to \mathcal{B}_{8}{D}{M_8}$ and $\mathcal{B}_{cc} \to \mathcal{B}_{10}{D}{M_8}$ modes}\label{doubly4}

The isospin rules for the $\mathcal{B}_{cc} \to \mathcal{B}_{8}{D}{M_8}$  modes derived via Eq.~\eqref{rule9} are
\begin{align}
     { SumI_-}\,[\Xi^{++}_{cc},\Sigma^{+},D^+,\pi^+]&=-\sqrt{2}\mathcal{A}(\Xi^{++}_{cc}\to\Sigma^0D^+\pi^+)-\mathcal{A}(\Xi^{++}_{cc}\to\Sigma^+D^0\pi^+)\nonumber\\&~~~~~-\sqrt{2}\mathcal{A}(\Xi^{++}_{cc}\to\Sigma^+D^+\pi^0)= 0,
\end{align}
\begin{align}
     { SumI_-}\,[\Xi^{++}_{cc},\Sigma^{+},D^0,\pi^+]&=-\sqrt{2}\mathcal{A}(\Xi^{+}_{cc}\to\Sigma^{0}D^0\pi^+)-\mathcal{A}(\Xi^{++}_{cc}\to\Sigma^{+}D^0\pi^+)\nonumber\\&~~~~~-\sqrt{2}\mathcal{A}(\Xi^{+}_{cc}\to\Sigma^{+}D^0\pi^0)= 0,
\end{align}
\begin{align}
     { SumI_-}\,[\Xi^{+}_{cc},\Sigma^{+},D^+,\pi^0]&=-\sqrt{2}\mathcal{A}(\Xi^{+}_{cc}\to\Sigma^{0}D^+\pi^0)-\mathcal{A}(\Xi^{+}_{cc}\to\Sigma^{+}D^0\pi^0)\nonumber\\&~~~~~+\sqrt{2}\mathcal{A}(\Xi^{+}_{cc}\to\Sigma^{+}D^+\pi^-)-\mathcal{A}(\Xi^{++}_{cc}\to\Sigma^{+}D^+\pi^0)= 0,
\end{align}
\begin{align}
     { SumI_-}\,[\Xi^{+}_{cc},\Sigma^{+},D^+,\eta_8]&=-\sqrt{2}\mathcal{A}(\Xi^{+}_{cc}\to\Sigma^0D^+\eta_8)-\mathcal{A}(\Xi^{+}_{cc}\to\Sigma^+D^0\eta_8)\nonumber\\&~~~~~-\mathcal{A}(\Xi^{++}_{cc}\to\Sigma^+D^+\eta_8)= 0,
\end{align}
\begin{align}
    { SumI_-}\,[\Xi^{+}_{cc},\Sigma^+,D^+_s,\overline{K}^0]&=-\mathcal{A}(\Xi^{++}_{cc}\to \Sigma^+D^+_s\overline{K}^0)-\sqrt{2}\mathcal{A}(\Xi^{+}_{cc}\to \Sigma^0D^+_s\overline{K}^0)\nonumber\\&~~~~~-\mathcal{A}(\Xi^{+}_{cc}\to \Sigma^+D^+_sK^-)= 0,
\end{align}
\begin{align}
    { SumI_-}\,[\Xi^{+}_{cc},p,D^+,\overline{K}^0]&=-\mathcal{A}(\Xi^{+}_{cc}\to pD^0\overline{K}^0)-\mathcal{A}(\Xi^{+}_{cc}\to pD^+K^-)\nonumber\\&~~~~~-\mathcal{A}(\Xi^{++}_{cc}\to pD^+\overline{K}^0)+\mathcal{A}(\Xi^{+}_{cc}\to nD^+\overline{K}^0)= 0,
\end{align}
\begin{align}
     { SumI_-}\,[\Xi^{+}_{cc},\Sigma^{0},D^+,\pi^+]&=-\mathcal{A}(\Xi^{+}_{cc}\to\Sigma^0D^0\pi^+)-\mathcal{A}(\Xi^{++}_{cc}\to\Sigma^0D^+\pi^+)\nonumber\\&~~~~~-\sqrt{2}\mathcal{A}(\Xi^+_{cc}\to\Sigma^0D^+\pi^0)+\sqrt{2}\mathcal{A}(\Xi^+_{cc}\to\Sigma^-D^+\pi^+)= 0,
\end{align}
\begin{align}
    { SumI_-}\,[\Xi^{+}_{cc},\Xi^{0},D^+,K^+]&=\mathcal{A}(\Xi^+_{cc}\to\Xi^0D^+K^0)-\mathcal{A}(\Xi^{++}_{cc}\to\Xi^0D^+K^+)\nonumber\\&~~~~~-\mathcal{A}(\Xi^+_{cc}\to\Xi^0D^0K^+)-\mathcal{A}(\Xi^+_{cc}\to\Xi^-D^+K^+)=0,
\end{align}
\begin{align}
    { SumI_-}\,[\Xi^{+}_{cc},\Xi^{0},D^+_s,\pi^+]&=-\mathcal{A}(\Xi^{++}_{cc}\to\Xi^0D^+_s\pi^+)-\sqrt{2}\mathcal{A}(\Xi^+_{cc}\to\Xi^0 D^+_s\pi^0)\nonumber\\&~~~~~-\mathcal{A}(\Xi^+_{cc}\to\Xi^-D^+_s\pi^+)=0,
\end{align}
\begin{align}
    { SumI_-}\,[\Xi^{+}_{cc},\Lambda^{0},D^+,\pi^+]&=-\mathcal{A}(\Xi^{+}_{cc}\to\Lambda^0D^0\pi^+)-\mathcal{A}(\Xi^{++}_{cc}\to\Lambda^0 D^+\pi^+)\nonumber\\&~~~~~-\sqrt{2}\mathcal{A}(\Xi^+_{cc}\to\Lambda^0D^+\pi^0)=0,
\end{align}
\begin{align}
     { SumI_-}\,[\Omega^{+}_{cc},\Sigma^{+},D^+,\overline{K}^0]&=-\sqrt{2}\mathcal{A}(\Omega^{+}_{cc}\to\Sigma^0D^+\overline{K}^0)-\mathcal{A}(\Omega^{+}_{cc}\to\Sigma^+D^0\overline{K}^0)\nonumber\\&~~~~~-\mathcal{A}(\Omega^+_{cc}\to\Sigma^+D^+K^-)= 0,
\end{align}
\begin{align}
    { SumI_-}\,[\Omega^{+}_{cc},\Xi^{0},D^+,\pi^+]&=-\mathcal{A}(\Omega^{+}_{cc}\to\Xi^0D^0\pi^+)-\sqrt{2}\mathcal{A}(\Omega^+_{cc}\to\Xi^0 D^+\pi^0)\nonumber\\&~~~~~-\mathcal{A}(\Omega^+_{cc}\to\Xi^-D^+\pi^+)=0,
\end{align}
\begin{align}
     { SumI_-^2}\,[\Xi^{+}_{cc},\Sigma^{+},D^+,\pi^+]&=2\big[\sqrt{2}\mathcal{A}(\Xi^+_{cc}\to\Sigma^0D^0\pi^+)+\sqrt{2}\mathcal{A}(\Xi^{++}_{cc}\to\Sigma^0D^+\pi^+)\nonumber\\&~~~~~+2\mathcal{A}(\Xi^+_{cc}\to\Sigma^0D^+\pi^0)+\mathcal{A}(\Xi^{++}_{cc}\to\Sigma^+D^0\pi^+)\nonumber\\&~~~~~+\sqrt{2}\mathcal{A}(\Xi^+_{cc}\to\Sigma^+D^0\pi^0)-\mathcal{A}(\Xi^+_{cc}\to\Sigma^-D^+\pi^+)\nonumber\\&~~~~~-\mathcal{A}(\Xi^+_{cc}\to\Sigma^+D^+\pi^-)+\sqrt{2}\mathcal{A}(\Xi^{++}_{cc}\to\Sigma^+D^+\pi^0)\big]=0,
\end{align}
\begin{align}
     { SumI_-^2}\,[\Xi^{+}_{cc},\Sigma^{+},D^+,K^+]&= 2\big[\sqrt{2}\mathcal{A}(\Xi^{+}_{cc}\to\Sigma^{0}D^0K^+)+\sqrt{2}\mathcal{A}(\Xi^{++}_{cc}\to\Sigma^{0}D^+K^+)\nonumber\\&~~~~~-\sqrt{2}\mathcal{A}(\Xi^{+}_{cc}\to\Sigma^{0}D^+K^0)+\mathcal{A}(\Xi^{++}_{cc}\to\Sigma^{+}D^0K^+)\nonumber\\&~~~~~-\mathcal{A}(\Xi^{+}_{cc}\to\Sigma^{+}D^0K^0)-\mathcal{A}(\Xi^{+}_{cc}\to\Sigma^{-}D^+K^+)\nonumber\\&~~~~~-\mathcal{A}(\Xi^{++}_{cc}\to\Sigma^{+}D^+K^0)\big]=0,
\end{align}
\begin{align}
   { SumI_-^2}\,[\Xi^{+}_{cc},\Sigma^{+},D^+_s,\pi^+]&=2\big[\sqrt{2}\mathcal{A}(\Xi^{++}_{cc}\to\Sigma^0D^+_s\pi^+)+2\mathcal{A}(\Xi^{+}_{cc}\to\Sigma^0D^+_s\pi^0)\nonumber\\&~~~~~-\mathcal{A}(\Xi^{+}_{cc}\to\Sigma^-D^+_s\pi^+)-\mathcal{A}(\Xi^{+}_{cc}\to\Sigma^+D^+_s\pi^-)\nonumber\\&~~~~~+\sqrt{2}\mathcal{A}(\Xi^{++}_{cc}\to\Sigma^+D^+_s\pi^0)\big]=0,
\end{align}
\begin{align}
    { SumI_-^2}\,[\Xi^{+}_{cc},p,D^+,\pi^+]&=2\big[\mathcal{A}(\Xi^{++}_{cc}\to pD^0\pi^+)+\sqrt{2}\mathcal{A}(\Xi^{+}_{cc}\to pD^0\pi^0) \nonumber\\&~~~~~-\mathcal{A}(\Xi^{+}_{cc}\to pD^+\pi^-)+\sqrt{2}\mathcal{A}(\Xi^{++}_{cc}\to pD^+\pi^0)\nonumber\\&~~~~~-\mathcal{A}(\Xi^{++}_{cc}\to nD^+\pi^+)-\mathcal{A}(\Xi^{+}_{cc}\to nD^0\pi^+)\nonumber\\&~~~~~-\sqrt{2}\mathcal{A}(\Xi^{+}_{cc}\to nD^+\pi^0)\big]=0,
\end{align}
\begin{align}
    { SumI_-^2}\,[\Omega^{+}_{cc},\Sigma^{+},D^+,\pi^+]&=2\big[\sqrt{2}\mathcal{A}(\Omega^{+}_{cc}\to\Sigma^0D^0\pi^+) +2\mathcal{A}(\Omega^{+}_{cc}\to\Sigma^0D^+\pi^0) \nonumber\\&~~~~~+\sqrt{2}\mathcal{A}(\Omega^{+}_{cc}\to\Sigma^+D^0\pi^0) -\mathcal{A}(\Omega^{+}_{cc}\to\Sigma^-D^+\pi^+)\nonumber\\&~~~~~- \mathcal{A}(\Omega^{+}_{cc}\to\Sigma^+D^+\pi^-)\big]=0,
\end{align}
\begin{align}
   { SumI_-^2}\,[\Xi^{+}_{cc},\Sigma^+,D^+_s,K^+]&=-2\big[\mathcal{A}(\Xi^{++}_{cc}\to \Sigma^+D^+_sK^0)  -\sqrt{2}\mathcal{A}(\Xi^{++}_{cc}\to \Sigma^0D^+_sK^+)\nonumber\\&~~~~~+\mathcal{A}(\Xi^{+}_{cc}\to \Sigma^-D^+_sK^+)  +\sqrt{2}\mathcal{A}(\Xi^{+}_{cc}\to \Sigma^0D^+_sK^0)\big]=0,
\end{align}
\begin{align}
   { SumI_-^2}\,[\Xi^{+}_{cc},p,D^+,K^+]&=2\big[\mathcal{A}(\Xi^{++}_{cc}\to pD^0K^+) -\mathcal{A}(\Xi^{+}_{cc}\to pD^0K^0)\nonumber\\&~~~~~-\mathcal{A}(\Xi^{++}_{cc}\to pD^+K^0)-\mathcal{A}(\Xi^{++}_{cc}\to nD^+K^+)\nonumber\\&~~~~~ -\mathcal{A}(\Xi^{+}_{cc}\to nD^0K^+)-\mathcal{A}(\Xi^{+}_{cc}\to nD^+K^0)\big]=0,
\end{align}
\begin{align}
   { SumI_-^2}\,[\Xi^{+}_{cc},p,D^+_s,\pi^+]&=-2\mathcal{A}(\Xi^{+}_{cc}\to pD^+_s\pi^-)  +2\sqrt{2}\mathcal{A}(\Xi^{++}_{cc}\to pD^+_s\pi^0)\nonumber\\&~~~~-2\mathcal{A}(\Xi^{++}_{cc}\to nD^+_s\pi^+)  -2\sqrt{2}\mathcal{A}(\Xi^{+}_{cc}\to nD^+_s\pi^0)=0,
\end{align}
\begin{align}
    { SumI_-^2}\,[\Omega^{+}_{cc},\Sigma^{+},D^+,K^+]&=2\sqrt{2}\mathcal{A}(\Omega^{+}_{cc}\to\Sigma^{0}D^0K^+)-2\big[\sqrt{2}\mathcal{A}(\Omega^{+}_{cc}\to\Sigma^{0}D^+K^0)\nonumber\\&~~~~~+\mathcal{A}(\Omega^{+}_{cc}\to\Sigma^{+}D^0K^0)+\mathcal{A}(\Omega^{+}_{cc}\to\Sigma^{-}D^+K^+)\big]=0,
\end{align}
\begin{align}
    { SumI_-^2}\,[\Omega^{+}_{cc},\Sigma^{+},D^+_s,\pi^+]&=4\mathcal{A}(\Omega^{+}_{cc}\to \Sigma^{0}D^+_s\pi^0)-2\big[\mathcal{A}(\Omega^{+}_{cc}\to \Sigma^{-}D^+_s\pi^+)\nonumber\\&~~~~~+\mathcal{A}(\Omega^{+}_{cc}\to \Sigma^{+}D^+_s\pi^-)\big]=0,
\end{align}
\begin{align}
    { SumI_-^2}\,[\Omega^{+}_{cc},p,D^+,\pi^+]&=2\sqrt{2}\mathcal{A}(\Omega^{+}_{cc}\to pD^0\pi^0)-2\mathcal{A}(\Omega^{+}_{cc}\to pD^+\pi^-)\nonumber\\&~~~~~-2\mathcal{A}(\Omega^{+}_{cc}\to nD^0\pi^+)-2\sqrt{2}\mathcal{A}(\Omega^{+}_{cc}\to nD^+\pi^0)=0.
\end{align}
The isospin rules for the $\mathcal{B}_{cc} \to \mathcal{B}_{10}{D}{M_8}$  modes derived via Eq.~\eqref{rule10} are
\begin{align}
     { SumI_-}\,[\Xi^{++}_{cc},\Delta^{++},D^+,\overline{K}^0]&=-\mathcal{A}(\Xi^{++}_{cc}\to\Delta^{++}D^0\overline{K}^0)-\mathcal{A}(\Xi^{++}_{cc}\to\Delta^{++}D^+K^-)\nonumber\\&~~~~~-\sqrt{3}\mathcal{A}(\Xi^{++}_{cc}\to\Delta^+D^+\overline{K}^0)= 0,
\end{align}
\begin{align}
     { SumI_-}\,[\Xi^{++}_{cc},\Sigma^{*+},D^+,\pi^+]&=\sqrt{2}\mathcal{A}(\Xi^{++}_{cc}\to\Sigma^{*0}D^+\pi^+)-\mathcal{A}(\Xi^{++}_{cc}\to\Sigma^{*+}D^0\pi^+)\nonumber\\&~~~~~-\sqrt{2}\mathcal{A}(\Xi^{++}_{cc}\to\Sigma^{*+}D^+\pi^0)= 0,
\end{align}
\begin{align}
     { SumI_-}\,[\Xi^{+}_{cc},\Delta^{++},D^0,\overline{K}^0]&=-\mathcal{A}(\Xi^{+}_{cc}\to\Delta^{++}D^0{K}^-)-\mathcal{A}(\Xi^{++}_{cc}\to\Delta^{++}D^0\overline{K}^0)\nonumber\\&~~~~~+\sqrt{3}\mathcal{A}(\Xi^{+}_{cc}\to\Delta^+D^0\overline{K}^0)= 0,
\end{align}
\begin{align}
     { SumI_-}\,[\Xi^{++}_{cc},\Delta^{++},D^+,{K}^-]&=-\mathcal{A}(\Xi^{+}_{cc}\to\Delta^{++}D^0{K}^-)-\mathcal{A}(\Xi^{++}_{cc}\to\Delta^{++}D^+K^-)\nonumber\\&~~~~~+\sqrt{3}\mathcal{A}(\Xi^{+}_{cc}\to\Delta^+D^+{K}^-)= 0,
\end{align}
\begin{align}
     { SumI_-}\,[\Xi^{+}_{cc},\Delta^{++},D^0,\overline{K}^0]&=-\mathcal{A}(\Xi^{+}_{cc}\to\Delta^{++}D^0{K}^-)-\mathcal{A}(\Xi^{++}_{cc}\to\Delta^{++}D^0\overline{K}^0)\nonumber\\&~~~~~+\sqrt{3}\mathcal{A}(\Xi^{+}_{cc}\to\Delta^+D^0\overline{K}^0)= 0,
\end{align}
\begin{align}
     { SumI_-}\,[\Xi^{+}_{cc},\Delta^{+},D^+,\overline{K}^0]&=-2\mathcal{A}(\Xi^{+}_{cc}\to\Delta^{0}D^+\overline{K}^0)-\mathcal{A}(\Xi^{+}_{cc}\to\Delta^{+}D^0\overline{K}^0)\nonumber\\&~~~~~-\mathcal{A}(\Xi^{+}_{cc}\to\Delta^+D^+K^-)-\mathcal{A}(\Xi^{++}_{cc}\to\Delta^+D^+\overline{K}^0)= 0,
\end{align}
\begin{align}
    { SumI_-}\,[\Xi^{+}_{cc},\Sigma^{*+},D^0,\pi^+]&=\sqrt{2}\mathcal{A}(\Xi^{+}_{cc}\to\Sigma^{*0}D^0\pi^+)-\mathcal{A}(\Xi^{++}_{cc}\to\Sigma^{*+}D^0\pi^+)\nonumber\\&~~~~~-\sqrt{2}\mathcal{A}(\Xi^{+}_{cc}\to\Sigma^{*+}D^0\pi^0)= 0,
\end{align}
\begin{align}
    { SumI_-}\,[\Xi^{+}_{cc},\Sigma^{*+},D^+,\pi^0]&=\sqrt{2}\mathcal{A}(\Xi^{+}_{cc}\to\Sigma^{*0}D^+\pi^0)-\mathcal{A}(\Xi^{+}_{cc}\to\Sigma^{*+}D^0\pi^0)\nonumber\\&~~~~~+\sqrt{2}\mathcal{A}(\Xi^{+}_{cc}\to\Sigma^{*+}D^+\pi^-)-\mathcal{A}(\Xi^{++}_{cc}\to\Sigma^{*+}D^+\pi^0)= 0,
\end{align}
\begin{align}
    {SumI_-}\,[\Xi^{+}_{cc},\Sigma^{*+},D^+,\eta_8]&=-\mathcal{A}(\Xi^{++}_{cc}\to\Sigma^{*+}D^+\eta_8)-\mathcal{A}(\Xi^{+}_{cc}\to\Sigma^{*+}D^0\eta_8)\nonumber\\&~~~~~+\sqrt{2}\mathcal{A}(\Xi^{+}_{cc}\to\Sigma^{*0}D^+\eta_8)=0,
\end{align}
\begin{align}
     { SumI_-}\,[\Xi^{+}_{cc},\Sigma^{*0},D^+,\pi^+]&=-\mathcal{A}(\Xi^{+}_{cc}\to\Sigma^{*0}D^0\pi^+)-\mathcal{A}(\Xi^{++}_{cc}\to\Sigma^{*0}D^+\pi^+)\nonumber\\&~~~~~-\sqrt{2}\mathcal{A}(\Xi^{+}_{cc}\to\Sigma^{*0}D^+\pi^0)+\sqrt{2}\mathcal{A}(\Xi^{+}_{cc}\to\Sigma^{*-}D^+\pi^+)= 0,
\end{align}
\begin{align}
     { SumI_-}\,[\Xi^{+}_{cc},\Xi^{*0},D^+,K^+]&=-\mathcal{A}(\Xi^{++}_{cc}\to\Xi^{*0}D^+K^+)-\mathcal{A}(\Xi^{+}_{cc}\to\Xi^{*0}D^0K^+)\nonumber\\&~~~~~+\mathcal{A}(\Xi^{+}_{cc}\to\Xi^{*0}D^+K^0)+\mathcal{A}(\Xi^{+}_{cc}\to\Xi^{*-}D^+K^+)= 0,
\end{align}
\begin{align}
     { SumI_-}\,[\Xi^{+}_{cc},\Xi^{*0},D^+_s,\pi^+]&=-\mathcal{A}(\Xi^{++}_{cc}\to\Xi^{*0}D^+_s\pi^+)-\sqrt{2}\mathcal{A}(\Xi^{+}_{cc}\to\Xi^{*0}D^+_s\pi^0)\nonumber\\&~~~~~+\mathcal{A}(\Xi^{+}_{cc}\to\Xi^{*-}D^+_s\pi^+)=0,
\end{align}
\begin{align}
{ SumI_-}\,[\Xi^{+}_{cc},\Sigma^{*+},D^+_s,\overline{K}^0]&=-\mathcal{A}(\Xi^{++}_{cc}\to\Sigma^{*+}D^+_s\overline{K}^0)-\mathcal{A}(\Xi^{+}_{cc}\to\Sigma^{*+}D^+_s{K}^-)\nonumber\\&~~~~~+\sqrt{2}\mathcal{A}(\Xi^{+}_{cc}\to\Sigma^{*0}D^+_s\overline{K}^0)=0,
\end{align}
\begin{align}
{ SumI_-}\,[\Omega^{+}_{cc},\Sigma^{*+},D^+,\overline{K}^0]&=-\mathcal{A}(\Omega^{+}_{cc}\to\Sigma^{*+}D^0\overline{K}^0)-\mathcal{A}(\Omega^{+}_{cc}\to\Sigma^{*+}D^+{K}^-)\nonumber\\&~~~~~+\sqrt{2}\mathcal{A}(\Omega^{+}_{cc}\to\Sigma^{*0}D^+\overline{K}^0)=0,
\end{align}
\begin{align}
     { SumI_-}\,[\Omega^{+}_{cc},\Xi^{*0},D^+,\pi^+]&=-\mathcal{A}(\Omega^{+}_{cc}\to\Xi^{*0}D^0\pi^+)-\sqrt{2}\mathcal{A}(\Omega^{+}_{cc}\to\Xi^{*0}D^+\pi^0)\nonumber\\&~~~~~+\mathcal{A}(\Omega^{+}_{cc}\to\Xi^{*-}D^+\pi^+)=0,
\end{align}
\begin{align}
      { SumI_-^2}\,[\Xi^{+}_{cc},\Sigma^{*+},D^+,\pi^+]&=-2\big[\sqrt{2}\mathcal{A}(\Xi^+_{cc}\to\Sigma^{*0}D^0\pi^+)+\sqrt{2}\mathcal{A}(\Xi^{++}_{cc}\to\Sigma^{*0}D^+\pi^+)\nonumber\\&~~~~~+2\mathcal{A}(\Xi^+_{cc}\to\Sigma^{*0}D^+\pi^0)-\mathcal{A}(\Xi^{++}_{cc}\to\Sigma^{*+}D^0\pi^+)\nonumber\\&~~~~~-\sqrt{2}\mathcal{A}(\Xi^+_{cc}\to\Sigma^{*+}D^0\pi^0)-\mathcal{A}(\Xi^+_{cc}\to\Sigma^{*-}D^+\pi^+)\nonumber\\&~~~~~+\mathcal{A}(\Xi^+_{cc}\to\Sigma^{*+}D^+\pi^-)-\sqrt{2}\mathcal{A}(\Xi^{++}_{cc}\to\Sigma^{*+}D^+\pi^0)\big]=0,
\end{align}
\begin{align}
     { SumI_-^2}\,[\Xi^{++}_{cc},\Delta^{++},D^+,\pi^+]&=-2\big[ -\sqrt{3}\mathcal{A}(\Xi^{++}_{cc}\to\Delta^{0}D^+\pi^+)+\sqrt{3}\mathcal{A}(\Xi^{++}_{cc}\to\Delta^{+}D^0\pi^+)\nonumber\\&~~~~~-\sqrt{2}\mathcal{A}(\Xi^{++}_{cc}\to\Delta^{++}D^0\pi^0)+\mathcal{A}(\Xi^{++}_{cc}\to\Delta^{++}D^+\pi^-)\nonumber\\&~~~~~+\sqrt{6}\mathcal{A}(\Xi^{++}_{cc}\to\Delta^{+}D^+\pi^0)\big]= 0,
\end{align}
\begin{align}
    { SumI_-^2}\,[\Xi^{+}_{cc},\Delta^{++},D^0,\pi^+]&=-2\big[ -\sqrt{3}\mathcal{A}(\Xi^{+}_{cc}\to\Delta^{0}D^0\pi^+)+\sqrt{3}\mathcal{A}(\Xi^{++}_{cc}\to\Delta^{+}D^0\pi^+)\nonumber\\&~~~~~+\mathcal{A}(\Xi^{+}_{cc}\to\Delta^{++}D^0\pi^-)-\sqrt{2}\mathcal{A}(\Xi^{++}_{cc}\to\Delta^{++}D^0\pi^0)\nonumber\\&~~~~~+\sqrt{6}\mathcal{A}(\Xi^{+}_{cc}\to\Delta^{+}D^0\pi^0)\big]= 0,
\end{align}
\begin{align}
    { SumI_-^2}\,[\Xi^{+}_{cc},\Delta^{+},D^+,\pi^+]&=2\big[ -2\mathcal{A}(\Xi^{+}_{cc}\to\Delta^{0}D^0\pi^+)
    -2\mathcal{A}(\Xi^{++}_{cc}\to\Delta^{0}D^+\pi^+)\nonumber\\&~~~~~
    -2\sqrt{2}\mathcal{A}(\Xi^{+}_{cc}\to\Delta^{0}D^+\pi^0)
    +\mathcal{A}(\Xi^{++}_{cc}\to\Delta^{+}D^0\pi^+)\nonumber\\&~~~~~
    +\sqrt{2}\mathcal{A}(\Xi^{+}_{cc}\to\Delta^{+}D^0\pi^0)
    +\sqrt{3}\mathcal{A}(\Xi^{+}_{cc}\to\Delta^{-}D^+\pi^+)\nonumber\\&~~~~~
    -\mathcal{A}(\Xi^{+}_{cc}\to\Delta^{+}D^+\pi^-)
    +\sqrt{2}\mathcal{A}(\Xi^{++}_{cc}\to\Delta^{+}D^+\pi^0)\big]= 0,
\end{align}
\begin{align}
    { SumI_-^2}\,[\Xi^{+}_{cc},\Sigma^{*+},D^+,K^+]&=-2\big[ -\mathcal{A}(\Xi^{++}_{cc}\to\Sigma^{*+}D^0K^+)+\mathcal{A}(\Xi^{++}_{cc}\to\Sigma^{*+}D^+K^0)\nonumber\\&~~~~~+ \sqrt{2}\mathcal{A}(\Xi^{++}_{cc}\to\Sigma^{*0}D^+K^+)+\mathcal{A}(\Xi^{+}_{cc}\to\Sigma^{*+}D^0K^0)\nonumber\\&~~~~~+\sqrt{2} \mathcal{A}(\Xi^{+}_{cc}\to\Sigma^{*0}D^0K^+)-\sqrt{2}\mathcal{A}(\Xi^{+}_{cc}\to\Sigma^{*0}D^+K^0)\nonumber\\&~~~~~ -\mathcal{A}(\Xi^{+}_{cc}\to\Sigma^{*-}D^+K^+)\big]=0,
\end{align}
\begin{align}
    { SumI_-^2}\,[\Xi^{+}_{cc},\Sigma^{*+},D^+_s,\pi^+]&= -2\big[2\mathcal{A}(\Xi^{+}_{cc}\to\Sigma^{*0}D^+_s\pi^0)-\mathcal{A}(\Xi^{+}_{cc}\to\Sigma^{*-}D^+_s\pi^+)\nonumber\\&~~~~~+\mathcal{A}(\Xi^{+}_{cc}\to\Sigma^{*+}D^+_s\pi^-)+\sqrt{2}\mathcal{A}(\Xi^{++}_{cc}\to\Sigma^{*0}D^+_s\pi^+)\nonumber\\&~~~~~-\sqrt{2}\mathcal{A}(\Xi^{++}_{cc}\to\Sigma^{*+}D^+_s\pi^0)\big]= 0,
\end{align}
\begin{align}
     { SumI_-^2}\,[\Omega^{+}_{cc},\Sigma^{*+},D^+,\pi^+]&=  -2\big[2\mathcal{A}(\Omega^{+}_{cc}\to\Sigma^{*0}D^+\pi^0)-\mathcal{A}(\Omega^{+}_{cc}\to\Sigma^{*-}D^+\pi^+)\nonumber\\&~~~~~+\mathcal{A}(\Omega^{+}_{cc}\to\Sigma^{*+}D^+\pi^-)+\sqrt{2}\mathcal{A}(\Omega^{+}_{cc}\to\Sigma^{*0}D^0\pi^+)\nonumber\\&~~~~~-\sqrt{2}\mathcal{A}(\Omega^{+}_{cc}\to\Sigma^{*+}D^0\pi^0)\big]=0,
\end{align}
\begin{align}
     { SumI_-^2}\,[\Xi^{++}_{cc},\Delta^{++},D^+,K^+]&=2\big[\sqrt{3}\mathcal{A}(\Xi^{++}_{cc}\to\Delta^{0}D^+K^+)-\sqrt{3}\mathcal{A}(\Xi^{++}_{cc}\to\Delta^{+}D^0K^+)\nonumber\\&~~~~~-\mathcal{A}(\Xi^{++}_{cc}\to\Delta^{++}D^0K^0)+\sqrt{3}\mathcal{A}(\Xi^{++}_{cc}\to\Delta^{+}D^+K^0)\big]=0,
\end{align}
\begin{align}
    { SumI_-^2}\,[\Xi^{++}_{cc},\Delta^{++},D^+_s,\pi^+]&=2\sqrt{3}\mathcal{A}(\Xi^{++}_{cc}\to\Delta^{0}D^+_s\pi^+)-2\big[\mathcal{A}(\Xi^{++}_{cc}\to\Delta^{++}D^+_s\pi^-)\nonumber\\&~~~~~+\sqrt{6}\mathcal{A}(\Xi^{++}_{cc}\to\Delta^{+}D^+_s\pi^0)\big]=0,
\end{align}
\begin{align}
    { SumI_-^2}\,[\Xi^{+}_{cc},\Delta^{++},D^0,K^+]&=2\big[\sqrt{3}\mathcal{A}(\Xi^{+}_{cc}\to\Delta^{0}D^0K^+)-\sqrt{3}\mathcal{A}(\Xi^{++}_{cc}\to\Delta^{+}D^0K^+)\nonumber\\&~~~~~-\mathcal{A}(\Xi^{++}_{cc}\to\Delta^{++}D^0K^0)+\sqrt{3}\mathcal{A}(\Xi^{+}_{cc}\to\Delta^{+}D^0K^0)\big]=0,
\end{align}
\begin{align}
    { SumI_-^2}\,[\Xi^{+}_{cc},\Delta^{+},D^+,K^+]&=-2\big[2\mathcal{A}(\Xi^{+}_{cc}\to\Delta^{0}D^0K^+)+2\mathcal{A}(\Xi^{++}_{cc}\to\Delta^{0}D^+K^+)\nonumber\\&~~~~~-2\mathcal{A}(\Xi^{+}_{cc}\to\Delta^{0}D^+K^0)-\mathcal{A}(\Xi^{++}_{cc}\to\Delta^{+}D^0K^+)\nonumber\\&~~~~~+\mathcal{A}(\Xi^{+}_{cc}\to\Delta^{+}D^0K^0)-\sqrt{3}\mathcal{A}(\Xi^{+}_{cc}\to\Delta^{-}D^+K^+)\nonumber\\&~~~~~+\mathcal{A}(\Xi^{++}_{cc}\to\Delta^{+}D^+K^0)\big]=0,
\end{align}
\begin{align}
     { SumI_-^2}\,[\Xi^{+}_{cc},\Delta^{+},D^+_s,\pi^+]&=-2\big[2\mathcal{A}(\Xi^{++}_{cc}\to\Delta^{0}D^+_s\pi^+)+2\sqrt{2}\mathcal{A}(\Xi^{+}_{cc}\to\Delta^{0}D^+_s\pi^0)\nonumber\\&~~~~~-\sqrt{3}\mathcal{A}(\Xi^{+}_{cc}\to\Delta^{-}D^+_s\pi^+)+\mathcal{A}(\Xi^{+}_{cc}\to\Delta^{+}D^+_s\pi^-)\nonumber\\&~~~~~-\sqrt{2}\mathcal{A}(\Xi^{++}_{cc}\to\Delta^{+}D^+_s\pi^0)\big]=0,
\end{align}
\begin{align}
 { SumI_-^2}\,[\Xi^{+}_{cc},\Sigma^{*+},D^+_s,K^+]&= 2\big[-\mathcal{A}(\Xi^{++}_{cc}\to\Sigma^{*+}D^+_sK^0)-\sqrt{2}\mathcal{A}(\Xi^{++}_{cc}\to\Sigma^{*0}D^+_sK^+)\nonumber\\&~~~~~+\sqrt{2}\mathcal{A}(\Xi^{+}_{cc}\to\Sigma^{*0}D^+_sK^0)+\mathcal{A}(\Xi^{+}_{cc}\to\Sigma^{*-}D^+_sK^+)\big]= 0,
 \end{align}
\begin{align}
  { SumI_-^2}\,[\Omega^{+}_{cc},\Delta^{++},D^0,\pi^+]&=  2\sqrt{3}\mathcal{A}(\Omega^{+}_{cc}\to\Delta^{0}D^0\pi^+)-2\mathcal{A}(\Omega^{+}_{cc}\to\Delta^{++}D^0\pi^-)\nonumber\\&~~~~~-2\sqrt{6}\mathcal{A}(\Omega^{+}_{cc}\to\Delta^{+}D^0\pi^0)=0,
\end{align}
\begin{align}
  { SumI_-^2}\,[\Omega^{+}_{cc},\Delta^{+},D^+,\pi^+]&= - 2\big[2\mathcal{A}(\Omega^{+}_{cc}\to\Delta^{0}D^0\pi^+)+2\sqrt{2}\mathcal{A}(\Omega^{+}_{cc}\to\Delta^{0}D^+\pi^0)\nonumber\\&~~~~~-\sqrt{2}\mathcal{A}(\Omega^{+}_{cc}\to\Delta^{+}D^0\pi^0)+\mathcal{A}(\Omega^{+}_{cc}\to\Delta^{+}D^+\pi^-) \nonumber\\&~~~~~-\sqrt{3}\mathcal{A}(\Omega^{+}_{cc}\to\Delta^{-}D^+\pi^+) \big]=0
\end{align}
\begin{align}
  { SumI_-^2}\,[\Omega^{+}_{cc},\Sigma^{*+},D^+,K^+]&=  2\big[-\mathcal{A}(\Omega^{+}_{cc}\to\Sigma^{*+}D^0K^0)-\sqrt{2}\mathcal{A}(\Omega^{+}_{cc}\to\Sigma^{*0}D^0K^+)\nonumber\\&~~~~~+\sqrt{2}\mathcal{A}(\Omega^{+}_{cc}\to\Sigma^{*0}D^+K^0)+\mathcal{A}(\Omega^{+}_{cc}\to\Sigma^{*-}D^+K^+)\big]=0,
\end{align}
\begin{align}
  { SumI_-^2}\,[\Omega^{+}_{cc},\Sigma^{*+},D^+_s,\pi^+]&=  2\mathcal{A}(\Omega^{+}_{cc}\to\Sigma^{*-}D^+_s\pi^+)-2\mathcal{A}(\Omega^{+}_{cc}\to\Sigma^{*+}D^+_s\pi^-)\nonumber\\&~~~~~-2\mathcal{A}(\Omega^{+}_{cc}\to\Sigma^{*0}D^+_s\pi^0)=0,
\end{align}
\begin{align}
     { SumI_-^3}\,[\Xi^{+}_{cc},\Delta^{++},D^+,\pi^+]&=6\big[-\sqrt{3}\mathcal{A}(\Xi^{+}_{cc}\to\Delta^{0}D^0\pi^+)-\sqrt{3}\mathcal{A}(\Xi^{++}_{cc}\to\Delta^{0}D^+\pi^+)\nonumber\\&~~~~~-\sqrt{6}\mathcal{A}(\Xi^{+}_{cc}\to\Delta^{0}D^+\pi^0)+\sqrt{3}\mathcal{A}(\Xi^{++}_{cc}\to\Delta^{+}D^0\pi^+)\nonumber\\&~~~~~+\mathcal{A}(\Xi^{+}_{cc}\to\Delta^{++}D^0\pi^-)-\sqrt{2}\mathcal{A}(\Xi^{++}_{cc}\to\Delta^{++}D^0\pi^0)\nonumber\\&~~~~~+\sqrt{6}\mathcal{A}(\Xi^{+}_{cc}\to\Delta^{+}D^0\pi^0)+\mathcal{A}(\Xi^{++}_{cc}\to\Delta^{++}D^+\pi^-)\nonumber\\&~~~~~+\mathcal{A}(\Xi^{+}_{cc}\to\Delta^{-}D^+\pi^+)-\sqrt{3}\mathcal{A}(\Xi^{+}_{cc}\to\Delta^{+}D^+\pi^-)\nonumber\\&~~~~~+\sqrt{6}\mathcal{A}(\Xi^{++}_{cc}\to\Delta^{+}D^+\pi^0)\big]=0,
\end{align}
\begin{align}
     { SumI_-^3}\,[\Xi^{+}_{cc},\Delta^{++},D^+,K^+]&=6\big[-\sqrt{3}\mathcal{A}(\Xi^{+}_{cc}\to\Delta^{0}D^0K^+)-\sqrt{3}\mathcal{A}(\Xi^{++}_{cc}\to\Delta^{0}D^+K^+)\nonumber\\&~~~~~+\sqrt{3}\mathcal{A}(\Xi^{+}_{cc}\to\Delta^{0}D^+K^0)+\sqrt{3}\mathcal{A}(\Xi^{++}_{cc}\to\Delta^{+}D^0K^+)\nonumber\\&~~~~~+\mathcal{A}(\Xi^{++}_{cc}\to\Delta^{++}D^0K^0)-\sqrt{3}\mathcal{A}(\Xi^{+}_{cc}\to\Delta^{+}D^0K^0)\nonumber\\&~~~~~+\mathcal{A}(\Xi^{+}_{cc}\to\Delta^{-}D^+K^+)-\sqrt{3}\mathcal{A}(\Xi^{++}_{cc}\to\Delta^{+}D^+K^0)\big]=0,
\end{align}
\begin{align}
     { SumI_-^3}\,[\Xi^{+}_{cc},\Delta^{++},D^+_s,\pi^+]&=6\big[-\sqrt{3}\mathcal{A}(\Xi^{++}_{cc}\to\Delta^{0}D^+_s\pi^+)-\sqrt{6}\mathcal{A}(\Xi^{+}_{cc}\to\Delta^{0}D^+_s\pi^0)\nonumber\\&~~~~~+\mathcal{A}(\Xi^{++}_{cc}\to\Delta^{++}D^+_s\pi^-)+\mathcal{A}(\Xi^{+}_{cc}\to\Delta^{-}D^+_s\pi^+)\nonumber\\&~~~~~-\sqrt{3}\mathcal{A}(\Xi^{+}_{cc}\to\Delta^{+}D^+_s\pi^-) +\sqrt{6}\mathcal{A}(\Xi^{++}_{cc}\to\Delta^{+}D^+_s\pi^0)\big]=0,
\end{align}
\begin{align}
     { SumI_-^3}\,[\Omega^{+}_{cc},\Delta^{++},D^+,\pi^+]&=6\big[-\sqrt{3}\mathcal{A}(\Omega^{+}_{cc}\to\Delta^{0}D^0\pi^+)-\sqrt{6}\mathcal{A}(\Omega^{+}_{cc}\to\Delta^{0}D^+\pi^0)\nonumber\\&~~~~~+\mathcal{A}(\Omega^{+}_{cc}\to\Delta^{++}D^0\pi^-)+\sqrt{6}\mathcal{A}(\Omega^{+}_{cc}\to\Delta^{+}D^0\pi^0)\nonumber\\&~~~~~+\mathcal{A}(\Omega^{+}_{cc}\to\Delta^{-}D^+\pi^+)-\sqrt{3}\mathcal{A}(\Omega^{+}_{cc}\to\Delta^{+}D^+\pi^-)\big]=0.
\end{align}

\end{appendix}

\end{document}